\documentclass[aps,prl,twocolumn]{revtex4-2}
\usepackage{graphicx,color}
\usepackage{amssymb}   % for math
\usepackage{amsmath}
\DeclareMathOperator{\sgn}{sgn}
\usepackage{epstopdf}
\usepackage{natbib}
\usepackage{multirow}
\usepackage[colorlinks,citecolor=green,urlcolor=blue,bookmarks=false,hypertexnames=true]{hyperref} 
\usepackage{bm}
\usepackage{color}
\usepackage{braket}

\begin{document}
\title{Non-Abelian topological superconductivity 
in maximally twisted double-layer spin-triplet valley-singlet superconductors}
\author{Benjamin T. Zhou} \thanks{Corresponding author: benjamin.zhou@ubc.ca}
\author{Shannon Egan} 
\author{Dhruv Kush}
\author{Marcel Franz} \thanks{Corresponding author: franz@phas.ubc.ca}

\affiliation{Department of Physics and Astronomy \& Stewart Blusson Quantum Matter Institute,
University of British Columbia, Vancouver BC, Canada V6T 1Z4}

\begin{abstract}
Recent theoretical and experimental studies point to a novel spin-triplet valley-singlet (STVS) superconducting phase in certain two-valley electron liquids, including rhombohedral trilayer graphene, Bernal bilayer graphene and ZrNCl. This fully gapped phase is exotic in that it combines into Cooper pairs same-spin electrons from valleys centered around the opposing corners of a hexagonal Brillouin zone, but is, nevertheless, topologically trivial.  Here, we predict that upon stacking two layers of an STVS material with an angular twist, a novel chiral topological phase -- an $f \pm if'$-wave superconductor -- emerges in the vicinity of the `maximal' twist angle of 30$^{\circ}$ where the system becomes an extrinsic quasi-crystal with 12-fold tiling. The resulting composite is a non-Abelian topological superconductor (TSC) with an odd number of chiral Majorana modes at its edges and a single Majorana zero mode (MZM) localized in the vortex core. Through symmetry analysis and detailed microscopic modelling based on a novel quasi-crystal band structure technique, we demonstrate that the non-Abelian TSC forms when the isolated Fermi pockets coalesce into a single connected Fermi surface around the center of the moir\'{e} Brillouin zone and is stable over a wide range of electron density. We further discuss how the energetics leading to the $f \pm if'$-wave phase results in anomalous $\pi$-periodic inter-layer Josephson effect, which can serve as a distinctive signature of the chiral phase. Distinct from the valley-preserving moir\'{e} physics in small-angle twisted graphene, our results establish the  large-angle moir\'{e} physics arising near maximal twist as a new avenue toward intrinsic TSC with non-Abelian excitations.
\end{abstract}

\maketitle

\section*{Introduction}

Being of fundamental interest and potential use for topological qubits, the search for topological superconductors hosting excitations with non-Abelian exchange statistics has been one of the central topics in condensed matter physics over the past two decades \cite{Wilczek, Alicea1, Beenakker, Marcel1, Sarma}. The simplest such particles -- Majorana zero modes (MZMs) -- were originally proposed as vortex core states in chiral $p$-wave superconductors \cite{Read&Green, Ivanov_2001, Volovik_1999}, but the lack of intrinsic $p$-wave superconductivity in nature has motivated worldwide effort to engineer synthetic platforms that emulate this behavior using more conventional ingredients \cite{Fu&Kane, Sau, Oreg, Alicea2, Kouwenhoven, Qi, He, Yazdani1, Wang}. Whether such effective $p$-wave superconductors and MZMs have been realized in recent experiments is still under debate \cite{JieLiu, HaoZhang, Kayyalha}. The quest for an intrinsic topological superconductor with non-Abelian excitations, on the other hand, remains an ongoing grand challenge to the condensed matter community.

Motivated by recent developments in twisted van der Waals materials \cite{BM, Cao1, Cao2, Wu, Yankowitz, Sharpe, Young1, Yazdani2, Andrei}, a new route toward topological superconductivity (TSC) has been proposed recently which takes two monolayers of high-$T_c$ cuprate superconductor with nodal $d_{x^2-y^2}$ pairing symmetry, such as Bi$_2$Sr$_2$CaCu$_2$O$_{8+\delta}$, stacked with a relative angular twist $\theta$ \cite{Marcel2, Pixley}. At $\theta \simeq 45^{\circ}$, the bilayer is predicted to enter a fully-gapped, topological chiral $d \pm id'$ phase with spontaneously broken time-reversal symmetry $\mathcal{T}$, which has been tentatively identified in a recent experimental study \cite{Kim0}. Despite its promise for realizing high-$T_c$ TSC, a chiral $d\pm id'$ superconductor cannot host truly non-Abelian excitations due to its spin-singlet pairing nature, which is always associated with an even number of MZMs that combine to form usual Abelian fermions. Generalizing the scheme to spin-triplet superconductors with nodal $p$-wave or $f$-wave order parameters one can create chiral SC phases with an odd number of MZMs in principle \cite{Tarun0}, while once again facing the scarcity of nodal $p$-wave and $f$-wave superconductors in nature.

\begin{figure}
\centering
\includegraphics[width=0.5\textwidth]{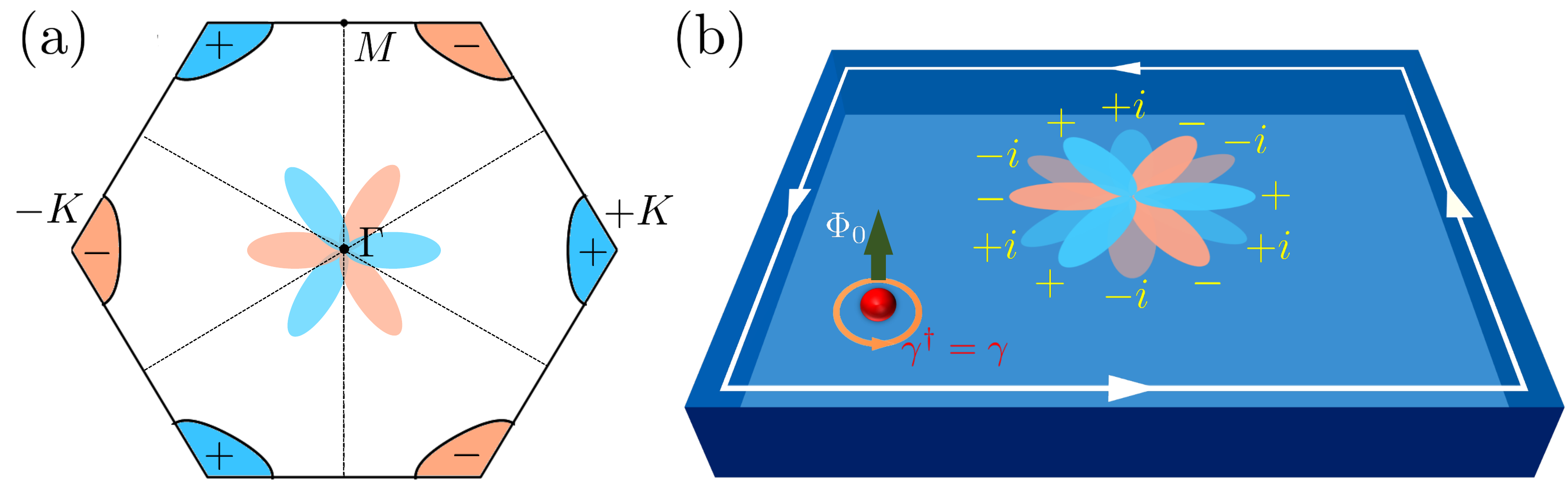}
\caption{(a) Structure of the $f_{x(x^2-3y^2)}$-wave order parameter in the spin-triplet valley-singlet (STVS) superconductor. Gap nodes along $\Gamma-M$  (dashed lines) are avoided by disconnected Fermi pockets around $\pm K$ resulting in a fully-gaped phase. (b) Schematic of a chiral $f \pm if'$ phase formed by twisting two layers of STVS superconductors at $\theta \simeq 30^{\circ}$. Chiral Majorana edge modes (white arrows) emerge on the edge and a single Majorana zero mode (red dot) forms at the vortex core. }
\label{FIG1}
\end{figure}

Recent progress in superconducting two-dimensional materials, however, has uncovered a growing amount of evidence for $f$-wave spin-triplet superconductivity, albeit in the disguise of a fully gapped phase: in a recent experiment, rhombohedral trilayer graphene (RTG) was found to superconduct in two different gate-tuned regions, where the peculiar SC2 superconducting phase was borne out of a spin-polarized, valley-unpolarized normal metal \cite{Young2}. Such an unusual normal-state fermiology strongly hints at its spin-triplet pairing nature, which is further supported by the observation of an in-plane critical field that far exceeds the Pauli paramagnetic limit. Similar results have also been reported in Bernal bilayer graphene (BBG) \cite{Young3}, and both experimental observations were interpreted theoretically as a signature of spin-triplet $f$-wave pairing \cite{Chou1, Chou2}. More recently, Cr\'{e}pel and Fu proposed that a novel `three-particle' mechanism involving virtual excitons generically gives rise to spin-triplet $f$-wave pairing in a two-valley electron liquid formed in doped insulators such as ZrNCl \cite{Crepel}, which provides a plausible explanation for the puzzling doping dependence of the gap structure revealed by early specific heat measurements on Li-doped ZrNCl \cite{Iwasa0}. 

In the scenarios described above, the parent normal-state Fermi surface (FS) consists of disconnected pockets enclosing the $+K$ and $-K$ corners of the hexagonal Brillouin zone (BZ), as illustrated in Fig.\ \ref{FIG1}a. Upon pairing electrons of the same spin and from opposite valleys, fermion exchange statistics require the order parameter to be odd under exchange of the valleys, which entails a spin-triplet valley-singlet (STVS) pairing. As shown schematically in Fig.\ \ref{FIG1}a, such an STVS superconductor has exactly the $f_{x(x^2-3y^2)}$-wave symmetry, while the excitation spectrum exhibits a full superconducting gap because the nodes of the $f$-wave gap function (located along $\Gamma-M$ lines) never intersect the disconnected FS. The gapped phase respects a spinless time reversal symmetry ${\cal T}'$ and particle-hole symmetry ${\cal P}$ (such that ${\cal T}'^2={\cal P}^2=+1$) and thus belongs to symmetry class BDI in Altland-Zirnbauer classification \cite{Altland-Zirnbauer}. In two space dimensions BDI class admits only trivial topology implying that the STVS superconductor is topologically trivial.

Here, we show that stacking two layers of STVS superconductor with an angular twist $\theta$ close to $30^{\circ}$ creates an intrinsic chiral $f \pm if'$-wave topological superconductor. This ${\cal T}'$-broken phase belongs to symmetry class D and admits nontrivial topology indicated by integer-valued Chern number $C$.
Our results, based on symmetry analysis and detailed microscopic modelling, show that for $\theta \simeq 30^{\circ}\pm 0.3^{\circ}$, the chiral $f\pm if'$ phase occurs robustly throughout a wide range of electron density and, at exactly $30^{\circ}$ twist where the system has a high 12-fold quasi-crystalline symmetry, extends up to the native critical temperature of the double-layer superconductor.  Within the chemical potential range where the disconnected $K$-pockets from the two layers merge into a single FS, we find $C=\pm 3$, indicating non-Abelian topology manifested through an odd number of chiral Majorana modes on its edge and a single MZM in the core of its superconducting vortex (Fig.\ \ref{FIG1}b). As $\theta$ deviates from $30^{\circ}$, the chiral topological phase evolves into a nodal topological $f_{x(x^2-3y^2)}$-wave superconductor, in which nodes of opposite chiralities in the bulk are connected by non-dispersive MZMs on one of the system edges, analogous to flat bands present on zigzag edges of monolayer graphene \cite{Fujita, Montambaux}. 

The emergence of non-Abelian $f \pm if'$-wave superconductivity in maximally twisted STVS superconductors, while sharing some similarities with twisted cuprates in its appearance, arises from an essentially different microscopic mechanism. One key difference lies in the absence of Dirac nodes in STVS superconductors due to small isolated FSs around $K$-points illustrated in Fig.\ \ref{FIG1}a. 
% prevents one from arguing for a $\mathcal{T}$-broken phase based on the Bogoliubov excitation spectrum as in twisted cuprates. 
In addition, due to formation of moir\'{e} patterns upon angular twist, the isolated FSs at zone boundaries are subject to significant reconstruction from band folding effects, which demands a new low-energy description. These features stand in sharp contrast to the situation in cuprates, where the initial {\em large and singly connected} FSs overlap strongly in a twisted bilayer and moir\'{e} effects are inessential \cite{Marcel2}. Finally, as we argue in the following, the existence of the non-Abelian TSC in bilayer STVS superconductors rests crucially upon the new type of {\em large-angle moir\'{e} physics} at maximal twist, which violates the $U_{v}(1)$ valley conservation symmetry that underpins the validity of the well-studied models of all small-twist-angle  two-valley systems. In particular, the novel large-angle moir\'{e} physics plays two important roles in creating the non-Abelian TSC: It first down-folds the isolated $K$-pockets into {\em a single connected} FS in the moir\'{e} BZ, and then reconstructs the pairing interactions in the twisted double-layer such that two orthogonal nodal $f$-wave order parameters are recovered in the moir\'{e} BZ to form the basis of the chiral $f \pm if'$ phase. Our results thus establish the novel large-twist-angle moir\'{e} physics, which is absent in twisted cuprates and is also fundamentally different from the small-angle moir\'{e} physics in magic-angle twisted graphene (see comparison in Supplementary Note 1), as a new route toward non-Abelian TSC.

\section*{Normal-state fermiology}

\begin{figure*}
\centering
\includegraphics[width=\textwidth]{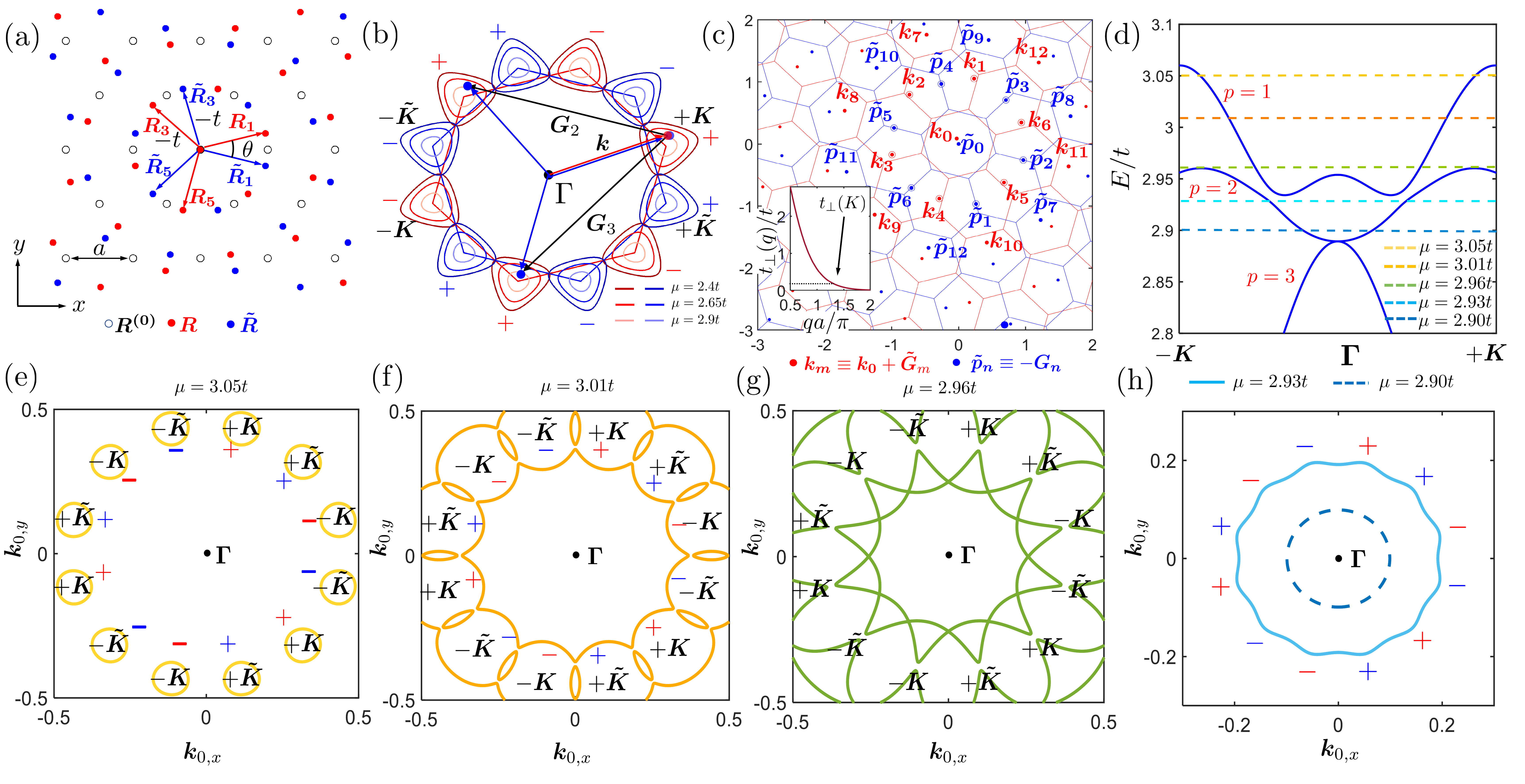}
\caption{(a) Real-space triangular lattices for two layers of STVS superconductors stacked with angular twist $\theta$. Dots in red (blue) denote lattice sites in layer 1 (2) with $a = 2.46 {\AA}$. Hopping parameter $t = 1$ eV in all figures. (b) Fermi surface (FS) contours of two decoupled layers at $\theta =30^{\circ}$ for chemical potential $2.4t < \mu < 3t$. $\bm{G}_2, \bm{G}_3$ denote reciprocal lattice vectors in layer 1. (c) Dual-momentum space lattice points $\bm{k}_m$ (red dots) and $\bm{\tilde{p}}_n$ (blue dots) for a fixed $\bm{k}_0$ in units of $2\pi/a$ (see subsection A in Methods). Red (blue) hexagons denote Wigner-Seitz cells in reciprocal lattice of layer 1 (2). Note that the original location of each $\bm{k}_m$ in the first Brillouin zone is exactly the location measured from the center of the red hexagon that contains it. Bloch momenta near $\pm K$ and $\pm \tilde{K}$ in (b) are covered by $m,n=1,2,...,6$ (encircled dots) as $\bm{k}_0$ varies in momentum-space. Inset: Fourier transform $t_{\perp}(\bm{q})$ of inter-layer coupling, $t_{\perp}(K)$ extrapolated at $Ka/\pi = 4/3$. (d) Band structure of $\mathcal{H}_{0, \rm eff}$ with topmost bands indexed by $p=1,2,3$. Dashed lines indicate chemical potentials of FS contours shown in panels (e-h). Units of ${k}_{0,x}, {k}_{0,y}$ axes are in ${\AA}^{-1}$. Red (blue) $\pm$ symbols indicate signs of pairing in layer 1 (2). }
\label{FIG2}
\end{figure*}

For the sake of concreteness and simplicity, we describe the normal state of a monolayer STVS superconductor by a triangular lattice tight-binding model with nearest-neighbor electron hopping $-t$ (Fig.\ \ref{FIG2}a). Such Hamiltonians are widely used to model systems with hexagonal symmetry whose FS consist of disconnected segments around $K$-points \cite{Crepel, Benjamin1} shown in Fig.\ \ref{FIG2}b. Motivated by the phenomenology observed in RTG and BBG systems \cite{Young2, Young3} and by theoretical ideas introduced in Ref.\ \cite{Crepel},  we focus here on equal-spin pairing between electrons belonging to opposite valleys and drop the spin index in our discussions. Moreover, we consider hole doping near $K$-points by setting $t>0$ with band maxima located at $\pm K$ given by $E_{\rm max} = 3t$ in each isolated layer (the case of electron doping can be covered by setting $t<0$ with $E_{\rm min} = 3t$). Throughout this work, we follow the convention that $\bm{k}$, $\bm{R}$, $\bm{G}$ respectively denote the Bloch momenta, real-space lattice vectors, and reciprocal lattice vectors in layer 1, while $\bm{\tilde{k}}$, $\bm{\tilde{R}}$, $\bm{\tilde{G}}$ stand for their counterparts in layer 2. Given lattice vectors $\bm{R}^{(0)}$ and reciprocal lattice vectors $\bm{G}^{(0)}$ in an unrotated triangular lattice (open circles in Fig.\ \ref{FIG2}a), we have $\bm{R} = R_z(\theta/2) \bm{R}^{(0)}$, $\bm{\tilde{R}} = R_z(-\theta/2) \bm{R}^{(0)}$ and $\bm{G} = R_z(\theta/2) \bm{G}^{(0)}$, $\bm{\tilde{G}} = R_z(-\theta/2) \bm{G}^{(0)}$, with $R_z(\theta)$ being the rotation of $\theta$ about the $z$-axis.

Upon an angular twist $\theta$, the normal-state Hamiltonians of the two decoupled layers are given by
\begin{equation}
    \begin{aligned}\label{eq:h0}
\mathcal{H}^{(1)}_{0} &= \sum_{\bm{k}} \xi_1 (\bm{k}) c_1^{\dagger}(\bm{k}) c_1(\bm{k}), \\
\mathcal{H}^{(2)}_{0} &= \sum_{\bm{\tilde{k}}} \xi_2 (\bm{\tilde{k}}) c_2^{\dagger}(\bm{\tilde{k}}) c_2(\bm{\tilde{k}}),
    \end{aligned}
\end{equation}
where $\xi_1(\bm{k}) = -2t \sum_{j=1,3,5} \cos(\bm{k}\cdot\bm{R}_j) -\mu$ is the kinetic energy term in layer 1, and $\xi_2(\bm{\tilde{k}}) = -2t \sum_{j=1,3,5} \cos(\bm{\tilde{k}}\cdot\bm{\tilde{R}}_j) -\mu$ in layer 2, $\mu$ is the chemical potential. For $\mu \in (2t, 3t)$, the triangular lattice model produces disconnected Fermi pockets around $K$ and $-K$ points  shown in Fig.\ \ref{FIG2}b.

Studies of twisted 2D materials have established that the interlayer coupling within a twisted bilayer structure has the general form \cite{BM, Wu, Mele, MKS, Benjamin2} 
\begin{equation}
\begin{aligned}\label{eq:Interlayer}
\mathcal{H}_{T} &= \sum_{\bm{k},\bm{\tilde{k}}} \left[c^{\dagger}_{1}(\bm{k}) T (\bm{k},\bm{\tilde{k}}) c_{2}(\bm{\tilde{k}}) + {\rm h.c.}\right] \\
T(\bm{k},\bm{\tilde{k}}) &=  -\sum_{\bm{G}, \bm{\tilde{G}}} t_{\perp} (\bm{k}+\bm{G}) \delta_{\bm{k}+\bm{G}, \bm{\tilde{k}}+\bm{\tilde{G}}},
\end{aligned}
\end{equation}
where $t_{\perp}(\bm{q})$ is the Fourier transform of the inter-layer coupling $t_{\perp}(\bm{r})$ as a function of spatial separation $\bm{r}$ between two atomic positions in different layers. Importantly, $t_{\perp}(\bm{q})$ decays rapidly as a function of $|\bm{q}|$ in general and becomes negligibly small on the scale of $q \simeq 2\pi/a$  where $a$ denotes the monolayer lattice constant. To be concrete, we model $t_{\perp}(\bm{r})$ by an empirical exponential formula describing $\sigma$-bonds formed by $p_z$-orbitals from the two layers (see Supplementary Note 2) and extrapolate an effective interlayer coupling strength of $t_{\perp}(K) \simeq 0.15t$ for states near $K$-points (inset of Fig.\ \ref{FIG2}c).

In twisted cuprates, superconductivity is borne out of large Fermi surfaces and Dirac nodes of the $d$-wave order parameter are located well inside the Brillouin zone of each layer. The leading-order inter-layer coupling according to Eq.\ \ref{eq:Interlayer} is simply the momentum-preserving term with $\bm{k} = \bm{\tilde{k}}$ and $\bm{G} =\bm{\tilde{G}} = 0$. This allows treating the inter-layer coupling as a constant in the continuum model in which moir\'{e} effects are inessential \cite{Marcel2}. In contrast, superconductivity in an STVS superconductor emerges from two disconnected pockets surrounding the $+K$ and $-K$ points, and, as indicated in Fig.\ \ref{FIG2}b, the leading-order inter-layer terms for a Bloch state with momentum $\bm{k} \simeq K$ in layer 1 include three different processes connecting it to states in layer 2 at $\bm{\tilde{k}} = (\bm{k},\bm{k} + \bm{G}_2,\bm{k} + \bm{G}_3$). 
The momentum transfer processes above are central ingredients in the celebrated Bistrizer-MacDonald (BM) model of small-angle twisted bilayer graphene \cite{BM} and its variants for other twisted materials with two $K$-valleys \cite{Wu, Benjamin2}, where the two valleys are modeled separately by valley-dependent low-energy effective Hamiltonians. However, the BM-type continuum model designed for the small twist-angle limit will fail to describe the maximally twisted double-layer STVS superconductor. This is because for $\theta \simeq 30^{\circ}$, the three different momenta $\bm{\tilde{k}}$ in layer 2 are located midway between the $+\tilde{K}$ and $-\tilde{K}$ points (Fig.\ \ref{FIG2}b), where the low-energy Hamiltonian defined for a single valley is no longer valid. 

To overcome this difficulty we apply a novel method (see subsection A of the Methods section) based on the dual momentum-space tight-binding (DMSTB) model introduced in a recent theoretical study of quasi-crystalline electronic bands in 30$^{\circ}$ twisted bilayer graphene \cite{MKS}. This approach was motivated and validated by experimental work \cite{Yao, Pezzini}. 
The authors showed that stacking two identical layers with honeycomb lattice geometries at an exact 30$^{\circ}$ twist results in an extrinsic quasi-crystal \cite{Note1} with 12-fold tiling but no exact crystalline symmetries. They further argued that owing to the limited number of leading-order inter-layer processes discussed above, it is possible to construct an effective momentum-space Hamiltonian of a relatively small size.  We adapted this method to our triangular lattice geometry to construct an effective Hamiltonian $\mathcal{H}_{0, \rm eff}$ based on the \textit{dual momentum-space lattice sites} $\bm{k}_{m=0,...,12}$ and $\bm{\tilde{p}}_{n=0,...,12}$ (see Methods section and Supplementary Note 3 for details) shown in Fig.\ \ref{FIG2}c and generalized it to arbitrary twist angles close to 30$^{\circ}$. Within this description moir\'{e} bands due to the large angular twist $\theta \simeq 30^{\circ}$ can be well defined up to the leading-order approximation, with $\bm{k}_0$ serving as the momentum in the moir\'{e} Brillouin zone.

The moir\'{e} bands of $\mathcal{H}_{0, \rm eff}$ at $\theta = 30^{\circ}$ are solved by exact numerical diagonalization (topmost bands labelled by $p=1,2,3$ shown in Fig.\ \ref{FIG2}d), and the evolution of the Fermi surface upon increasing the doping level is presented in Fig.\ \ref{FIG2}e-h. Due to level repulsion caused by inter-layer coupling, the band maxima at all $K$-points are shifted to $E_{\rm max} \simeq 3.06 t$. At light hole doping $\mu > 3.01 t$, the Fermi surface consists of 12 disconnected pockets stemming from the 12 dual momentum-space sites in Fig.\ \ref{FIG2}c, and resembles the disconnected Fermi surfaces in the decoupled limit (Fig.\ \ref{FIG2}b). It is important to note that due to the moir\'{e} band-folding effects introduced by large angle twist, the Fermi surface in the twisted double layer quickly undergoes a Liftshitz transition as doping level increases and becomes connected already at $\mu \simeq 3.01t$ (Fig.\ \ref{FIG2}f), at which point the Fermi pockets in the decoupled limit would still remain well isolated (Fig.\ \ref{FIG2}b). 

Upon further doping, the Fermi surface undergoes a second Liftshitz transition at $\mu \simeq 2.96t$ (Fig.\ \ref{FIG2}g) and the system enters a regime with a {\em single connected Fermi surface} centered at the $\Gamma$ point (Fig.\ \ref{FIG2}h). Such FS then remains stable over a wide range of chemical potentials with higher hole doping (Fig.\ \ref{FIG2}d). Crucially, as we show in the next section, at doping levels where a single connected FS exists, the twisted double-layer STVS material at $\theta \simeq 30^{\circ}$ becomes an intrinsic chiral $f\pm if'$ superconductor with non-Abelian excitations.\\

\section*{Chiral $f \pm if'$-wave superconductivity at $\theta \simeq 30^{\circ}$}
\subsection*{A. Microscopic model}

\begin{figure*}
\centering
\includegraphics[width=\textwidth]{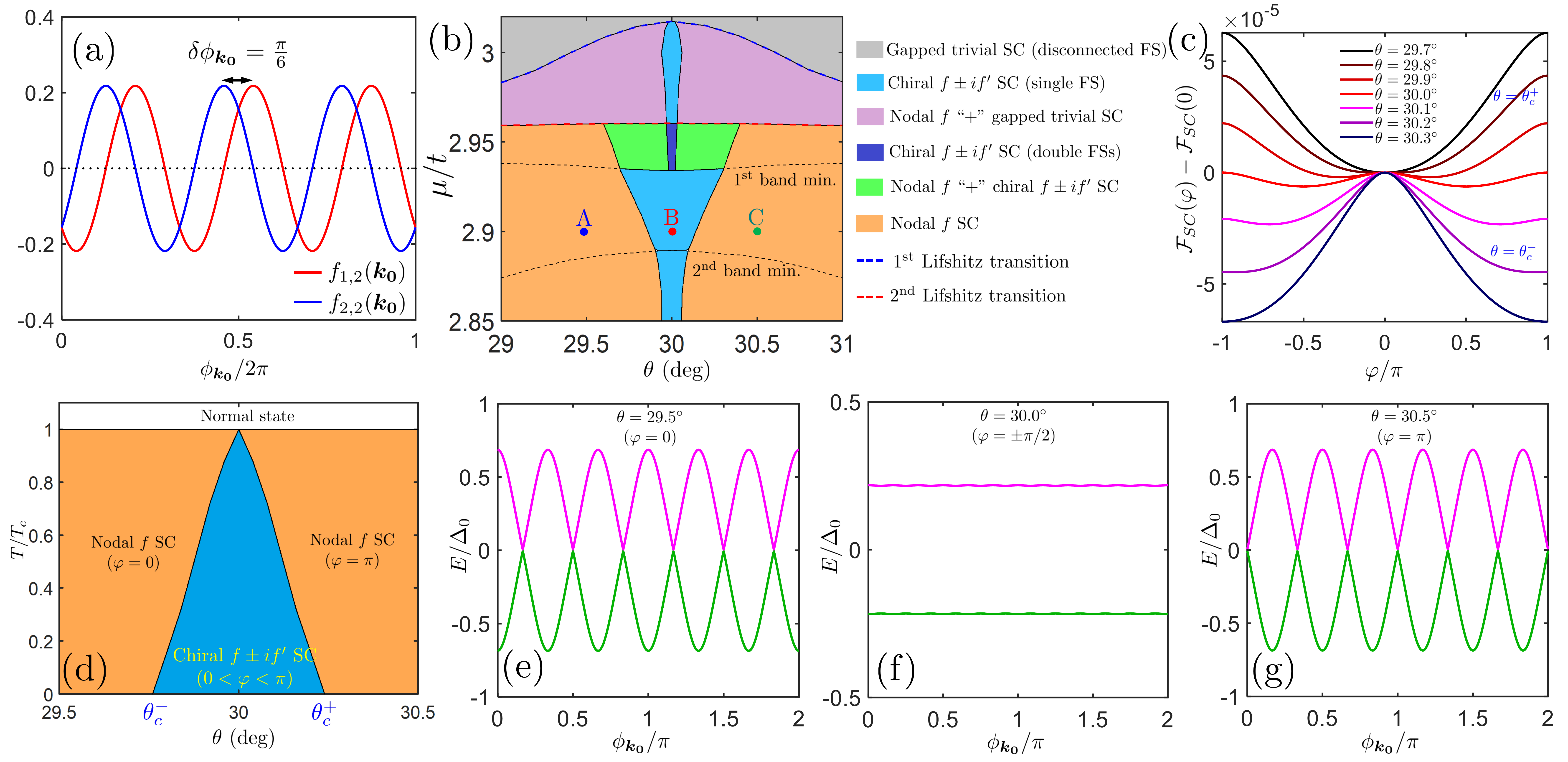}
\caption{Chiral $f\pm if'$ superconductivity at 30$^{\circ}$ twist. (a) Basis functions $f_{1,p}(\bm{k}_0), f_{2,p}(\bm{k}_0)$ of the projected pair wavefunctions in band $p=2$ along the circular Fermi surface at $\mu = 2.9t$ with $|\bm{k}_0|\simeq 0.11$ {\AA}$^{-1}$ shown in Fig.\ \ref{FIG2}d. $\phi_{\bm{k}_0}$ is the polar angle of $\bm{k}_0$ in the 2D plane. (b) Phase diagram of a twisted double-layer STVS superconductor in the $\mu$-$\theta$ plane. A robust chiral $f \pm if'$ phase (regions depicted in blue) is found over the entire chemical potential range for $\theta \simeq 30^{\circ}$. (c) Evolution of the free energy landscape as a function of $\theta$ at $\mu = 2.9t$. Unit of $y$-axis set in eV. (d) Phase diagram in the $T$-$\theta$ plane obtained at $\mu = 2.9t$ and $U_0 = 0.013t$, corresponding to $T_c\simeq 3K$. 
(e-g) Bulk Bogoliubov excitation gap at (e) $\theta = 29.5^{\circ}$, (f) $\theta = 30.0^{\circ}$, and (g) $\theta = 30.5^{\circ}$, corresponding to dots A, B and C in panel (b), respectively.  $\Delta_0 = 1$ meV is used here in line with values of $\mu, U_0$ used in (d).}
\label{FIG3}
\end{figure*}

While microscopic mechanisms leading to STVS superconductivity may vary across materials such as RTG/BBG \cite{Chou1, Chou2} and ZrNCl \cite{Crepel}, on general grounds the interaction responsible for STVS pairing boils down to an effective attraction between electrons in the spin-triplet $f$-wave channel. In the momentum-space representation, the interaction within each isolated layer of STVS superconductor has the form
\begin{equation}
    \begin{aligned}\label{eq:intralayerinteraction}
\mathcal{V}^{(1)} &= -U_0 \sum_{\bm{k},\bm{k}'} f_{1}(\bm{k}) f_{1}(\bm{k}') c^{\dagger}(\bm{k})c^{\dagger}(-\bm{k}) c(-\bm{k}')c(\bm{k}'),\\
\mathcal{V}^{(2)} &= -U_0 \sum_{\bm{\tilde{k}},\bm{\tilde{k}}'} f_{2}(\bm{\tilde{k}}) f_{2}(\bm{\tilde{k}}') c^{\dagger}(\bm{\tilde{k}})c^{\dagger}(-\bm{\tilde{k}}) c(-\bm{\tilde{k}}')c(\bm{\tilde{k}}'),
    \end{aligned}
\end{equation}
where $U_0$ denotes the interaction strength, and $f_{1,2}$ are the basis functions, $f_{1}(\bm{k}) = \sum_{j=1,3,5} \sin(\bm{k}\cdot\bm{R}_j)$ and $f_{2}(\bm{\tilde{k}}) = \sum_{j=1,3,5} \sin(\bm{\tilde{k}}\cdot\bm{\tilde{R}}_j)$, with exact $f$-wave symmetries shown in Fig.\ \ref{FIG1}a.

For isolated layer 1, one can define the self-consistent mean field $\Delta_{1} = -U_0 \sum_{\bm{k}'} f_{1}(\bm{k}') \braket{c(-\bm{k}')c(\bm{k}')}$, and the corresponding gap function  $\Delta_1(\bm{k}) = \Delta_{1} f_1(\bm{k})$. Similarly, $\Delta_2(\bm{\tilde{k}}) = \Delta_{2} f_2(\bm{\tilde{k}})$ for isolated layer 2. Note that $\Delta_1(\bm{k})$ and $\Delta_2(\bm{\tilde{k}})$ are almost constant near $\pm K$ and $\pm \tilde{K}$ but exhibit a valley-dependent sign, as $f_1(\bm{k} \simeq \pm K), f_2(\bm{\tilde{k}} \simeq \pm \tilde{K}) \simeq \mp \frac{3\sqrt{3}}{2}$.

As the disconnected $K$-pockets in two layers merge into a single connected Fermi surface in the twisted double-layer (Fig.\ \ref{FIG2}h), the piecewise constant gap functions with alternating signs are transformed into continuous functions along the single circular Fermi contour in $\bm{k}_0$-space.  Signs of $\Delta_1(\bm{k})$ and $\Delta_2(\bm{\tilde{k}})$, indicated by red and blue ``$\pm$'' symbols in Fig.\ \ref{FIG2}h, are seen to resemble two orthogonal $f$-wave components superimposed on top of each other. Notably, as the Fermi surface gets reconnected in $\bm{k}_0$-space via inter-layer coupling, it is easy to see that nodes in each of the $f$-wave components are recovered -- if time-reversal remains unbroken and hence the order parameters are real. This happens because the projected pairing on the Fermi surface from each layer must change continuously and a nodal point is mandated whenever a sign change in real order parameter occurs. In analogy with twisted $d$-wave superconductors \cite{Marcel2}, this suggests that, in order to avoid node formation and thus lower the overall superconducting free energy, the twisted STVS double-layer may develop a spontaneous complex phase difference between the order parameters of the two layers. This realizes the chiral $f \pm if'$ phase.

In the following we support this intuitive picture of the chiral ${\cal T}$-broken phase formation with an explicit microscopic calculation. We note that the self-consistency of $\Delta_1(\bm{k})$ and $\Delta_2(\bm{\tilde{k}})$ in isolated layer 1 and layer 2 implicitly rests upon the translational invariance within each decoupled layer. Upon introducing the inter-layer coupling, this translational symmetry is strongly modified. This forces us to reformulate the superconducting gap equations in terms of wave functions and energy bands derived from the DMSTB model $\mathcal{H}_{0, \rm eff}$ (see subsection B of Methods section), so that effects from inter-layer coupling are properly incorporated. In the following, we focus on the topmost three bands indexed by $p=1,2,3$ (inset of Fig.\ \ref{FIG2}d) that are accessible by experimentally relevant doping levels.

In terms of fermionic operators $a^{\dagger}_p(\bm{k}_0)$ which create electrons at $\bm{k}_0$ in band $p$, the Bogoliubov-de Gennes (BdG) Hamiltonian for the superconducting state in the twisted double-layer 
\begin{eqnarray}\label{eq:HBdG}
\mathcal{H}_{\rm BdG} &=& \sum_{p,\bm{k}_0} \xi_p(\bm{k}_0) a^{\dagger}_p(\bm{k}_0) a_p(\bm{k}_0) \\\nonumber
&+& \sum_{p, \bm{k}_0} \left[\Delta_{p}(\bm{k}_0) a^{\dagger}_p(\bm{k}_0) a^{\dagger}_p(-\bm{k}_0) + {\rm h.c.}\right],
\end{eqnarray}
where $\xi_p(\bm{k}_0) = E_p(\bm{k}_0) - \mu$ with $E_p(\bm{k}_0)$ the kinetic energy of band $p$, and $\Delta_{p}(\bm{k}_0)$ the pairing in band $p$: $\Delta_{p}(\bm{k}_0) = \Delta_{1,p} f_{1,p}(\bm{k}_0) + \Delta_{2,p} f_{2,p}(\bm{k}_0)$, where $f_{l,p}(\bm{k}_0)$ are dimensionless basis functions characterizing the projected pairings in the moir\'{e} Brillouin zone for layer $l$ and band $p$. The relations between $f_{l,p}(\bm{k}_0)$ and $f_1(\bm{k}), f_2(\bm{\tilde{k}})$ in Eq.\ \ref{eq:intralayerinteraction} are explicitly given in subsection B of Methods section. Note that mean-fields $\Delta_{1,p}, \Delta_{2,p}$ serve as the superconducting order parameters in moir\'{e} band $p$ of layer 1 and layer 2, respectively. 

To demonstrate that the projected pairings from layer 1 and 2 form two orthogonal $f$-wave components at $\theta = 30^{\circ}$, we plot the dimensionless basis functions $f_{l,p}(\bm{k}_0)$ for projected pairing in band $p=2$ along the circular Fermi surface in the moir\'{e} Brillouin zone (Fig.\ \ref{FIG2}d) at $\mu = 2.9t$ in Fig.\ \ref{FIG3}a. Clearly, $f_{1,2}(\bm{k}_0)$ and $f_{2,2}(\bm{k}_0)$ have $f$-wave symmetries with 6 nodes, and the relative phase shift between the two is exactly $\delta\phi_{\bm{k}_0} = \pi/6$ as in two orthogonal $f$-wave components. Thus, the down-folded pairing interaction in the moir\'{e} bands of the twisted double-layer leads to the reconstruction of two orthogonal nodal $f$-wave order parameters in the moir\'{e} BZ, which provides the basis for the $\mathcal{T}$-broken chiral phase. Next, we solve for $\Delta_{1,p}$ and $\Delta_{2,p}$ by minimizing the free energy density
\begin{align}\label{eq:FreeEnergy}
\mathcal{F}_{SC} = \sum_{l,p}\frac{|\Delta_{l,p}|^2}{U_0} - \frac{1}{V}\sum_{s, \bm{k}_0} \frac{1}{2\beta}\ln(1 + e^{-\beta E_{s,\bm{k}_0}}),
\end{align}
where $V$ is the volume of the system, $\beta = 1/k_B T$ ($T$: temperature), $E_{s,\bm{k}_0}$ are the eigenvalues of $\mathcal{H}_{\rm BdG}$ (Eq.\ \ref{eq:HBdG}). Given two identical layers of STVS superconductors, we have $|\Delta_{1,p}| = |\Delta_{2,p}|$, and the general solution up to an overall phase is given by $\Delta_{1,p} = \Delta_0$, $\Delta_{2,p} = \Delta_0 e^{i \varphi_p}$ where we take $\Delta_0$ to be real. To explore the resulting phase diagram in a concrete setting we set $U_0 = 0.013t$ in Eq.\ \ref{eq:FreeEnergy} and $\mu =2.9t$. With $t = 1$ eV this yields $T_c \simeq 3K$ and $\Delta_0 \simeq 1$ meV at $T=0$, almost independent of $\theta$.

The complete superconducting phase diagram in the $\mu - \theta$ space is shown in Fig.\ \ref{FIG3}b. For $\theta$ in close vicinity of $30^{\circ}$, the system develops a spontaneous $\mathcal{T}$-broken phase characterized by $0<\varphi_{p}<\pi$ and becomes an intrinsic chiral $f \pm if'$-wave superconductor. Notably, the chiral phase persists over almost the entire chemical potential range that produces a connected FS, as shown in blue-shaded regions in Fig.\ \ref{FIG3}b. At $\theta = 30^{\circ}$, the free energy $\mathcal{F}_{SC}$ is minimized exactly at $\varphi_p = \pm \pi/2$ (red solid line in Fig.\ \ref{FIG3}c) with $\Delta_{2,p} = \pm i \Delta_{1,p}$, which corresponds to a perfect $f \pm if'$-wave symmetry. As $\theta$ deviates from $30^{\circ}$, $\varphi_p$ gradually evolves towards $0$ or $\pi$ for $\theta \leq \theta^{-}_c$ and  $\theta \geq \theta^{+}_c$, respectively (Fig.\ \ref{FIG3}c), and the two layers of STVS superconductors eventually form a $0 (\pi)$-phase junction.  

The bulk Bogoliubov excitation energy gaps along the circular Fermi surface at $\mu = 2.9t$ in different superconducting phases are shown in Fig.\ \ref{FIG3}e-g. We find that for $\theta < \theta^{c}_{-}$ and $\theta > \theta^{c}_{+}$, the system in the $\mathcal{T}$-preserving phase is a nodal $f$-wave superconductor with 6 nodes along the Fermi surface. While in the $\mathcal{T}$-broken chiral regime, the system exhibits a full superconducting gap.  In the next section, we demonstrate the nontrivial topological properties of the chiral $f\pm if'$ phase as well as the nodal $f$-wave SC phase by studying their boundary and vortex core excitations.

\subsection*{B. Ginzburg-Landau theory}

To understand why a robust chiral $f \pm if'$ phase emerges at $\theta \simeq 30^{\circ}$, we construct a phenomenological Ginzburg-Landau (GL) theory in terms of the reconstructed $\psi_1 \equiv \Delta_{1,p}$, $\psi_2 \equiv \Delta_{2,p}$ in moir\'{e} band $p$. Note that due to the reconstruction of pairing interactions in the moir\'{e} bands, the symmetry properties of $\psi_1$, $\psi_2$, as summarized in Table \ref{Tab:IR}, do not directly follow from the $f$-wave symmetries in each isolated layer, but need to be \textit{derived} from the defining equations for the basis functions $f_{1,2}(\bm{k}_0)$ and $f_{2,2}(\bm{k}_0)$ formulated in the moir\'{e} bands (Eq.\ \ref{eq:basisfunc} in the Methods section). For a general twist angle $\theta$, the double-layer has $D_6$ point group symmetry, which dictates the form of the GL free energy 
\begin{align}\label{eq:GLenergy}
\mathcal{F}_{GL}[\psi_1, \psi_2] &= \sum_{l=1,2} \alpha_0 |\psi_l|^2 + \frac{\beta_0}{2} |\psi_l|^4 + a_0 |\psi_1|^2 |\psi_2|^2 \nonumber \\
&+ b_0 (\psi_1 \psi^{\ast}_2 + \textrm{c.c.}) + c_0 (\psi^2_1 \psi^{\ast 2}_2 + \textrm{c.c.})
\end{align}
where the coefficients $b_0$, $c_0$ characterize the coherent tunneling of single and double Cooper pairs between the layers, respectively. Taking $\psi \equiv |\psi_1| = |\psi_2|$ (see detailed analysis on the validity of this choice in Supplementary Note 4 and 5), this becomes 
$\mathcal{F}_{GL}(\varphi)= \mathcal{F}_0 + 2 b_0 \psi^2 \cos(\varphi) + 2 c_0 \psi^4 \cos(2\varphi)$, where $\varphi$ is the phase difference between $\psi_1$ and $\psi_2$, and $\mathcal{F}_0 = 2\alpha_0\psi ^2 + (\beta_0 + a_0) \psi^4$.

\begin{table}[tp]
\caption{Irreducible representations (ireps) of point group $D_6$ ($\theta \neq 30^{\circ}$) and $D_{6d}$ ($\theta = 30^{\circ}$) with cubic functions and reconstructed $f$-wave order parameters $\psi_1$, $\psi_2$ in the moir\'{e} bands in twisted double-layer. Symmetry properties of $\psi_1$, $\psi_2$ are derived from Eq.\ \ref{eq:basisfunc}. $A_1$ is the trivial irep in both cases.}
\centering
\begin{tabular}{c|c|c|c}
\hline\hline
Group & IR & cubic functions & SC order parameters\\
\hline
\multirow{3}{4em} {$D_6$} & $A_1$ & --- & \scriptsize{$\psi_1 \psi^{\ast}_{2} + \psi^{\ast}_1 \psi_2$, $\psi^2_1 \psi^{\ast 2}_2 + \psi^{\ast 2}_1 \psi^2_2$} \\ 
                          & $B_1$ & \scriptsize{$x^3-3xy^2$} & \scriptsize{$\psi_1 + \psi_2$}\\ 
                          & $B_2$ & \scriptsize{$3x^2y-y^3$} & \scriptsize{$\psi_1 - \psi_2$}\\ 
\hline
\multirow{2}{4em} {$D_{6d}$} & $A_1$ & --- &  \scriptsize{$\psi^2_1 \psi^{\ast 2}_2 + \psi^{\ast 2}_1 \psi^2_2$}\\ 
                             & $E_3$ & \scriptsize{$\{ x^3-3xy^2, 3x^2y-y^3\}$} & \scriptsize{$\{\psi_1 + \psi_2, \psi_1 - \psi_2 \}$} \\ 
\hline\hline
\end{tabular}
\label{Tab:IR}
\end{table}

At the maximal twist $\theta = 30^{\circ}$, the twisted double-layer becomes a quasi-crystal with 12-fold tiling as discussed in Section II. The symmetry of the quasi-crystal is described by the non-crystallographic $D_{6d}$ point group, which includes an extra improper rotation $S_{12} \equiv \sigma_h \otimes  C_{12}$, \textit{i.e.}, a 12-fold rotation about the $z$-axis combined with the reflection about the horizontal mirror plane lying mid-way between the two layers (see Fig.\ \ref{FIG2}a). We deduce that under $S_{12}: \psi_1 \rightarrow -\psi_2, \psi_2 \rightarrow \psi_1$. For the GL free energy in Eq.\ \ref{eq:GLenergy} to be invariant under $S_{12}$, the single-pair tunneling term must vanish: $b_0 = 0$, and the only $\varphi$-dependent term is proportional to coefficient $c_0$ and exhibits  $\cos(2\varphi)$ dependence. As argued in the twisted cuprate case, the coefficient $c_0$ associated with the double-pair tunneling is generally positive. In Supplementary Note 4 we further verify that $c_0 > 0$ generally holds in the case of twisted STVS superconductors by expanding the microscopic free energy Eq.\ \ref{eq:FreeEnergy} in terms of $\Delta_{1,2}$ and $\Delta_{2,2}$ using the imaginary-time path integral formalism. Thus, the quasi-crystalline $D_{6d}$ symmetry dictates that the free energy is always minimized for  $\varphi_{\rm min} = \pm \pi/2$ implying a state with spontaneously broken $\mathcal{T}$-symmetry.

The distinctive quasi-crystalline $D_{6d}$ point group  at $\theta = 30^{\circ}$ has important consequences for the temperature dependence of the chiral $f \pm if'$ phase. It is worth noting that by construction for general $\theta$ (Fig.\ \ref{FIG2}a-b), the basis functions $f_1(\bm{k}_0)$ and $f_2(\bm{k}_0)$ are always symmetric about the $\bm{k}_{0,y} = 0$ axis when $\phi_{\bm{k}_0} = 0, \pi$, and anti-symmetric about $\bm{k}_{0,x} = 0$ when $\phi_{\bm{k}_0} = \pm \pi/2$ (see Fig.\ \ref{FIG3}a). As such, the $\varphi = 0$ phase is essentially the $f_{x(x^2-3y^2)}$ pair function formed by the real combination $f_1(\bm{k}_0) + f_2(\bm{k}_0) \propto k_{x,0}(k_{x,0}^2-3k_{y,0}^2)$, which belongs to the 1D irreducible representation (irep) of $D_6$ labeled $B_1$ in Table \ref{Tab:IR}. On the other hand, the $\varphi = \pi$ phase corresponds to the  $f_{y(3x^2-y^2)}$ pair function formed by an orthogonal real linear combination $f_1(\bm{k}_0) - f_2(\bm{k}_0) \propto k_{y,0}(3k_{x,0}^2-k_{y,0}^2)$, which belongs to another 1D irep of $D_6$ labeled $B_2$. Thus, for $\theta$ away from 30$^{\circ}$, the two phases belonging to distinct ireps of $D_6$ have different ground state energies (see free energy landscapes at $T=0$ in Fig.\ \ref{FIG3}c for $\theta \neq 30^{\circ}$) and correspond to different $T_c$.

At $\theta = 30^{\circ}$, however, the two orthogonal $f$-wave states with $\varphi = 0$ and $\varphi=\pi$ are degenerate (see free energy landscape in Fig.\ \ref{FIG3}c for $\theta = 30^{\circ}$), and form a 2D representation of $D_{6d}$, labeled $E_3$ in Table \ref{Tab:IR}, with both components having the same $T_c$. Accordingly, as shown in the $T - \theta$ phase diagram Fig.\ \ref{FIG3}d, obtained by minimizing $\mathcal{F}_{SC}$ at finite $T$, the chiral $f \pm if'$ phase extends all the way to $T = T_c$ for $\theta = 30^{\circ}$ because both orthogonal $f$-wave components condense simultaneously when superconductivity sets in at $T_c$. For $\theta$ away from $30^{\circ}$, the component with higher $T_c$ (either $\varphi = 0$ or $\varphi = \pi$) sets in first, and one needs to further lower the temperature to access the other orthogonal component with lower $T_c$ to form the chiral $f \pm if'$ phase.

In the zero-temperature limit, the chiral $f \pm if'$ phase extends between  critical angles $\theta^{-}_c \simeq 29.7^{\circ} - 29.8^{\circ}$ and $\theta^{+}_c \simeq 30.2^{\circ} - 30.3^{\circ}$ (Fig.\ \ref{FIG3}b-d). The overall twist angle range of $\delta \theta \simeq 0.4 -  0.6^{\circ}$ is well within reach of twist angle engineering precision $\sim 0.1^{\circ}$ now common in the state-of-the-art sample fabrication technique for twisted materials \cite{Cao1, Cao2, Yankowitz, Sharpe, Young1}. Notably, the twist angle range predicted here is narrower than the chiral $d \pm id'$ phase found in twisted cuprates which can span several degrees \cite{Marcel2}. As we explain in Supplementary Note 6, the relatively narrow twist angle range originates from the nontrivial $\theta$-dependence of the reciprocal lattice vectors $\bm{\tilde{G}}_m, \bm{G}_n$, for $m,n\neq 0$ that determine the interlayer coupling strength, as opposed to the coupling dominated by $\bm{\tilde{G}}_0 = \bm{G}_0 = \bm{0}$ in cuprates which is, to good approximation, $\theta$-independent.

%-----------------------------------
\section*{Topological properties}

\subsection*{A. Nodal $f$-wave  phase }

\begin{figure}
\centering
\includegraphics[width=0.5\textwidth]{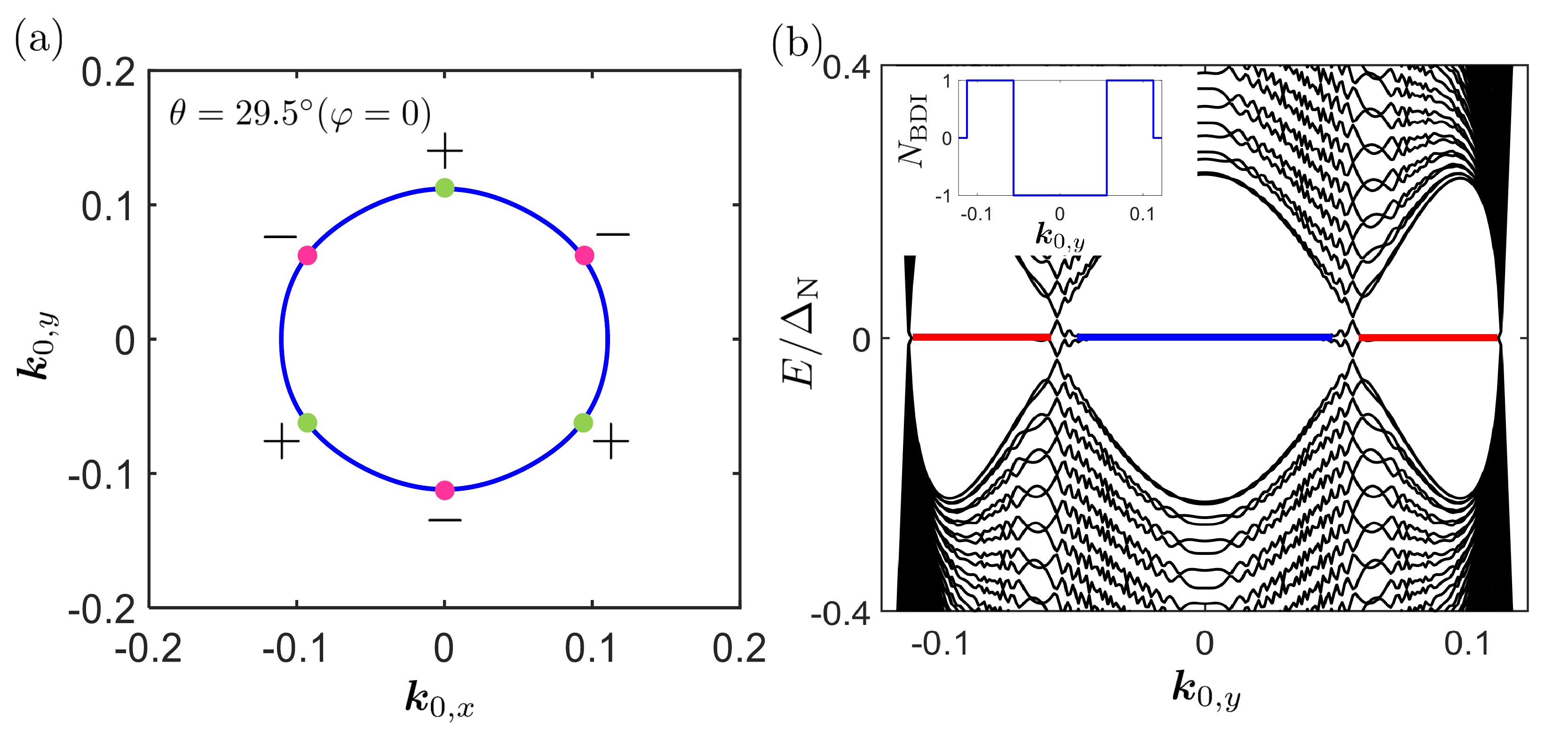}
\caption{(a) Chiralities of nodes indicated by ``$\pm$" signs in the $\varphi = 0$ phase with $f_{x(x^2-3y^2)}$-wave pairing symmetry. (b) Edge spectrum of $H_{\rm BdG}$ as a function of $k_{y,0}$ along edges oriented in the $y$-direction, calculated using the lattice model in Eq.\ \ref{eq:latticemodel}-\ref{eq:lattgap} with $\Delta_{\rm NN} = 0$ such that only the $f_{x(x^2-3y^2)}$-wave component is present. Inset: BDI invariant $N_{\rm BDI}$ as a function of $k_{y,0}$. The regions with $N_{\rm BDI} = +1 (-1)$ correspond to regions with MZMs highlighted in red (blue) in (b). ${k}_{0,x}, {k}_{0,y}$ are given in units of  $\rm {\AA}^{-1}$.}
\label{FIG4}
\end{figure}

As we discussed in previous sections, for $\theta < \theta^{-}_{c}$ ($\theta > \theta^{+}_{c}$), the twisted double-layer favors the $\varphi = 0$ ($\varphi = \pi$) phase and becomes a nodal $f_{x(x^2-3y^2)}$-wave ($f_{y(3x^2-y^2)}$-wave) superconductor. This nodal phase is topologically nontrivial in the sense that the $f$-wave nodes are characterized by chirality numbers \cite{Goerbig}, and, in a geometry with edges, nodes of opposite chirality are connected by protected non-dispersive Majorana edge modes.

To understand the nontrivial topological property of the nodal $f$-wave phase, we consider the specific case with $\theta = 29.5^{\circ}$ and $\varphi = 0$ in band $p=2$ corresponding to Fig.\ \ref{FIG3}e. In the Nambu basis $\psi(\bm{k}_0) = \big( a_{2}(\bm{k}_0), a_{2}^{\dagger}(\bm{k}_0) \big)^{T}$, the bulk BdG Hamiltonian is written as $\mathcal{H}_{\rm{BdG}} = \sum_{\bm{k}_0} \psi^{\dagger}(\bm{k}_0) H_{\rm{BdG}}(\bm{k}_0) \psi(\bm{k}_0) $, where $H_{\rm{BdG}}(\bm{k}_0) = \xi_{2}(\bm{k}_0) \tau_3 + \Delta_{2}(\bm{k}_0) \tau_1$ with $\tau_{\alpha = 1,2,3}$ as Pauli matrices acting on particle-hole space, and $\Delta_2(\bm{k}_0) = \Delta_0 (f_{1,2}(\bm{k}_0) + f_{2,2}(\bm{k}_0))$.

The Hamiltonian $H_{\rm{BdG}}$ respects a chiral symmetry $\mathcal{C} H_{\rm{BdG}} \mathcal{C}^{-1} = -H_{\rm{BdG}}$ where $\mathcal{C} = \tau_2$. It can be represented in the eigenbasis of $\mathcal{C}$, by translating  $\tau_{3} \mapsto \tau_{1}, \tau_{1} \mapsto \tau_{2}$. Then, in the vicinity of a nodal point $\bm{k}_{0,N}$, the excitations can be described as 2D massless Dirac fermions
\begin{equation}
h_N(\bm{p}_0) \simeq v_1 p_{0,1} \tau_1 + v_2 p_{0,2} \tau_2, 
\end{equation}
where $p_{0,1}$ and $p_{0,2}$ denote the normal and tangential components along the Fermi surface of a momentum $\bm{p}_0=\bm{k}_0-\bm{k}_{0,N}$. Dirac velocities $v_1 \equiv \nabla_{\bm{k}_0}\xi_2(\bm{k}_0) \cdot \hat{n}_1$ and $v_2 \equiv \nabla_{\bm{k}_0}\Delta_2(\bm{k}_0) \cdot \hat{n}_2$ are evaluated at $\bm{k}_0 = \bm{k}_{0,N}$. The chirality at $\bm{k}_{0,N}$ can then be defined as $C(\bm{k}_{0,N}) = \sgn (v_1 v_2) \hat{z}\cdot(\hat{n}_1 \times \hat{n}_2)$. 

The chiralities $C(\bm{k}_{0,N})$ of the 6 nodes in the $\varphi =0$ phase are calculated from $H_{\rm{BdG}}(\bm{k}_0)$ and indicated in Fig.\ \ref{FIG4}a. Clearly, nodes with opposite chiralities come in three pairs, reflecting the underlying $f_{x(x^2-3y^2)}$-symmetry. It is worth noting that by projecting the bulk spectrum onto the edges oriented along certain high-symmetry directions, e.g.\ the $y$-direction, nodes with opposite chiralities do not cancel out. In this situation we expect protected non-dispersive edge modes to appear in analogy with the flat bands on zigzag edges of monolayer graphene \cite{Montambaux}.

\begin{figure*}
\centering
\includegraphics[width=\textwidth]{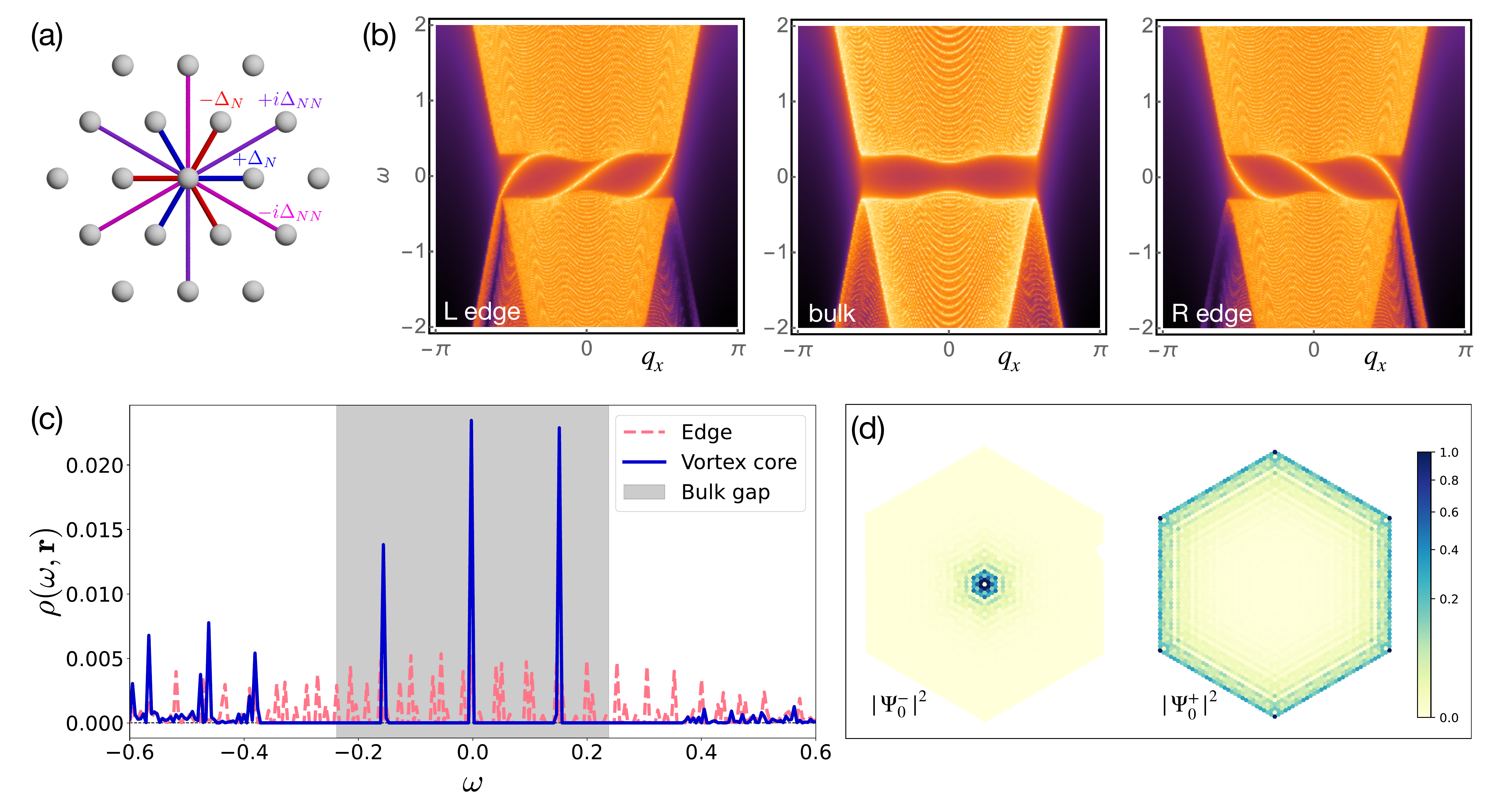}
\caption{Exact diagonalization results for edge and vortex core modes. (a) Pairing amplitudes for the $f$-wave order parameter on the simplified triangular lattice model with lattice constant $a_0$. (b) Spectral function $A(q_x, \omega)$ evaluated at the edges and in the bulk of an infinite strip with width of $L=200$ sites. $\omega$ is shown in units of $t_0$. (c) Local density of states $\rho(\omega,{\bm r})$ computed on a hexagon-shaped sample with $N = 3169$ sites and a vortex located at its center.   (d) Real-space probability distribution of zero-energy eigenstates $\Psi_0^{\pm}=(\Psi_0\pm\Psi_0')/\sqrt{2}$, normalized to the maximum site amplitude for each state. In all panels parameters $(\mu_0, \Delta_{N}, \Delta_{NN}) = (0.8t_0, 0.5t_0, \Delta_{N}/3\sqrt{3})$ with $t_0 = -0.01$ eV were used.}
\label{FIG5}
\end{figure*}

To demonstrate the existence of edge states in the nodal $f$-wave phase,
we now introduce a simplified triangular lattice model with lattice constant $a_0$ (shown schematically in Fig.\ \ref{FIG5}a) for band $p=2$ derived from the DMSTB model above (Fig.\ \ref{FIG2}d). The lattice model captures both the parabolic dispersion near $\Gamma$ and the $f$-wave pairing symmetry, thus facilitating explicit calculcations of edge states as well as vortex excitations of the chiral superconductor presented below. The lattice model is defined by
\begin{align}\label{eq:latticemodel}
    \mathcal H_{\rm LAT} = \sum_{\bm{q}}\left[  \xi_{\bm{q}}{c^{\dagger}_{\bm{q}}c_{\bm{q}} +
    (\Delta_{\bm{q}}c^{\dagger}_{\bm{q}}c^{\dagger}_{-\bm{q}} + {\rm h.c.}})\right],
\end{align}
where $\xi_{\bm{q}} = -2t_0 \sum_{j=1, 3, 5} \cos(\bm{q} \cdot \bm{\delta}_{j}) -\mu_0$ denotes the band energy from effective electron hopping $-t_0$, where we set $t_0 \simeq 0.01t$ by fitting the parabolic dispersion at $\Gamma$ in Fig.\ \ref{FIG2}d. The gap function is given by 
\begin{align}\label{eq:lattgap}
\Delta_{\bm{q}} = 2\sum_{j=1,3,5}\left[\Delta_{\rm N}  \sin(\bm{q} \cdot \bm{\delta}_{j})+i\Delta_{\rm NN}  \sin(\bm{q} \cdot \bm{\delta}'_{j})\right]
\end{align}
with $\Delta_{\rm N}$ ($\Delta_{\rm NN}$) denoting the first (second) nearest-neighbor pairing amplitudes as shown in Fig.\ \ref{FIG5}a, while $\bm{\delta}_{j}$ and $\bm{\delta}'_{j}$ are the corresponding bond vectors indicated by red and purple lines, respectively. It is straightforward to check that in the small $\bm{q}$ expansion, the two pairing terms produce two orthogonal $f_{x(x^2-3y^2)}$- and $f_{y(3x^2-y^2)}$-wave components, respectively.

We use the lattice model in Eq.\ \ref{eq:latticemodel} with $\Delta_{\rm NN}=0$  to calculate the edge spectrum of the nodal $f_{x(x^2-3y^2)}$-wave superconductor as a function of $k_{0,y}$ (Fig.\ \ref{FIG4}b), where we identify $k_{0,y}$ as $q_y$ in Eq.\ \ref{eq:latticemodel}-\ref{eq:lattgap}. As anticipated the bulk nodes with opposite chiralities are connected by non-dispersive zero energy modes on the edge (highlighted by red and blue lines in Fig.\ \ref{FIG4}b). As we explain in Supplementary Note 7, for each fixed $k_{0,y}$, $H_{\rm BdG}({k}_{0,x})$ describes a one-dimensional BDI class topological superconductor oriented in the $x$-direction, with its topological property characterized by a winding number $N_{\rm BDI}$ \cite{Altland-Zirnbauer, Schnyder1, Chiu1} (inset of Fig.\ \ref{FIG4}b). The bulk-edge correspondence between $N_{\rm BDI}$ and the number of zero energy modes allows us to establish the edge state associated with each $k_{y,0}$ as a MZM. Importantly, as long as the chiral symmetry $\mathcal{C}$ is respected, chiralities of nodes are well defined and act as topological charges that only annihiliate when opposite charges merge in the bulk. Therefore, the large number of non-dispersive MZMs on the edge are protected by the nontrivial bulk topology against $\mathcal{C}$-preserving perturbations such as charge disorder \cite{Sato, Schnyder2, Chiu2, Wenyu}.

\subsection*{B. Chiral $f\pm if'$ phase}

In the ${\cal T}$-broken $f\pm if'$ phase, that we shall model by the lattice Hamiltonian Eq.\ \ref{eq:latticemodel} with $\Delta_{\rm NN}\neq 0$, the 6 Dirac nodes are gapped out by alternating mass terms produced by the imaginary $f_{y(3x^2-y^2)}$-wave component of the order parameter. This gapped phase belongs to symmetry class D in Altland-Zirnbauer classification \cite{Altland-Zirnbauer, Schnyder1, Chiu1} and its topology is therefore characterized by the Chern number $C$. In analogy with the $d+id'$ phase in cuprates we expect each gapped Dirac point to contribute ${1\over 2}\sgn(\Delta_{\rm NN})$ to the total Chern number, suggesting that the system will have $C=\pm 3$ in the gapped chiral phase. Importantly, because the BdG representation of the spinless superconductor is redundant, the Chern number here determines the number of chiral Majorana edge modes with central charge $1/2$ (as opposed to complex fermion modes whose central charge would be 1).       

We can now confirm the existence of edge states by placing $\mathcal{H}_{\rm LAT}$ on a strip geometry with periodic boundary conditions in the $x$-direction, and $L$ rows of atoms in the $y$-direction. 
The spectral function in 1D momentum space, plotted in Figure \ref{FIG5}b, allows us to visualize the excitations present in each row of the strip.
It is defined as
\begin{equation}
    A(q_x, \omega) = {\rm Im}[(\omega + i\eta) - \mathcal{H}_{\rm LAT}(q_x)]^{-1}
\end{equation}
where $\eta$ is a positive infinitesimal, and $\mathcal{H}_{\rm LAT}(q_x)$ is the $L \times L$ Hamiltonian on the strip with $q_x$ the lattice momentum along $x$. The spectral function reveals a fully gapped bulk and three distinct edge modes traversing the bulk gap, propagating in opposite directions at each edge. This confirms the Chern number $C=3$ deduced above from general considerations. 

In addition to gapless edge modes, chiral $p$-wave superconductors threaded with unit magnetic flux are predicted to host unpaired MZMs obeying non-Abelian exchange statistics, which are localized at vortex cores \cite{Ivanov_2001, Volovik_1999}.  To model the effect of an Abrikosov vortex in the $f+if^{'}$ superconductor, we adopt a real-space representation of the lattice model in Eq.\ \ref{eq:latticemodel}.  
We consider a hexagonal domain with open boundary conditions, and place a vortex at the origin.  This induces a phase winding on the order parameter for each bond
\begin{align}
    \Delta_{\mathbf{r}, \bm{\delta}_{j}} &= (-1)^{j-1}\Delta_{\rm N} \exp{(-i n\theta_{\mathbf{r},\bm{\delta}_{j}})} \nonumber\\
    \Delta_{\mathbf{r},\bm{\delta}_{j}^{'}} &= (-1)^{j-1}\Delta_{\rm NN} \exp{(-i n\theta_{\mathbf{r},\bm{\delta}_{j}^{'}})} 
    \label{eq:vortexphase}
\end{align}
where $\mathbf{r}$ denote lattice sites in real-space; $n$ is the vorticity; and $\theta_{\mathbf{r},\bm{\delta}_{j}}$ is the angle subtended by the midpoint of the bond ($\mathbf{r} + \frac{1}{2}\bm{\delta}_{j}$), the origin, and the $x$-axis.  We then numerically diagonalize the $2N\times2N$ matrix representing $\mathcal{H}_{\rm LAT}$, where $N$ is the number of lattice sites. 

For a single-quantum vortex solution with $n=1$, we indeed find a single zero-energy mode, which manifests as a zero-bias peak in the local density of states (LDOS) at the vortex core. Its partner lives at the sample edge, as shown in Fig.\ \ref{FIG5}c.
%this single-quantum vortex in the spinless chiral superconductor traps a single MZM, which was shown to exhibit non-Abelian MZMs \cite{Ivanov_2001}.
The LDOS is given by
\begin{equation}
    \rho(\omega, \bm{r}) = \sum_{E_n > 0} \left\{ | u_n(\bm{r}) |^2 \delta(\omega - E_n) + |v_n(\bm{r})|^2 \delta(\omega + E_n) \right\}
\end{equation}
where $E_n$ are eigenvalues of $\mathcal{H}_{\rm LAT}$ with eigenstates $\Psi_n(\bm{r}) = (u_n(\bm{r}), v_n(\bm{r}))^T$.  
To confirm the nature of the two zero-energy states, denoted $\Psi_0$ and $\Psi_0^{'}$, we plot their real-space wavefunctions in Fig.\ \ref{FIG5}d.  The symmetric and anti-symmetric linear combination of these states are self-conjugate eigenstates of the charge conjugation operator, and represent Majorana zero modes localized at the sample edge and vortex core, respectively. 
%The small energy splitting $\varepsilon_0 \simeq (2.6\times10^{-4})t_0$ occurs due to finite size effects.  

\section{Experimental relevance and signatures}

Here we discuss how the exotic non-Abelian TSC phase can be realized in twisted double layers formed by the promising candidate materials, rhombohedral graphene \cite{Young2, Young3} and ZrNCl \cite{Crepel}, which are thought to be STVS superconductors. It is worth noting that the spinless-fermion triangular lattice model used for our theoretical considerations captures most of the essential features of the spin-triplet SC2 phase in rhobohedral trilayer graphene (RTG): (i) the SC2 phase emerges under strong displacement fields which polarize the layer and sublattice degrees of freedom, such that electrons involved in superconducting pairing actually live on an effective triangular lattice formed by the A (or B) sublattice \cite{Young3}; (ii) the Fermi surface of the parent spin-polarized valley-unpolarized normal state underlying the SC2 phase is well reproduced by our model (Fig.\ \ref{FIG2}).

To realize the topological phase with non-Abelian excitations, it is crucial that the isolated $K$-pockets from each layer coalesce into a single connected Fermi surface in the moir\'{e} Brillouin zone. Our results based on the DMSTB model suggest that a minimal Fermi momentum $k_F a \sim 0.2$ ($a = 2.46 {\rm \AA}$ for graphene) measured from $\pm K (\tilde{K})$ is required for a single connected Fermi surface to emerge (see Fig.\ \ref{FIG2}f). The typical value of Fermi momentum in RTG, corresponding to a low doping level with carrier density $n_{2D} \approx 0.5 \times 10^{12}$ cm$^{-2}$, was found to be of order $k_F a \sim 0.1$ \cite{Young2}, which is almost on par with the minimal requirement according to our theory. We note that higher doping levels, up to $n_{2D} \sim 1 \times 10^{12}$ cm$^{-2}$, are indeed accessible through electrostatic gating \cite{Young2, Young3}, and the non-Abelian topological superconductivity could thus be achieved by further raising the doping level in maximally twisted double-layer RTG.   

The triangular lattice model also captures the parabolic dispersion near $\pm K$ of the doped band insulator ZrNCl \cite{Heid, Yin} and was in fact used to study the STVS pairing \cite{Crepel}. In particular, ZrNCl superconducts within a wide range of electron doping $x \sim 0.01 - 0.3$ \cite{Tou1, Tou2, Iwasa} ($x$: number of electrons per unit cell). The simple parabolic dispersion near $\pm K$ allows us to extract a Fermi momentum $k_F a = \sqrt{x \pi} \simeq 0.2 - 1.0$ with lattice constant $a = 3.663 {\rm \AA}$ for ZrNCl, which suggests that the condition for a single connected FS in maximally twisted double-layer is readily fulfilled. 

The $f$-wave pairing interaction in the non-magnetic ZrNCl may generally involve all three triplet channels given its spin-degenerate band structure, with the spinor part of the pair function characterized by a three-component $\bm{d}$-vector $\hat{\Delta}_{t} = (\bm{d}\cdot\bm{\sigma}) i\sigma_y$ ($\sigma_{\alpha =x,y,z}$: Pauli matrices for spins). Given the $D_6$-point group of the twisted double-layer, however, the doublet formed by equal-spin states $\{ \ket{\uparrow \uparrow}, \ket{\downarrow \downarrow} \}$ and the anti-parallel  state $ \ket{\uparrow \downarrow} + \ket{\downarrow \uparrow}$ will in general be distinct in energies as they belong to different $E_1$ and $A_2$ ireps of $D_6$, respectively. In the $E_1$ phase with a two-component order parameter $\bm{d} =(d_x, d_y,0)$, the total BdG Hamiltonian is decomposed into two independent spin sectors, each one described by the spinless model discussed in this work. Thus, our analysis directly applies and the system near the maximal twist becomes a \emph{spinful} chiral $f \pm if'$ superconductor. As degrees of freedom are doubled, there would be two species of chiral Majorana modes on the edge corresponding to three complex fermions; as well as two MZMs, one from each spin sector, localized at the single-vortex core. Notably, as the two-component $\bm{d}$-vector can rotate around the vortex core, the spinful chiral $f \pm if'$ state admits half-quantum vortex (HQV) solutions trapping a $\pi$-flux. Following the analysis developed for spinful chiral $p$-wave superconductors in the context of Sr$_2$RuO$_4$ and $^3$He-A phase \cite{Ivanov_2001, Leggett}, the HQV is equivalent to a single-quantum vortex in one of the effective spin sectors and thus hosts a non-Abelian MZM.  

In obtaining the phase diagrams in Fig\ \ref{FIG3}, we considered a relatively strong pairing interaction which yields a native $T_c \simeq 3K$ and sizable pairing amplitude $\Delta_0 \simeq 1$ meV for the twisted double-layer. While such temperature and energy scales are directly relevant to ZrNCl \cite{Iwasa}, the spin-triplet superconductivity in RTG and BBG is found with a much lower $T_c \simeq 50$ mK \cite{Young2, Young3}. It is important to note, however, that 
the emergence of the chiral $f\pm if'$ phase follows from general symmetry principles as illustrated in our Ginzburg-Landau analysis. The proposed mechanism should therefore be largely insensitive to microscopic details and remains applicable to RTG and BBG. We further note that phase diagrams in Fig.\ \ref{FIG3} do not change qualitatively when the inter-layer coupling strength is varied, as long as it remains on the scale of $t_{\perp}(K) \sim 0.1t$. For weaker interlayer coupling, $t_{\perp}(K) \lesssim 0.01t$, the energy bands become dense in energy space and the parameter regime with a single connected Fermi surface in Fig.\ \ref{FIG3}b is reduced.

To detect the $\mathcal{T}$-broken chiral $f \pm if'$ phase a suite of  spectroscopic and transport experiments proposed for the chiral $p+ip'$ and $d + id'$ phases in Sr$_2$RuO$_4$ and twisted cuprates can be applied.  The nonzero orbital angular momentum $L_z = \pm 3$ in the chiral $f \pm if'$ phase can be probed by polar Kerr effect measurements \cite{Yip, Oguz}, and the two-minimum free energy landscape near $\theta = 30^{\circ}$ shown in Fig.\ \ref{FIG3}c will give rise to anomalous $\pi$-periodic inter-layer Josephson current-phase relation $I_J =(2e/\hbar) \partial \mathcal{F}/\partial \varphi \propto \sin(2\varphi)$ \cite{Marcel2}. Upon tuning $\theta$ and $T$, the transition from fully gapped chiral phase to nodal $f$-wave phase (Fig.\ \ref{FIG3}d-h) can be detected by a change from $U$-shaped to $V$-shaped spectra in the bulk LDOS, which can be probed by scanning tunneling microscopy (STM) measurements \cite{Fischer}. Moreover, STM can be used to detect and resolve the spatial profile of the zero bias peaks induced by the MZM localized at the vortex core  \cite{Wang, Yazdani1}, as well as the non-dispersive edge MZMs in the nodal $f$-wave phase.

\section*{Conclusions and outlook}

We established a new avenue, through twist-angle engineering, toward intrinsic chiral $f\pm if'$ TSC with non-Abelian excitations. In particular, the emergence of non-Abelian TSC in maximally twisted STVS superconductors relies on a novel large-angle moir\'{e} physics which is absent in twisted cuprates and is fundamentally different from the moir\'{e} physics in small-angle twisted graphene (see detailed comparison between these systems in Supplementary Note 1). Our Ginzburg-Landau analysis reveals that the energetics leading to the exotic chiral $f \pm if'$ phase are governed by a non-crystallographic $D_{6d}$ symmetry, which emerges generically in the 12-fold quasi-crystalline structure formed at $30^{\circ}$ twist. By virtue of adiabatic continuity we expect the gapped topological phase to persist in a finite range $\delta\theta$ of twist angles around $30^{\circ}$ and our microscopic model indeed indicates stability for $\delta\theta\simeq 0.4^{\circ}-0.6^{\circ}$, well within the capability of current sample fabrication techniques.

The proposed mechanism applies in general to any two-valley material with hexagonal symmetry that exhibits gapped $f$-wave superconductivity in its monolayer form. Importantly, the formation of a single connected FS in the twisted double-layer - an important prerequisite for non-Abelian excitations - requires states from the two different valleys $\pm K (\pm\tilde{K})$ to hybridize (Fig.\ \ref{FIG2}c-h). Therefore, the novel large-angle moir\'{e} physics necessarily violates the fundamental valley conservation symmetry $U_v(1)$ that underpins the well-established moir\'{e} physics in small-angle-twisted graphene and transition-metal dichalcogenides. This provides an alternative symmetry perspective on why the BM-type continuum models applicable in the small-angle limit, in which the $U_v(1)$ is built-in by construction, fail to describe the moir\'{e} physics at maximal twist. As we discuss in detail in Supplementary Note 1, the $U_v(1)$-violation in maximally twisted double-layer is crucial for non-Abelian TSC because any description respecting the  $U_v(1)$ symmetry would necessarily lead to Abelian excitations in a zero-momentum-paired superconductor, regardless of its pairing symmetry. This $U_{v}(1)$-based criterion reveals a profound connection between the large-angle moir\'{e} physics established in this work and non-Abelian TSC.

While our assumption of a dominant pairing interaction in the $f$-wave channel for spin-triplet SC2 phase of RTG/BBG is supported by proposals based on acoustic phonons \cite{Chou1, Chou2} and renormalization group analysis \cite{Roy}, some recent works suggest an alternative possibility of chiral $p \pm ip'$ pairing symmetry \cite{Chatterjee, Berg, Levitov}, which is also compatible with the phenomenology of the SC2 phase. In Supplementary Note 8 we present a detailed analysis of pairing instabilities in all possible pairing channels in maximally twisted double-layer RTG/BBG. Our microscopic calculations reveal that the chiral $f$-wave phase takes up the vast majority of the superconducting phase diagram, which lends strong support to the central idea developed in this work. In particular, we find that the chiral $p$-wave phase is favored only when the $p$-wave coupling constant is overwhelmingly larger than the coupling constant in the $f$-wave channel.

Our detailed calculations in addition show that under a dominant chiral $p$-wave interaction, the superconducting free energy and inter-layer Josephson current at $\theta = 30^{\circ}$ exhibits a $2\pi$-periodicity in its $\varphi$-dependence, as opposed to the anomalous $\pi$-periodic dependence in the case of $f$-wave pairing (see Fig.\ \ref{FIG3}c and Fig.S5 in Supplementary Note 8). These two contrasting behaviors can serve as experimental signatures discriminating between $f$-wave and chiral $p$-wave order parameters proposed for the SC2 phase in RTG. These results suggest novel applications of large-twist-angle engineering in probing the pairing symmetries of unconventional superconductors.

While the exact nature of the pairing symmetry of the SC2 spin-triplet phase remains to be settled by future experiments, it is important to note that even under the topological chiral $p\pm ip'$ pairing symmetry, an isolated monolayer still cannot support non-Abelian excitations due to the  disconnected nature of its Fermi surface. Interestingly,  maximally twisted double layer favors a configuration with the same $p$-wave chiralities in both layers. The resulting composite system then becomes a standard spinless chiral $p \pm ip'$ superconductor hosting non-Abelian MZMs when the disconnected FS in each layer coalesce into a single FS -- a detailed discussion of this is given in Supplementary Note 8. Thus, the alternative assumption of chiral $p$-wave pairing does not alter our conclusion that a maximal twist is required to turn the system into a non-Abelian topological phase, which further fortifies the connection between the novel and relatively unexplored large-angle moir\'{e} physics and non-Abelian topological superconductivity.

\section*{Methods}
\subsection{A. Dual momentum-space tight-binding (DMSTB) model}

Here we briefly outline the basic idea behind our generalized DMSTB model and present a detailed derivation in Supplementary Note 3. 

The DMSTB model is rooted in the observation that for any given momentum $\bm{k}_0$, the sets of Bloch states involved in the inter-layer coupling Hamiltonian Eq.\ \ref{eq:Interlayer} are $\mathcal{S}_1(\bm{k}_0 ) =  \{ \ket{\bm{k}_0 + \bm{\tilde{G}} ,1}, \forall \bm{\tilde{G}} \}$ in layer 1, and $\mathcal{S}_2(\bm{k}_0) = \{ \ket{\bm{k}_0 + \bm{G} ,2}, \forall \bm{G} \}$ in layer 2. By viewing the Bloch states $\ket{\bm{k}_m = \bm{k}_0 + \bm{\tilde{G}}_m, 1} \in \mathcal{S}_1(\bm{k}_0)$ and $\ket{\bm{\tilde{k}}_n = \bm{k}_0 + \bm{G}_n, 2} \in \mathcal{S}_2(\bm{k}_0)$ as ``Wannier orbitals" localized at the \emph{dual momentum-space lattice sites} $\bm{k}_m$ and $\bm{\tilde{p}}_n \equiv \bm{k}_0 - \bm{\tilde{k}}_n \equiv -\bm{G}_n$, the inter-layer coupling in Eq.\ \ref{eq:Interlayer} can be regarded as `inter-site hopping' between $\bm{k}_m$ and $ \bm{\tilde{p}}_n $ with `hopping strength' $t_{\perp, mn}(\bm{k}_0) = t_{\perp}(\bm{k}_m - \bm{\tilde{p}}_n)$ determined precisely by the geometric distance $|\bm{k}_m - \bm{\tilde{p}}_n|$ (see Supplementary Note 3). Importantly, the rapidly decaying character of $t_{\perp}(\bm{q})$ implies weak hybridization among Wannier orbitals (Bloch states) at $\bm{k}_m$ and $\bm{\tilde{p}}_n$, and states in the twisted double-layer live predominantly only on a small number of $\bm{k}_m$ and $\bm{\tilde{p}}_n$ points. Thus, there exists an approximately \textit{closed} finite subspace over which an effective Hamiltonian $\mathcal{H}_{0, \rm eff}$ can be constructed and the mapping from $\bm{k}_0$ to the set of eigenvalues $E_p(\bm{k}_0)$ ($p$: band index) of $\mathcal{H}_{0, \rm eff}(\bm{k}_0)$ then defines the band structure and the Fermi surface reformulated in the $\bm{k}_0$-space.

For $\theta \simeq 30^{\circ}$, states near $\pm K$ $(\tilde{K})$ in layer 1 (2) (Fig.\ \ref{FIG2}b) are well covered by the 12 dual momentum-space lattice points $\bm{k}_m,\ m=1,2,...,6$ and $\bm{\tilde{p}}_n,\ n = 1,2,...,6$ indicated in Fig.\ \ref{FIG2}c. 
This motivates us to consider $\mathcal{H}_{0, \rm eff}(\bm{k}_0)$ which includes these 12 sites together with leading-order corrections from their nearest-neighboring points $\bm{k}_{m}, \bm{\tilde{p}}_{n}$ with $m,n =0, 7,...,12$ as illustrated in Fig.\ \ref{FIG2}c. As further justified in Supplementary Note 3, the leading nearest-neighbor hopping terms between $\bm{k}_{m}, \bm{\tilde{p}}_{n}$ in such approximation scheme accounts exactly for the leading-order inter-layer terms depicted in Fig.\ \ref{FIG2}b. 
The effective normal-state Hamiltonian of a near-30$^{\circ}$-twisted double-layer then reads $\mathcal{H}_{0, \rm eff} = \sum_{\bm{k}_0} \mathcal{H}_{0}(\bm{k}_0)$ with 
\begin{align}\label{eq:normalstateH}
\mathcal{H}_{0}(\bm{k}_0) &= \sum^{12}_{m=0} \xi_1(\bm{k}_m) c^{\dagger}_{1}( \bm{k}_m) c_{1}( \bm{k}_m) \nonumber \\
&+  \sum^{12}_{n=0}\xi_2(\bm{\tilde{k}}_n)c^{\dagger}_{2}(\bm{\tilde{k}}_n) c_{2}(\bm{\tilde{k}}_n) \\ \notag
&- \sum^{12}_{m,n=0} \left[t_{\perp}(\bm{k}_m - \bm{\tilde{p}}_n) c^{\dagger}_1( \bm{k}_m) c_{2}(\bm{\tilde{k}}_n) + {\rm h.c.}\right].
\end{align}

To verify that $\mathcal{H}_{0, \rm eff}$ in Eq.\ \ref{eq:normalstateH} provides an accurate description of the normal-state fermiology near the maximal twist, we note that the approach above applies to any twist angle close to  $30^{\circ}$, including commensurate twist angles where the bilayer forms a periodic moir\'{e} superlattice and the band structure can be computed {\em exactly} via a real-space lattice model. As a convenient test case we consider commensurate angle $\theta_{c} = 2\sin^{-1}(\sqrt{3}/(2\sqrt{13})) \approx 27.8^{\circ}$ which gives rise to a moir\'{e} unit cell with 26 sites. As we demonstrate in Supplementary Note 3, $\bm{k}_0$ becomes exactly the crystal momentum at $\theta_{c}$, and $\mathcal{H}_{0, \rm eff}$ reproduces the electronic bands of the moir\'{e} lattice model with excellent accuracy. We further note that the summation over $\bm{k}_0$ in Eq.\ \ref{eq:normalstateH} should be restricted to within an area the size of the moir\'{e} Brillouin zone at $\theta = \theta_{c}$ to avoid over-counting of the degrees of freedom.

We further note that inter-layer spatial displacements between two layers due to angular twist can generally introduce nonzero phases in the inter-layer tunneling term in Eq.\ \ref{eq:normalstateH} \cite{Wu, Benjamin2}. As we explain in Supplementary Note 9, such phases cancel out in the total phase of Cooper pairs due to the spinless time-reversal symmetry $\mathcal{T}'$ and thus do not affect our analysis on the superconducting phase.
 
\subsection{B. Derivation of mean-field gap equation for twisted double-layer STVS superconductors}

To derive the mean-field gap equations for the twisted double-layer, we first rewrite the total interaction $\mathcal{V}_{\rm tot} = \mathcal{V}^{(1)} + \mathcal{V}^{(2)}$ in the band basis. Note that the fermionic operator creating an electron at $\bm{k}_0$ in band $p$ is given by
\begin{equation}
\label{eq:eigenvectors}
a^{\dagger}_p(\bm{k}_0) = \sum_{m} u_{pm}(\bm{k}_0) c^{\dagger}(\bm{k}_m) + \sum_{n} u_{pn}(\bm{k}_0) c^{\dagger}(\bm{\tilde{k}}_n),
\end{equation}
where the coefficients $u_{pm}(\bm{k}_0), u_{pn}(\bm{k}_0)$ can be found by exact numerical diagonalization of $\mathcal{H}_{0, \rm eff}$ in Eq.\ \ref{eq:normalstateH}. Fermionic operators in $\mathcal{V}^{(1)}$ and $\mathcal{V}^{(2)}$ can be rewritten as $c^{\dagger}(\bm{k}_m) = \sum_{p} u^{\ast}_{mp}(\bm{k}_0)a^{\dagger}_p(\bm{k}_0)$, $c^{\dagger}(\bm{\tilde{k}}_n) = \sum_{p} u^{\ast}_{np}(\bm{k}_0)a^{\dagger}_p(\bm{k}_0)
$. Using the one-to-one correspondence $\bm{k}_m \equiv \bm{k}_{0} + \bm{\tilde{G}}_m$ and $\bm{\tilde{k}}_n \equiv \bm{k}_{0} + \bm{G}_n$, the pairing interaction in the band basis becomes
\begin{eqnarray}\label{eq:totV}
\mathcal{V}_{\rm eff} = -U_0 (F^{\dagger}_{1}F_{1} + F^{\dagger}_{2}F_{2}), 
\end{eqnarray}
where $F^{\dagger}_{l} \equiv \sum_{\bm{k}_0, p} f_{l, p}(\bm{k}_0) a_p^{\dagger}(\bm{k}_0) a_p^{\dagger}(-\bm{k}_0)$ is the pair creation operator in layer $l = 1,2$, with
\begin{equation}
\begin{aligned}\label{eq:basisfunc}
f_{1, p}(\bm{k}_0) &= \sum_{m} \Lambda_{m,p}(\bm{k}_0) f_1(\bm{k}_0 + \bm{\tilde{G}}_m), \\
f_{2, p}(\bm{k}_0) &= \sum_{n} \Lambda_{n,p}(\bm{k}_0)f_2(\bm{k}_0 + \bm{G}_n).
\end{aligned}
\end{equation}
Here, $\Lambda_{m,p}(\bm{k}_0) \equiv u^{\ast}_{m,p}(\bm{k}_0) u^{\ast}_{-m,p}(-\bm{k}_0)$ and $\Lambda_{n,p}(\bm{k}_0) \equiv u^{\ast}_{n,p}(\bm{k}_0) u^{\ast}_{-n,p}(-\bm{k}_0)$ denote the form factors arising generally from projecting the interaction onto the band basis \cite{Note}, and we introduced the shorthand notation $u^{\ast}_{-m,p}(-\bm{k}_0)$ to denote the coefficient associated with $c^{\dagger}(-\bm{k}_m)$ and $a^{\dagger}_p(-\bm{k}_0)$. The standard mean-field reduction for Eq.\ \ref{eq:totV} leads to the gap equation $\Delta_{l,p} \equiv -U_0 \braket{F_{l,p}}$ for the pairing $\Delta_{l,p}$ in the BdG Hamiltonian in Eq.\ \ref{eq:HBdG} for the superconducting state in the twisted double-layer, and $\Delta_{l,p}$ are obtained by minimizing $\mathcal{F}_{SC}$ in Eq.\ \ref{eq:FreeEnergy}.

\section*{Acknowledgements}

The authors thank Oguzhan Can, Rafael Haenel, Christine Au-Yeung, Andrea Damascelli, Andreas Schnyder for illuminating discussions and communications. This work was supported by NSERC and the Canada First Research Excellence Fund, Quantum Materials and Future Technologies Program. B.T.Z. acknowledges the support of the Croucher Foundation.\\

\section*{Note}

During the preparation of this manuscript, we became aware of a recent preprint \cite{Liu} which suggested the possibility of chiral $f \pm if'$ pairing in the context of high-angular momentum-superconductivity in large-angle-twisted homobilayer systems.

\clearpage

\onecolumngrid
\begin{center}
\textbf{\large Supplementary Information for ``Non-Abelian topological superconductivity 
in maximally twisted double-layer spin-triplet valley-singlet superconductors"} \\[.2cm]
Benjamin T. Zhou$^{1}$, Shannon Egan$^{1}$, Dhruv Kush$^1$, Marcel Franz$^{1}$\\[.1cm]
{\itshape ${}^1$Department of Physics and Astronomy \& Stewart Blusson Quantum Matter Institute,
University of British Columbia, Vancouver BC, Canada V6T 1Z4}	\\
\end{center}
\setcounter{equation}{0}
\setcounter{section}{0}
\setcounter{figure}{0}
\setcounter{table}{0}
\setcounter{page}{1}
\renewcommand{\theequation}{S\arabic{equation}}
\renewcommand{\thetable}{S\arabic{table}}
\renewcommand{\thesection}{\arabic{section}}
\renewcommand{\thefigure}{S\arabic{figure}}
\makeatletter

\maketitle

\section*{Supplementary Note 1: Comparison among twisted STVS superconductors, twisted cuprates and small-angle twisted graphene/TMDs} \label{AppendixA}

As mentioned in the main text, the physics underlying the non-Abelian topological superconductivity (TSC) in maximally twisted STVS superconductors is essentially different from both twisted cuprates and small-angle twisted graphene/transition-metal dichalcogenides (TMDs). Here, we illustrate these differences in terms of their distinctive symmetry, topology, and microscopic descriptions.

We first compare the twisted STVS superconductors to the case of twisted cuprates studied in Ref. \cite{Marcel2}. Apart from important  differences already mentioned in the main text, \textit{i.e.}, the statistics of low-energy excitations (non-Abelian versus Abelian) and microscopic modeling (26-band DMSTB model versus continuum model based on parabolic bands and constant inter-layer coupling), we further note that microscopic mechanisms behind the emergence of chiral superconducting phases in these two cases are also \textbf{different}, despite their formal similarities in the Ginzburg-Landau descriptions. In particular, in the case of twisted STVS superconductors, the two orthogonal nodal $f$-wave order parameters in the moir\'{e} Brillouin zone of the twisted double-layer, which form the basis of the chiral $f$-wave phase, are \textbf{reconstructed} in the basis of moir\'{e} bands in maximally twisted double-layer. Microscopically, the form of the two orthogonal reconstructed nodal $f$-wave components, as well as their symmetry properties, \textbf{cannot be directly inferred from the $f$-wave symmetry of the order parameter in each isolated layer}. Instead, they must be \textbf{re-derived} through the projected pairing interactions in the moir\'{e} bands in the twisted double-layer, as outlined in Eqs.\ 14-16 in subsection B of the Methods section.
%In other words, without the large-angle moir\'{e} physics established through the microscopic DMSTB model, it is not possible to classify the point group symmetries of $\psi_1$ and $\psi_2$ listed in Table I of the main text, or to construct the Ginzburg-Landau functional in Eq.\ 6. 
This exemplifies the indispensable role of large-angle moir\'{e} physics in the chiral $f \pm if'$ phase in twisted STVS superconductor.

On the other hand, moir\'{e} effects play no essential role in twisted bilayer cuprates, and the twisted double-layer inherits two orthogonal $d$-wave components directly from the $d$-wave order parameters in each isolated layer \cite{Marcel2}. The absence of moir\'{e} physics in twisted cuprates allows a relatively simple Ginzburg-Landau analysis based on the $d$-wave symmetry of the order parameters in each layer. Overall, the microscopic description of the twisted STVS superconductor is at a different level of complexity from that in twisted cuprates.

Next, we point out that the large-angle moir\'{e} physics behind the maximally twisted STVS superconductor is also \textbf{fundamentally different} from the well-established moir\'{e} physics in small-angle twisted graphene and TMD systems, which can be well captured by the Bistritzer-MacDonald(BM)-type continuum models \cite{BM}. The BM-type Hamiltonians are defined within a single valley $+K$ or $-K$ only, which allows one to treat the two valleys separately. This approximation, which is associated with an emergent valley conservation symmetry $U_v(1)$ as discussed in the literature on small-angle twisted graphene \cite{Adrian1S} and TMDs \cite{Wu}, is valid only in the small-angle limit because the dominant inter-layer tunneling occurs between the same $K$-points in two layers \cite{BM}. In contrast, a key feature in the maximally twisted STVS superconductors is that states from the two different $\pm K$ valleys \textbf{hybridize}, as we demonstrate explicitly in Fig.\ 2 of the main text. The BM-type models fail to describe the physics in the large-angle limit where the two valleys must be put on an equal footing. From the symmetry point of view, the intervalley hybridization \textbf{violates} the fundamental $U_v(1)$-symmetry that underpins the well-established small-angle moir\'{e} physics in twisted graphene and TMDs.

It is important to note that the difference concerning the $U_{v}(1)$-symmetry discussed above has profound implications for the Abelian/non-Abelian nature of the low-energy excitations of the superconducting state in twisted two-valley systems. For any description respecting the  $U_{v}(1)$ symmetry,  the normal-state Hamiltonian necessarily involves two nearly degenerate moir\'{e} bands from two different valleys $\xi = \pm$, which are related by time-reversal symmetry. The general form of the BdG Hamiltonian with zero-momentum-pairing (\textit{i.e.}, inter-valley pairing) is given by $\mathcal{H}_{SC} = \sum_{\bm{k}_0 \in MBZ} \Psi^{\dagger}_{\bm{k}_0} H_{BdG}(\bm{k}_0) \Psi_{\bm{k}_0}$ with the Nambu basis $\Psi(\bm{k}_0) = \big(c_{+} (\bm{k}_0), c_{-} (\bm{k}_0), c^{\dagger}_{+}(-\bm{k}_0), c^{\dagger}_{-}(-\bm{k}_0) \big)^T$ and 
\begin{eqnarray}\label{eq:BdGSmallAngle}
H_{BdG}(\bm{k}_0) &=& 
\begin{pmatrix}
H_{0,+}(\bm{k}_0) & 0 & 0 & \Delta_{+-}(\bm{k}_0)\\
0 & H_{0,-}(\bm{k}_0) &  \Delta_{-+}(\bm{k}_0) & 0\\
0 &  \Delta^{\ast}_{-+}(\bm{k}_0) & -H^{\ast}_{0,+}(-\bm{k}_0) & 0\\
\Delta^{\ast}_{+-}(\bm{k}_0) & 0 & 0 & -H^{\ast}_{0,-}(-\bm{k}_0)
\end{pmatrix}.
\end{eqnarray}
Here, $\bm{k}_0$ is the momentum in the moir\'{e} Brillouin zone,  $H_{0, \xi = \pm}$ are the normal-state Hamiltonians for valley $\xi = \pm$, and $\Delta_{\xi \xi'}$ is the inter-valley pairing matrix. The Bogoliubov quasi-particles of $\mathcal{H}_{SC}$ above are generally in the form of $\gamma = u c_{+} + v c^{\dagger}_{-}$ (or alternatively, $\gamma' = u' c_{-} + v' c^{\dagger}_{+}$). Notably, due to the different valley flavors $+$, $-$ involved in the electron and hole part of $\gamma$, even if the pairing $\Delta_{+-}$ is topologically nontrivial, \textit{e.g.}, a chiral $p$-wave pairing with $\Delta_{+-}(\bm{k}_0) \simeq \Delta_0 (k_{0,x} + i k_{0,y})$, the topologically protected zero modes at system boundaries or vortex core can only arise as $\gamma_0 = (c_{+} + c^{\dagger}_{-})/\sqrt{2}$ , which \textit{cannot be self-conjugate} as $\gamma^{\dagger}_0 =  (c_{-} + c^{\dagger}_+)/\sqrt{2} \neq \gamma_0$. Instead, $\gamma_0$ must be regarded as a complex fermion that can be written as a combination of two \textit{different} Majorana operators: $\gamma_0 = \gamma_1 + i\gamma_2$ with $\gamma_1 = (c_{+} + c_{-} + c^{\dagger}_{+} + c^{\dagger}_{-})/2\sqrt{2}$ and $\gamma_2 = -i (c_{-} - c^{\dagger}_{-} - c_{+} + c^{\dagger}_{+} )/2\sqrt{2}$. Therefore, the zero-momentum-paired superconducting state in small-angle twisted graphene or TMDs can only support Abelian excitations. We note that the analysis above on the statistics of low-energy excitations also applies to an inter-valley paired chiral $p$-wave superconductor, a possibility for the SC2 phase in RTG that we discuss in detail in Supplementary Note 8.

In contrast, as we demonstrated in the main text, in the maximal twist limit of $\theta = 30^{\circ}$, the single connected Fermi surface in the moir\'{e} Brillouin zone is formed by hybridizing states from different valleys and layers ($U_v(1)$-violation) which enables a description without any band degeneracy enforced by flavor symmetries. This provides the crucial condition for the emergence of non-Abelian Majorana modes as we demonstrated in the main text. Our analysis above reveals a profound connection between the internal $U_{v}(1)$-symmetry in twisted two-valley systems and non-Abelian topological superconductivity. 

Some key differences between twisted STVS superconductors, twisted cuprates and small-angle twisted graphene/TMDs are summarized in Table \ref{tab:comparison} below. In summary, our finding of the non-Abelian topological superconductivity in twisted STVS superconductors serves as a notable example of how a change in Fermi surface topology driven by angular twist can lead to profound consequences for the symmetry, topology, and microscopic descriptions of twisted systems. 

\begin{table}[tp]
\caption{Comparison among twisted STVS superconductors, twisted cuprates, and small-angle twisted graphene/TMDs.\\}
\centering
\begin{tabular}{c|c|c|c}
\hline\hline
Twisted system & STVS superconductors & Cuprates & Graphene/TMDs\\
\hline
Twist angle & $\theta\sim 30^{\circ}$ & $\theta\sim 45^{\circ}$ & $\theta\sim 1^{\circ} - 2^{\circ}$ 
\\

FS topology in monolayer & 2 disconnected pockets ($\pm K$) &  Single connected pocket ($\Gamma$) & 2 disconnected pockets ($\pm K$)\\

Microscopic model & 26-site DMSTB model & Continuum model with parabolic bands & BM-type continuum model\\

Moir\'{e} physics & \textcolor{red}{Essential} & Inessential & \textcolor{red}{Essential}\\

Valley $U_v(1)$-symmetry & \textcolor{red}{Violated} & Not applicable & \textcolor{blue}{Preserved}\\

Topology & Chiral topological SC ($f \pm if'$) & Chiral topological SC ($d \pm id'$) & Possible topological SC\\

Statistics of excitations & \textcolor{red}{Non-Abelian} & Abelian & Abelian\\
\hline\hline
\end{tabular}
\label{tab:comparison}
\end{table}

\section*{Supplementary Note 2: Exponential decay model of inter-layer coupling} \label{AppendixB}

Here, we present details of the model we used to obtain the Fourier transform $t_{\perp}(\bm{q})$ shown in the in-set of Fig.\ 2c of the main text. For concreteness, we consider $p_z$-orbitals in the topmost layer of rhombohedrally stacked graphene and assume that the inter-layer tunneling is dominated by $\sigma$-bonds with an exponential decay as a function of geometric distance between atomic sites in two different layers. The tunneling strength is given by
\begin{eqnarray}\label{eq:interlayertunneling}
t_{\perp}(r) = g_0 e^{-\lambda \sqrt{r^2 + d^2}} \cos^2{\theta(r)},
\end{eqnarray}
where $\bm{r}$ is the in-plane separation, and  $\cos{\theta(r)}= \frac{d}{\sqrt{d^2 + r^2}}$ is the directional cosine for $\sigma$-bonding between $p_z$-orbitals. $\lambda$ characterizes the exponential decay length. Using the empirical relation $t_{\perp}(2a) \simeq 0.1 t_{\perp}(a)$, we obtain $\lambda = \log(10)/a$.

The Fourier transform of $t_{\perp}(r)$ is given by
\begin{eqnarray}\label{eq:FTT}
t_{\perp}(q) &=& \int_{0}^{\infty} dr r t_{\perp}(r) \int_{0}^{2\pi} d\theta e^{-i qr \cos \theta} \\\nonumber
               &=& 2\pi \int_{0}^{\infty} dr r J_0(qr) t_{\perp}(r),
\end{eqnarray}
where $J_0(x)$ is the Bessel function of order zero. Throughout our calculations, $t_{\perp}(q)$ is obtained by numerically integrating Eq.\ \ref{eq:FTT}. The parameters in $t_{\perp}(r)$ are set to be $g_0 = 45$ eV, $d = 3.4$ {\AA}, $a = 2.46$ {\AA}, such that the effective inter-layer tunneling strength for states near $K$-points reproduce the literature value $t_{\perp}(K) \approx 150$ meVs \cite{BM}, as shown in the inset of Fig.\ 2c of the main text with hopping parameter $t=1$ eV.

\section*{Supplementary Note 3: Details of DMSTB model} \label{AppendixC}

\begin{figure}
\centering
\includegraphics[width=0.5\textwidth]{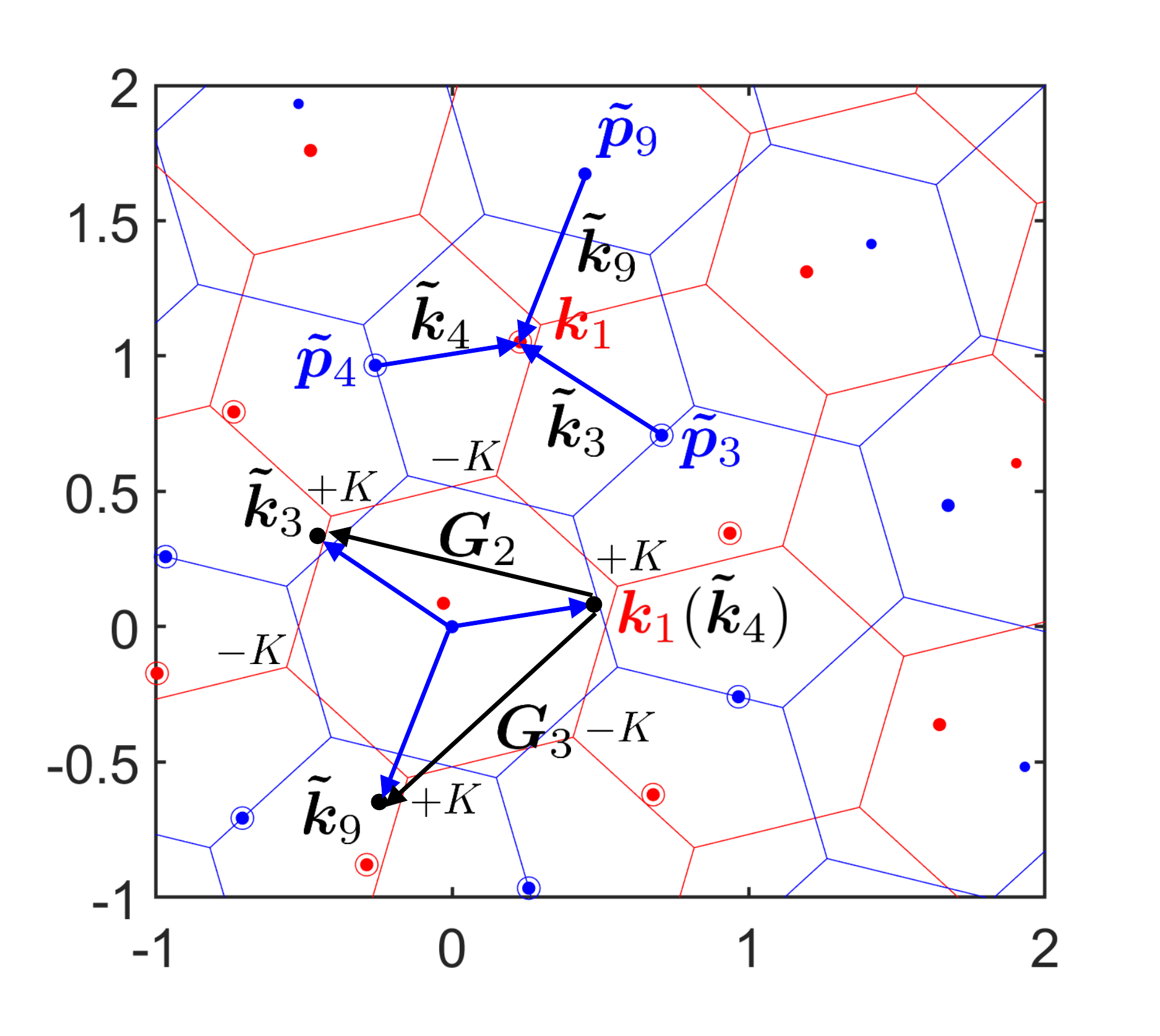}
\caption{A zoomed-in view of Fig.\ 2c of the main text with focus on $\bm{k}_1 =\bm{k}_0 + \bm{\tilde{G}}_1$ and its nearest-neighbors $\bm{\tilde{p}}_4, \bm{\tilde{p}}_3, \bm{\tilde{p}}_9$ in dual momentum-space lattice. $\bm{\tilde{k}}_4, \bm{\tilde{k}}_3, \bm{\tilde{k}}_9$ are the original Bloch momenta in layer 2 corresponding to $\bm{\tilde{p}}_4, \bm{\tilde{p}}_3, \bm{\tilde{p}}_9$, with $\bm{\tilde{k}}_n = \bm{k}_0 - \bm{\tilde{p}}_n$ as defined in the main text.}
\label{FIGB1}
\end{figure}

In the Methods section, we outlined the essential ingredients of the dual momentum-space tight-binding (DMSTB) model $\mathcal{H}_{0, \rm eff}$ (Eq.13 of the main text), and argued that the model provides a reliable description of the twisted double-layer STVS superconductors for $\theta$ around 30$^{\circ}$. Here, we support these claims by presenting relevant technical details of the DMSTB model. Our derivation here parallels the derivation given in Ref.\ \cite{MKS} for the honeycomb lattice.  

First, we mentioned in the Methods section that the coupling strength between states at $\bm{k}_m = \bm{k}_0 + \bm{\tilde{G}}_m$ and $\bm{\tilde{k}}_n = \bm{k}_0 + \bm{G}_n $ is given by the geometric distance between the dual momentum-space lattice sites: $|\bm{k}_m - \bm{\tilde{p}}_n|$. Here, we present a detailed proof: according to Eq.\ 2 of the main text, for any given $\bm{k}_0$ the Bloch state $\ket{\bm{k}_0,1}$ in layer 1  couples to the set of states $\mathcal{S}_2(\bm{k}_0) = \{ \ket{\bm{k}_0 + \bm{G} - \bm{\tilde{G}},2} \}$ in layer 2 for all pairs of $\bm{G}, \bm{\tilde{G}}$. As momentum is conserved modulo $\bm{\tilde{G}}$ in layer 2, $\mathcal{S}_2(\bm{k}_0) = \{ \ket{\bm{k}_0 + \bm{G} - \bm{\tilde{G}},2} \} \equiv \{ \ket{\bm{k}_0 + \bm{G},2} \}$. Similarly, given any $\ket{\bm{k}_0 + \bm{G}_0, 2}$ in $\mathcal{S}_2(\bm{k}_0)$:  the set of Bloch states in layer 1 which couple to $\ket{\bm{k}_0 + \bm{G}_0, 2}$ is given by $\mathcal{S}_1(\bm{k}_0 + \bm{G}_0) = \{ \ket{\bm{k}_0 + \bm{G}_0 + \bm{\tilde{G}} - \bm{G},1} \}$ for all pairs of $\bm{G}, \bm{\tilde{G}}$. Note again that $\mathcal{S}_1(\bm{k}_0 + \bm{G}_0) \equiv \{ \ket{\bm{k}_0 + \bm{\tilde{G}},1} \}$ as momentum is conserved modulo $\bm{G}$ in layer 1, and $\bm{G}_0 \in \{ \bm{G} \}$. This indicates that all inter-layer processes involving $\bm{k}_0$ occur between $\mathcal{S}_1(\bm{k}_0) \equiv \{ \ket{\bm{k}_0 + \bm{\tilde{G}},1} \}$ and $\mathcal{S}_2(\bm{k}_0) \equiv \{ \ket{\bm{k}_0 + \bm{G},2} \}$. The states $\ket{\bm{k}_0, 1}$ and $\ket{\bm{k}_0, 2}$ are automatically included in $\mathcal{S}_1(\bm{k}_0)$ and $\mathcal{S}_2(\bm{k}_0)$ for $\bm{G} = \bm{\tilde{G}} = \bm{0}$.

To work out the coupling strengths between $\ket{\bm{k}_m,1}$ and $\ket{\bm{\tilde{k}}_n,2}$, we consider two arbitrary states $\ket{\bm{k}_0 + \bm{\tilde{G}_0}, 1} \in \mathcal{S}_1(\bm{k}_0)$ and $\ket{\bm{k}_0 + \bm{G}_0, 2} \in \mathcal{S}_2(\bm{k}_0)$, and identify $\bm{k}_0 + \bm{\tilde{G}_0} \equiv \bm{k}$ and $\bm{k}_0 + \bm{G}_0 \equiv \bm{\tilde{k}}$ as the pair of Bloch momenta involved in $T(\bm{k}, \bm{\tilde{k}})$ (Eq.\ 2 of the main text). Our goal here is to find $\bm{G}_1$ and $\bm{\tilde{G}}_1$ such that $\bm{k}_0 + \bm{\tilde{G}_0} + \bm{G}_1 = \bm{k}_0 + \bm{G}_0 + \bm{\tilde{G}}_1$, and the coupling strength is then given by $t_{\perp}(\bm{k}_0 + \bm{\tilde{G}_0} + \bm{G}_1) $. Obviously, the condition is met only when $\bm{G}_1 = \bm{G}_0, \bm{\tilde{G}}_1 = \bm{\tilde{G}}_0$, and the coupling strength between $\ket{\bm{k}_0 + \bm{\tilde{G}_0}, 1} $ and $\ket{\bm{k}_0 + \bm{G}_0, 2} $ is precisely $t_{\perp}(\bm{k}_0 + \bm{\tilde{G}}_0 + \bm{G}_0)$. Given the definition of the dual momentum-space sites $\bm{k}_m = \bm{k}_0 + \bm{\tilde{G}}_m$ and $\bm{\tilde{p}}_n \equiv \bm{k}_0 - \bm{\tilde{k}}_n \equiv -\bm{G}_n$, one can rewrite $t_{\perp}(\bm{k}_0 + \bm{\tilde{G}}_m + \bm{G}_n) = t_{\perp}(\bm{k}_m - \bm{\tilde{p}}_n)$, which completes the proof.

Next, we show that by including nearest-neighbor sites $\bm{k}_{m=7,...,12}$ and $\bm{\tilde{p}}_{n=7,...,12}$ surrounding the 12-sites $\bm{k}_{m=1,...,6}$ and $\bm{\tilde{p}}_{n=1,...,6}$ as depicted in Fig.\ 2c of the main text, all leading-order inter-layer coupling terms involving states near $\pm K (\pm \tilde K)$ are incorporated in $\mathcal{H}_{0, \rm eff}$. Without loss of generality, we consider 
a representative point $\bm{k}_1 \equiv \bm{k}_0 + \bm{\tilde{G}}_1$ in Fig.\ 2c of the main text which is located near $+K$ point in layer 1 and thus can be easily compared to the scenario depicted in Fig.\ 2b of the main text. For illustrative purposes, a zoomed-in view of Fig.\ 2c of the main text with focus on $\bm{k}_1$ and its neighboring sites is presented in Fig.\ \ref{FIGB1}.

\begin{figure}
\centering
\includegraphics[width=0.8\textwidth]{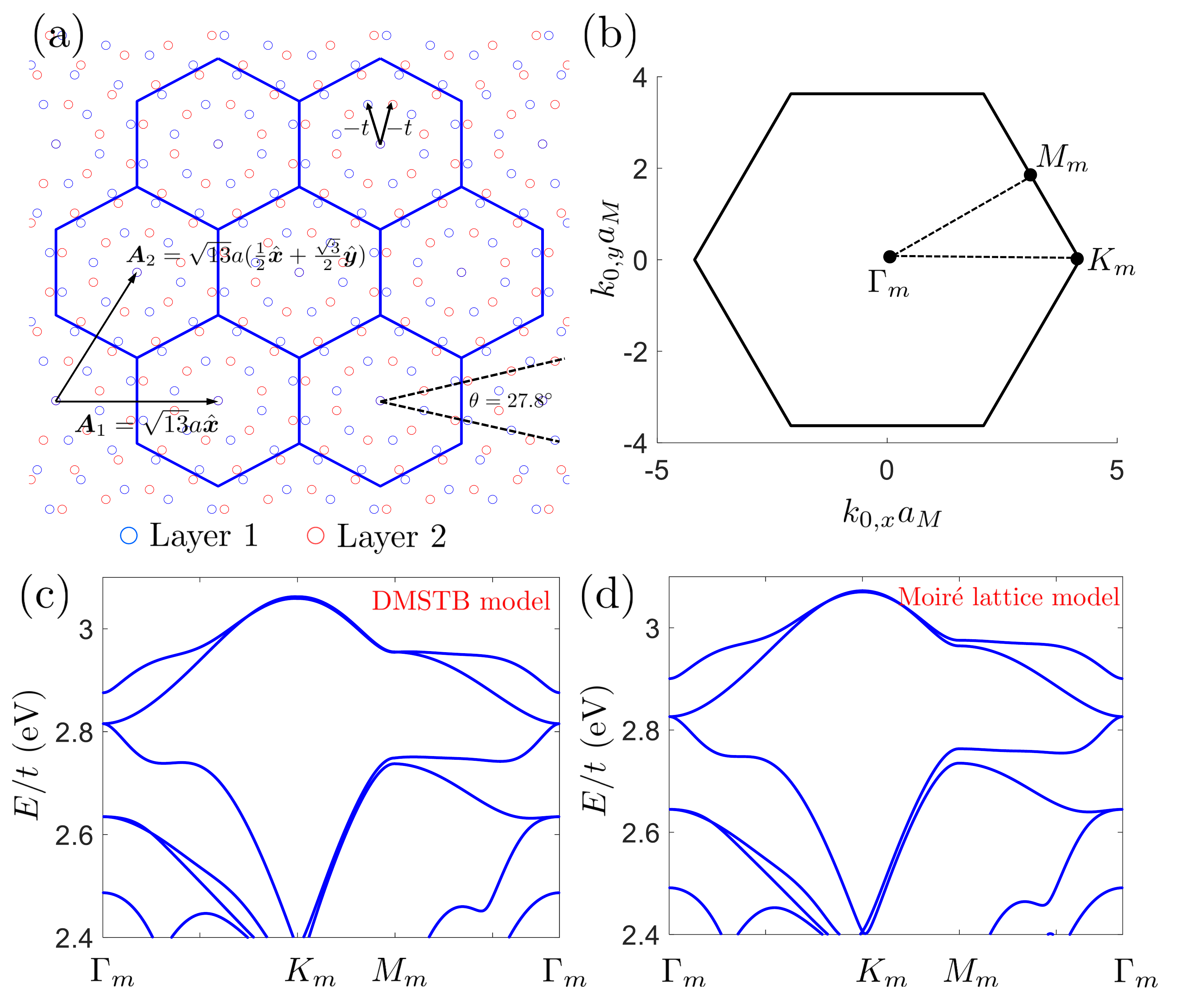}
\caption{(a) Formation of moir\'{e} superlattice from stacking two triangular lattices at twist angle $\theta_{c} = 2\sin^{-1}(\sqrt{3}/(2\sqrt{13})) \approx 27.8^{\circ}$. The moir\'{e} pattern forms a triangular superlattice with spacing $a_M = \sqrt{13} a$, and each hexagonal supercell contains 26 atomic sites, 13 from each constituent layer. (b) Moir\'{e} Brillouin zone of the moir\'{e} superlattice formed at $\theta = \theta_{c}$. (c-d) Energy bands obtained from the DMSTB model ($\mathcal{H}_{0, \rm eff}$ in Eq.\ 3 of the main text) and the moir\'{e} lattice model $\mathcal{H}_{\rm M}$ in Eq.\ \ref{eq:Hmoire}. }
\label{FIGB2}
\end{figure}

Using the defining relations $\bm{\tilde{p}}_n = \bm{k}_0 - \bm{\tilde{k}}_n$ and $\bm{k}_m = \bm{k}_0 + \bm{\tilde{G}}_m, \bm{\tilde{k}}_n = \bm{k}_0 + \bm{G}_n$, we note that the inter-site separation between any pair of $\bm{k}_m$ and $\bm{\tilde{p}}_n$ is precisely $\bm{k}_m - \bm{\tilde{p}}_n = \bm{k}_0 + \bm{\tilde{G}}_m + \bm{G}_n = \bm{\tilde{k}}_n + \bm{\tilde{G}}_m$. Note that $\bm{\tilde{G}}_m$ is a reciprocal lattice vector in layer 2, hence $\ket{\bm{\tilde{k}}_n + \bm{\tilde{G}}_m, 2} \equiv \ket{\bm{\tilde{k}}_n, 2}$. In other words, $\bm{k}_m - \bm{\tilde{p}}_n$ can be used to map out the original $\bm{\tilde{k}}_n$ corresponding to $\bm{\tilde{p}}_n$. As shown in Fig.\ \ref{FIGB1}, we map out the original Bloch momenta $\bm{\tilde{k}}_4, \bm{\tilde{k}}_3, \bm{\tilde{k}}_9$ in layer 2 from subtracting $\bm{\tilde{p}}_4, \bm{\tilde{p}}_3, \bm{\tilde{p}}_9$ from $\bm{k}_1$. Notably, $\bm{\tilde{k}}_4$ overlaps with $\bm{k}_1$, while $\bm{\tilde{k}}_3$ and $\bm{\tilde{k}}_9$ are related to $\bm{k}_1$ by exactly $\bm{G}_2$ and $\bm{G}_3$ as shown in Fig.\ 2b of the main text. Thus, the nearest-neighbor hopping from dual momentum-space lattice sites $\bm{\tilde{p}}_4, \bm{\tilde{p}}_3, \bm{\tilde{p}}_9$ to $\bm{k}_1$ correspond exactly to the three leading-order inter-layer tunneling processes with $\bm{\tilde{k}} = (\bm{k}, \bm{k}+\bm{G}_2, \bm{k} + \bm{G}_3)$ discussed in the main text. This confirms that all leading-order inter-layer processes for states near $\pm K$ and $\pm \tilde{K}$ are incorporated in $\mathcal{H}_{0, \rm eff}$.

Finally, we demonstrate the accuracy of the DMSTB model. As mentioned in the main text, the DMSTB model applies to a general $\theta$ close to 30$^{\circ}$, which includes a commensurate angle $\theta_{c}= 2\sin^{-1}(\sqrt{3}/(2\sqrt{13})) \approx 27.8^{\circ}$. The lattice geometry at $\theta_{c}$, shown in Fig.\ \ref{FIGB2}a, indicates a triangular moir\'{e} superlattice with spacing $a_M = \sqrt{13} a$ and each hexagonal super-cell containing 26 atomic sites (13 sites per layer). Considering the same electron hopping $-t$ in the monolayer, and the same form of $t_{\perp}(\bm{r})$ as in Eq.\ \ref{eq:interlayertunneling} with $\bm{r}$ discretized as $\bm{r} \rightarrow \bm{R} - \bm{\tilde{R}}$, one can directly construct a real-space moir\'{e} lattice model for the twisted double-layer
\begin{eqnarray}\label{eq:Hmoire}
\mathcal{H}_{\rm M} &=& -t \sum_{\braket{\bm{R}, \bm{R}'}} c^{\dagger}(\bm{R}) c(\bm{R}') + {\rm h.c.} - \mu \hat{n}(\bm{R}) \nonumber\\
&-& t \sum_{\braket{\bm{\tilde{R}}, \bm{\tilde{R}}'}} c^{\dagger}(\bm{\tilde{R}}) c(\bm{\tilde{R}}') + {\rm h.c.} - \mu \hat{n}(\bm{\tilde{R}}) \\\nonumber
&-& \sum_{\braket{\bm{R}, \bm{\tilde{R}}}'} t_{\perp}(\bm{R} - \bm{\tilde{R}}) c^{\dagger}(\bm{R}) c(\bm{\tilde{R}}) + {\rm h.c.},
\end{eqnarray}
where $\hat{n}(\bm{R}) = c^{\dagger}(\bm{R}) c(\bm{R})$ denotes the occupancy at site $\bm{R}$. The symbol $\braket{\bm{R},\bm{R}'}$ in the first two rows indicates the summation is taken up to nearest-neighboring sites $\bm{R},\bm{R}'$ within the triangular lattice in each layer, while $\braket{\bm{R},\bm{\tilde{R}}}'$ in the last row indicates the summation is taken over all sites $\bm{R}$ and $\bm{\tilde{R}}$ up to nearest-neighboring \emph{super-cells} in the moir\'{e} superlattice.  

The commensurate superlattice structure allows us to Fourier transform $\mathcal{H}_{\rm M}$ in Eq.\ \ref{eq:Hmoire} and obtain the band structure in the moir\'{e} Brillouin zone (Fig.\ \ref{FIGB2}b) by appealing to the Bloch theorem. On the other hand, one can obtain the electronic bands alternatively by setting $\theta = \theta_c \simeq 27.8^{\circ}$ in the DMSTB model ($\mathcal{H}_{0, \rm eff}$ in Eq.\ 13 of the main text) by identifying $\bm{k}_0$ as the lattice momentum defined on the moir\'{e} superlattice. For a direct comparison of the two approaches, we plot the topmost bands along the $\Gamma_m - K_m - M_m - \Gamma_m$ line in the moir\'{e} Brillouin zone obtained from both $\mathcal{H}_{0, \rm eff}$ in Eq.\ 13 of the main text and $\mathcal{H}_{\rm M}$ in Eq.\ \ref{eq:Hmoire} as shown in Fig.\ \ref{FIGB2}c-d. Clearly, the energy bands obtained from the two methods agree very well with each other. This establishes the DMSTB model as a valid approximation for the normal-state fermiology of near-maximally-twisted STVS superconductors.

\section*{Supplementary Note 4: Derivation of Ginzburg-Landau coefficients from imaginary-time path integral} \label{AppendixD}

In this section, we provide a microscopic derivation of the Ginzburg-Landau (GL) coefficients and their relations using the imaginary-time path integral formalism based on the mean-field Hamiltonian $\mathcal{H}_{\rm BdG}$ in Eq.\ 4 of the main text. In particular, as we mentioned in the main text, at $\theta = 30^{\circ}$, the Ginzburg-Landau (GL) coefficient $b_0$ associated with single-pair tunneling vanishes, and the $\varphi$-dependent part of the free energy $\mathcal{F}_{SC}(\varphi) -\mathcal{F}_0 = 2 c_0 \psi^4 \cos(2\varphi)$ is minimized at a complex phase $\varphi = \pm \pi/2$ if $c_0 >0$ holds. In the following, we prove that $c_0 > 0$ strictly holds and the coefficients $\beta_0$, $a_0$ and $c_0$ in Eq.\ 6 of the main text satisfy the relations (i) $a_0 = 4c_0$, (ii) $\beta_0 > 2c_0$. As discussed in Supplementary Note 5 below, these relations govern the energetics of the superconductor and guarantee that the solution with $|\psi_1| = |\psi_2| \neq 0$ corresponds to the global minimum of the free energy. Without loss of generality, we focus on band $p = 2$ and drop the band index in the following. To set up the path integral we introduce grassmann fields ($\bar{\psi}_{\bm{k}_0}, \psi_{\bm{k}_0}$) corresponding to the fermionic operators $\hat{a}^{\dagger}(\bm{k}_0), \hat{a}(\bm{k}_0)$ in Eq.\ 4 of the main text, and bosonic fields ($\bar{\Delta}_{l}, \Delta_{l}$) representing the pair creation operators $F^{\dagger}_{l}, F_{l}$ in Eq.\ 15 in the Methods section of the main text. 
In terms of ($\bar{\psi}, \psi$) and ($\bar{\Delta}_{l}, \Delta_{l}$) the  partition function $\mathcal{Z}$ can be expressed as an imaginary-time path integral
\begin{equation}\label{eq:partitionfnc}
\mathcal{Z} = \int \prod_{n, \bm{k}_0} {\cal D}(\bar{\psi}_{n,\bm{k}_0},  \psi_{n,\bm{k}_0}) \prod_{l=1,2} {\cal D}(\bar{\Delta}_{l}, \Delta_{l}) e^{-\beta\mathcal{S}_{\rm eff}},
\end{equation}
where $n$ is the index of the fermionic Matsubara frequency $\omega_n = (2n+1)\pi/\beta$,  and $\mathcal{S}_{\rm eff}$ is the effective action given by
\begin{equation}
\mathcal{S}_{\rm eff} =  -\frac{|\Delta_1|^2 +|\Delta_2|^2}{U_0} -\frac{1}{2} \sum_{n,\bm{k}_0} \bar{\Psi}_{n,\bm{k}_0} \mathcal{G}^{-1}(\bm{k}_0, i \omega_n) \Psi_{n,\bm{k}_0},
\end{equation}
with the fermionic fields arranged in the Nambu basis $\Psi_{n,\bm{k}_0} \equiv (\psi_{n,\bm{k}_0}, \bar{\psi}_{_{-n,-\bm{k}_0}})^{T}$, $\bar{\Psi}_{n,\bm{k}_0} \equiv (\bar{\psi}_{n,\bm{k}_0}, \psi_{-n,-\bm{k}_0})^{T}$. $\mathcal{G}(\bm{k}_0, i \omega_n)$ is the Gor'kov Green's function defined by
\begin{align}
\mathcal{G}^{-1}(\bm{k}_0, i \omega_n) = i \omega_n I_{2\times2} - 
\begin{pmatrix}
\xi(\bm{k}_0) & \Delta(\bm{k}_0)\\
\Delta(\bm{k}_0)^\ast & -\xi(-\bm{k}_0)
\end{pmatrix},
\end{align}
where $\Delta(\bm{k}_0) =\Delta_{1} f_{1}(\bm{k}_0) + \Delta_{2} f_{2}(\bm{k}_0)$ and $f_{1}(\bm{k}_0)$, $f_{2}(\bm{k}_0)$ are the basis functions shown in Fig.\ 3a of the main text. By carrying out the Gaussian integral over fermionic fields in Eq.\ \ref{eq:partitionfnc}, we derive the mean-field free energy as
\begin{equation}\label{eq:FMatsubara}
\mathcal{F}_{SC} = \frac{|\Delta_1|^2 +|\Delta_2|^2}{U_0} -\frac{1}{2\beta} \sum_{n, \bm{k}_0} \log(\det[\mathcal{G}^{-1}(\bm{k}_0, i \omega_n)]).
\end{equation}

To expand the free energy in Eq.\ \ref{eq:FMatsubara} in terms of $\Delta_1, \Delta_2$, we note that the inverse of Gor'kov Green's function can be rewritten as: $\mathcal{G}^{-1}(\bm{k}_0, i \omega_n) = \mathcal{G}_0^{-1}(\bm{k}_0, i \omega_n)[1-D(\bm{k}_0, i \omega_n)]$, where $\mathcal{G}_{0}^{-1}(\bm{k}_0, i\omega_n) = \textrm{diag}[i\omega_n - \xi(\bm{k}_0), i\omega_n + \xi(-\bm{k}_0)]$ and 
\begin{align}
D(\bm{k}_0, i \omega_n) = 
\begin{pmatrix}
0 & \frac{\Delta(\bm{k}_0)}{i\omega_n - \xi(\bm{k}_0)}\\
\frac{\Delta(\bm{k}_0)^\ast}{i\omega_n + \xi(-\bm{k}_0)} & 0
\end{pmatrix}.
\end{align}
Up to fourth-order terms involving $\Delta_1$ and $\Delta_2$, the GL free energy can be found by expanding
\begin{align}
\log(\det\mathcal{G}^{-1}) &= \textrm{Tr}\log[\mathcal{G}_0^{-1}(1-D)]\\\nonumber
&= \textrm{Tr} \log[\mathcal{G}_0^{-1}] - \frac{1}{2} \textrm{Tr}(D^2) - \frac{1}{4}\textrm{Tr}(D^4) + ...
\end{align}
in Eq.\ \ref{eq:FMatsubara}. Note that $D^2$, $D^4$ are $2 \times 2$ identity matrices: $D^2 = A(\bm{k}_0, i \omega_n) I_{2 \times 2}$, $D^4 = B(\bm{k}_0, i \omega_n) I_{2 \times 2}$ with
\begin{eqnarray}
A(\bm{k}_0, i \omega_n) &=& \frac{ \big(\Delta_{1} f_{1}(\bm{k}_0) + \Delta_{2} f_{2}(\bm{k}_0)\big) \big(\Delta^{\ast}_{1} f_{1}(\bm{k}_0) + \Delta^{\ast}_{2} f_{2}(\bm{k}_0)\big)  }{(i\omega_n -\xi(\bm{k}_0)) (i\omega_n +\xi(-\bm{k}_0))}, \\\nonumber
B(\bm{k}_0, i \omega_n) &=& \frac{ [\Delta_{1} f_{1}(\bm{k}_0) + \Delta_{2} f_{2}(\bm{k}_0)]^2 [\Delta^{\ast}_{1} f_{1}(\bm{k}_0) + \Delta^{\ast}_{2} f_{2}(\bm{k}_0)]^{2}  }{[i\omega_n -\xi(\bm{k}_0)]^2 [i\omega_n +\xi(-\bm{k}_0)]^2 }.
\end{eqnarray}

To derive the GL coefficients $\alpha_0, \beta_0, a_0, c_0$ in Eq.\ 6 of the main text, we expand the numerator in $A(\bm{k}_0, i \omega_n), B(\bm{k}_0, i \omega_n)$ and identify the corresponding GL coefficient $\alpha_0, \beta_0$ as the following Matsubara sums:
\begin{align}\label{eq:GLBeta}
\alpha_0 &= \frac{1}{U_0} + \frac{1}{\beta} \sum_{n,\bm{k}_0} \frac{f^2_1(\bm{k}_0)}{[i\omega_n -\xi(\bm{k}_0)] [i\omega_n +\xi(-\bm{k}_0)] } = \frac{1}{U_0} + \frac{1}{\beta} \sum_{n,\bm{k}_0} \frac{f^2_2(\bm{k}_0)}{[i\omega_n -\xi(\bm{k}_0)] [i\omega_n +\xi(-\bm{k}_0)] } \\\nonumber
&= \frac{1}{U_0} - \frac{1}{\beta} \sum_{n,\bm{k}_0} \frac{f^2_1(\bm{k}_0) +  f^2_2(\bm{k}_0)}{2[\omega^2_n +\xi^2(\bm{k}_0)] },
\\\nonumber
\beta_0 &= \frac{1}{\beta} \sum_{n,\bm{k}_0} \frac{f^4_1(\bm{k}_0)}{[i\omega_n -\xi(\bm{k}_0)]^2 [i\omega_n +\xi(-\bm{k}_0)]^2 } = \frac{1}{\beta} \sum_{n,\bm{k}_0} \frac{f^4_2(\bm{k}_0)}{[i\omega_n -\xi(\bm{k}_0)]^2 [i\omega_n +\xi(-\bm{k}_0)]^2 } \\\nonumber
&= \frac{1}{\beta} \sum_{n,\bm{k}_0} \frac{f^4_1(\bm{k}_0) +  f^4_2(\bm{k}_0)}{2[\omega^2_n +\xi^2(\bm{k}_0)]^2 } > 0.
\end{align}
In arriving at the second line in both $\alpha_0$ and $\beta_0$ above, we made use of the relation $\xi(\bm{k}_0) = \xi(-\bm{k}_0)$ imposed by $\mathcal{T}$-symmetry in the normal state. We note that while for a general $\bm{k}_0$, $f_1(\bm{k}_0) \neq f_2(\bm{k}_0)$ as shown in Fig.\ 3a of the main text, we have the relation $f_1(-\bm{k}_{0,x}, \bm{k}_{0,y}) = f_2(\bm{k}_{0,x}, \bm{k}_{0,y})$, and $\xi(-\bm{k}_{0,x}, \bm{k}_{0,y}) = \xi(\bm{k}_{0,x}, \bm{k}_{0,y})$ due to the $C_{2y}$-symmetry of the $D_6$ point group. Upon summmation over all $\bm{k}_0$ the two different Matsubara sums become equal (first line of Eq.\ \ref{eq:GLBeta}), which allows us to rewrite $\alpha_0, \beta_0$ as the average of the two equal sums in the second line. 

It is worth noting that the coefficient $\alpha_0$ can be written as $\alpha_0(T) = 1/U_0 - \chi_{SC}(T)$, where the quantity
\begin{eqnarray}
\chi_{SC}(T) = - \frac{1}{\beta} \sum_{n,\bm{k}_0} \frac{f^2_{l = 1,2}(\bm{k}_0)}{[i\omega_n -\xi(\bm{k}_0)] [i\omega_n +\xi(-\bm{k}_0)] } = \frac{1}{\beta} \sum_{n,\bm{k}_0} \frac{f^2_1(\bm{k}_0) +  f^2_2(\bm{k}_0)}{2[\omega^2_n +\xi^2(\bm{k}_0)] } >0
\end{eqnarray}
is known as the \textbf{pairing susceptibility} which measures the tendency of the normal-state Fermi surface to develop electron pairing in a particular pairing channel. According to the general result of GL theory, for $U_0 > \chi_{SC}(T)$, we have $\alpha_0(T) < 0$ and the system favors the superconducting state, while for $U_0 < \chi_{SC}(T)$, we have $\alpha_0(T) > 0$ and the system favors the normal state. At the superconductor-normal phase transition point where $a_0(T_c) = 0$, we obtain the \textbf{linearized gap equation}:
\begin{eqnarray}\label{eq:LinearizedGapEqnF}
1 = U_0 \chi_{SC}(T = T_c)
\end{eqnarray}
which determines the $T_c$ under a given coupling constant $U_0$. As we discussed in the main text, the maximal twist $\theta = 30^{\circ}$ renormalizes the Fermi surface as well as the $f$-wave basis functions (Eq.\ 14-16 of the main text), which would generically lead to a different $\chi_{SC}(T)$ from the value $\chi_{SC,0}(T)$ for each isolated layer. Assuming the intralayer coupling constant $U_0$ remains unaffected by the angular twist, this would imply differences between the native $T_c$ of the twisted double-layer and the $T_{c,0}$ of an isolated monolayer. In our calculations, we indeed find minor corrections in $T_c$ with a typical $T_c \simeq 0.8-0.9 T_{c,0}$ in the weak coupling regime with $U_0 \sim 1 - 10$ meV.

The coefficient $c_0$ associated with the double-pair tunneling $(\Delta_1^2 \Delta_2^{\ast 2} + {\rm c.c.})$ can be similarly derived as:
\begin{align}\label{eq:GLC0}
c_0 &= \frac{1}{\beta} \sum_{n,\bm{k}_0} \frac{f^2_1(\bm{k}_0) f^2_2(\bm{k}_0)}{2[i\omega_n -\xi(\bm{k}_0)]^2 [i\omega_n +\xi(-\bm{k}_0)]^2 } \\\nonumber
&= \frac{1}{\beta} \sum_{n,\bm{k}_0} \frac{f^2_1(\bm{k}_0) f^2_2(\bm{k}_0)}{2[\omega^2_n +\xi^2(\bm{k}_0)]^2 } > 0.
\end{align}
This confirms that the phase-dependent part of the free energy $2c_0\psi^4 \cos(2\varphi)$ is minimized at $\varphi = \pm \pi/2$.

Similarly, the coefficient $a_0$ associated with $|\Delta_1|^2 |\Delta_2|^2$ is given by
\begin{align}\label{eq:GLA0}
a_0 &= \frac{1}{\beta} \sum_{n,\bm{k}_0} \frac{2 f^2_1(\bm{k}_0) f^2_2(\bm{k}_0)}{[i\omega_n -\xi(\bm{k}_0)]^2 [i\omega_n +\xi(-\bm{k}_0)]^2 } \\\nonumber
&= \frac{1}{\beta} \sum_{n,\bm{k}_0}  \frac{2 f^2_1(\bm{k}_0) f^2_2(\bm{k}_0)}{[\omega^2_n +\xi^2(\bm{k}_0)]^2 } > 0.
\end{align}

Notably, the GL coefficients above satisfy the following relations: (i) $a_0 = 4c_0$; (ii) $\beta_0 > 2c_0$. The factor of 4 in relation (i) can be easily seen in the expansion of the numerator in $B(\bm{k}_0, i\omega_n)$, where the term $|\Delta_1|^2|\Delta_2|^2$ results from the product of $2\Delta_1\Delta_2 \times 2\Delta^{\ast}_1\Delta^{\ast}_2$. On the other hand, the inequality in relation (ii)  essentially follows from $2 f^2_{1}(\bm{k}_0) f^2_{2}(\bm{k}_0) \leq  f^4_{1}(\bm{k}_0) + f^4_{2}(\bm{k}_0)$. Since $f_1(\bm{k}_0) \neq f_2(\bm{k}_0)$ for general $\bm{k}_0$, $\beta_0 > 2c_0$ strictly holds.

\section*{Supplementary Note 5: General solutions of Ginzburg-Landau equations at maximal twist} \label{AppendixE}

In the main text, we considered solution of the form $|\psi_1| = |\psi_2| \neq 0$ that minimizes the GL energy functional $\mathcal{F}_{GL}[\psi_1, \psi_2]$ in Eq.\ 6. Here, we provide a rigorous proof that such solutions indeed correspond to the global minimum of $\mathcal{F}_{GL}$ at the maximal twist of $\theta = 30^{\circ}$, given the relations: (i) $a_0 = 4c_0$; (ii) $\beta_0 > 2c_0$ derived in Supplementary Note 4 above.

The local extrema of $\mathcal{F}_{GL}$  Eq.\ 6  can be found as zeros of the functional derivatives $\partial \mathcal{F}_{GL} /\partial \psi^{\ast}_l = 0$, with $l=1,2$ labelling layer 1 and 2, respectively. This leads to the following coupled GL equations:
\begin{equation}\label{eq:GLEq}
\begin{cases}
        \alpha_0 \psi_1 + \beta_0 |\psi_1|^2 \psi_1 + a_0 \psi_1 |\psi_2|^2 + 2c_0 \psi^{\ast}_1 \psi^2_2 = 0, \\
\alpha_0 \psi_2 + \beta_0 |\psi_2|^2 \psi_2 + a_0 \psi_2 |\psi_1|^2 + 2c_0 \psi^{\ast}_2 \psi^2_1 = 0,
\end{cases}
\end{equation}
with $\alpha_0<0$ for $T<T_c$, $\beta_0>0$, and $a_0 = 4c_0 > 0$ as we derived explicitly in Supplementary Note 4 above. Note that Eq.\ \ref{eq:GLEq} above admits two classes of nontrivial solutions: (I) $|\psi_1| \neq 0, \psi_2 = 0$ (or equivalently, $|\psi_2| \neq 0, \psi_1 = 0$ by the symmetry of $\mathcal{F}_{GL}[\psi_1, \psi_2]$ upon exchanging the order parameters $\psi_1 \leftrightarrow \psi_2$); (II) $|\psi_1|, |\psi_2| \neq 0$. Physically, class I solutions describe the scenario where only one layer is superconducting while the other layer is not. Without loss of generality, taking $\psi_2 = 0$ in Eq.\ \ref{eq:GLEq} we obtain $|\psi_1|^2 = -\frac{\alpha_0}{\beta_0}$ and the total free energy is simply given by $\mathcal{F}_I = -\frac{\alpha_0^2}{2\beta_0}$. 

For class II solutions with $|\psi_1|, |\psi_2| \neq 0$, we multiply both sides of the upper (lower) equation in Eq.\ \ref{eq:GLEq} by $\psi^{\ast}_1$ ($\psi^{\ast}_2$):
\begin{equation}\label{eq:GLEq2}
\begin{cases}
        \alpha_0 |\psi_1|^2 + \beta_0 |\psi_1|^4 + a_0 |\psi_1|^2 |\psi_2|^2 + 2c_0 \psi^{\ast 2}_1 \psi^2_2 = 0, \\
\alpha_0 |\psi_2|^2 + \beta_0 |\psi_2|^4 + a_0 |\psi_2|^2 |\psi_1|^2 + 2c_0 \psi^{\ast 2}_2 \psi^2_1 = 0.
\end{cases}
\end{equation}
We further add and subtract the two equations in Eq.\ \ref{eq:GLEq2} to obtain a more elegant form:
\begin{equation}\label{eq:GLEq3}
\begin{cases}
        \alpha_0 (|\psi_1|^2 - |\psi_2|^2) + \beta_0 (|\psi_1|^4 - |\psi_2|^4) + 4 i c_0 |\psi_1|^2 |\psi_2|^2\sin(2\varphi) = 0, \\
\alpha_0 (|\psi_1|^2 + |\psi_2|^2) + \beta_0 (|\psi_1|^4 + |\psi_2|^4) + 2 a_0 |\psi_1|^2 |\psi_2|^2 + 4 c_0 |\psi_1|^2 |\psi_2|^2 \cos(2\varphi) = 0.
\end{cases}
\end{equation}
Here, $\varphi \equiv \varphi_2 - \varphi_1$ is the phase difference between the order parameters $\psi_2 = |\psi_2| e^{i \varphi_2}$ and $\psi_1 = |\psi_1| e^{i \varphi_1}$.

It is important to note that all the GL coefficients $\alpha_0, \beta_0, a_0, c_0$ must be real numbers for the overall free energy $\mathcal{F}_{GL}[\psi_1, \psi_2]$ (Eq.\ 6 of the main text) to be real. Therefore, for the upper equation in Eq.\ \ref{eq:GLEq3} to hold, both the real and imaginary parts on the left hand side must be zero. For the imaginary part to be zero, we must have $\sin(2\varphi) = 0$ and the only possible solutions are $\varphi = 0, \pi, \pm \pi/2$. For the real part to be zero, there are two possible solutions: (i) $|\psi_1| \neq |\psi_2|$, which necessarily requires $|\psi_1|^2 + |\psi_2|^2 = -\frac{\alpha_0}{\beta_0}$; (ii) $|\psi_1| = |\psi_2|$.

Let us first consider case (i) with $|\psi_1| \neq |\psi_2|$ and $|\psi_1|^2 + |\psi_2|^2 = -\frac{\alpha_0}{\beta_0}$. Substituting $|\psi_1|^2 + |\psi_2|^2 = -\frac{\alpha_0}{\beta_0}$ into the lower equation in Eq.\ \ref{eq:GLEq3} and rewrite $|\psi_1|^4 + |\psi_2|^4 = (|\psi_1|^2 + |\psi_2|^2 )^2 - 2|\psi_1|^2|\psi_2|^2$ we obtain:
\begin{equation}
[-\beta_0 + a_0 + 2 c_0 \cos(2\varphi)]|\psi_1|^2|\psi_2|^2  = 0,   
\end{equation}
which implies $\beta_0 = a_0 + 2 c_0 \cos(2\varphi)$ as $|\psi_1|, |\psi_2| \neq 0$. As $a_0 = 4c_0 >0$ (Supplementary Note 4), we have $\beta_0 = c_0 (4+2\cos(2\varphi))$, where $\varphi$ can only take on $\varphi = 0, \pi, \pm \pi/2$. It is straightforward to see that with $\varphi = \pm \pi/2$, we have $\beta_0 = 2c_0$ which contradicts with the relation $\beta_0 > 2c_0$ proven in Supplementary Note 4 above. For $\varphi = 0, \pi$, we must have $\beta_0 = 4c_0$. While this may occur for some special choice of parameters, we note that the $\varphi$-dependent term in $\mathcal{F}_{GL}$ is given by $2c_0 \cos(2\varphi)$ and $\partial^2 \mathcal{F}_{GL} /\partial^2 \varphi \propto -\cos(2\varphi) < 0$ for $\varphi$ near $0$ or $\pi$. Thus, the solution with $|\psi_1| \neq |\psi_2|$, which is only allowed for $\varphi = 0, \pi$, corresponds to a local maximum of $\mathcal{F}_{GL}$ according to the second derivative test. Therefore, we discard the solution with $|\psi_1| \neq |\psi_2|$ which is both fortuitous and unstable.

We are then left with the only possibility of case (ii) with $|\psi_1| = |\psi_2|$ in class II solutions. Define $\psi \equiv |\psi_1| = |\psi_2|$, the overall free energy can be rewritten as $\mathcal{F}_{GL} = 2\alpha_0 \psi^2 + (\beta_0 + a_0 + 2c_0 \cos(2\varphi)) \psi^4$ and $\mathcal{F}_{GL}$ is minimized at $\varphi_{min} = \pm \pi/2$ with $\cos(2\varphi_{min}) = -1$ as we discussed in the main text. Given $a_0 = 4c_0$ and $\varphi_{min} = \pm \pi/2$, we note that the total free energy at the minimum $\varphi_{min} = \pm \pi/2$ is exactly given by $\mathcal{F}_{GL} = 2\alpha_0 \psi^2 + (\beta_0 + 2 c_0)  \psi^4$, which is minimized at $\psi^2 = -\alpha_0/(\beta_0 + 2c_0)$ with a total free energy $\mathcal{F}_{II} = -\frac{\alpha_0^2}{\beta_0 + 2c_0}$. Note that $\beta_0 > 2c_0$ implies $\mathcal{F}_{II} = -\frac{\alpha_0^2}{\beta_0 + 2c_0} <  -\frac{\alpha_0^2}{2\beta_0} = \mathcal{F}_{I}$. Hence, we conclude that the solution $|\psi_1| = |\psi_2|$ always corresponds to the global minimum of $\mathcal{F}_{GL}$ and our choice of the solution in the main text is valid.

\section*{Supplementary Note 6: Twist-angle dependence of chiral $f \pm if'$ phase} \label{AppendixF}
\begin{figure}
\centering
\includegraphics[width=0.8\textwidth]{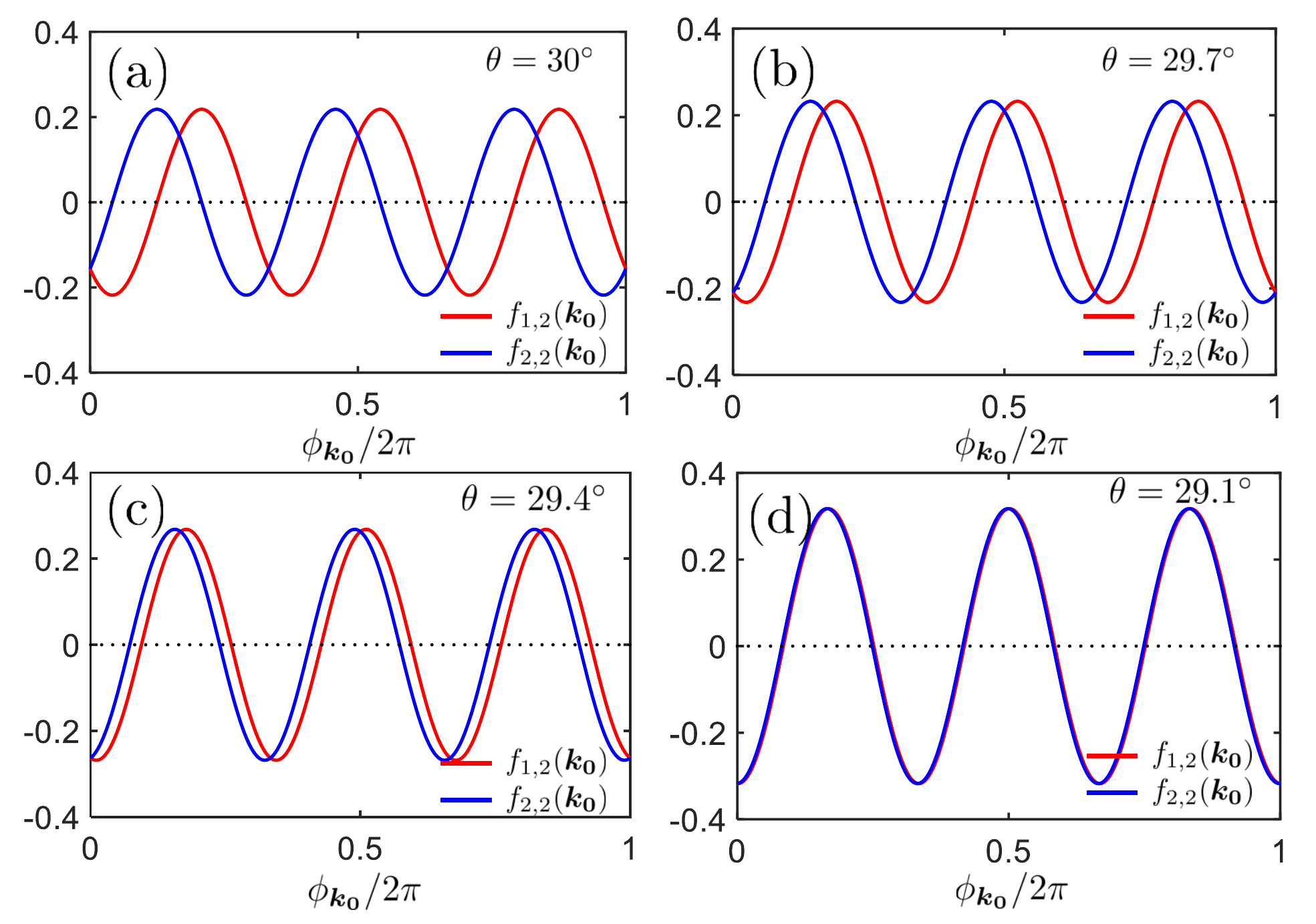}
\caption{Evolution of dimensionless basis functions $f_{1,2}(\bm{k}_0), f_{2,2}(\bm{k}_0)$ of projected pairings in band $p=2$ as $\theta$ decreases in steps of $0.3^{\circ}$ from $30^{\circ}$ to $29.1^{\circ}$. The two $f$-wave components become almost in-phase at $\theta = 29.1^{\circ}$.}
\label{FIGD1}
\end{figure}
As mentioned in the main text, the twist angle range of the chiral $f \pm if'$ phase in STVS superconductors is narrower than the chiral $d \pm id'$ phase found in twisted cuprates. Here, we show that the relatively narrow twist angle range originates from the nontrivial $\theta$-dependence governing the behavior of $\bm{k}_m$ and $\bm{\tilde{k}}_n$ for $m,n \neq 0$, which causes the two out-of-phase orthogonal $f$-wave components present at $\theta = 30^{\circ}$ to evolve rapidly into two nearly in-phase $f$-wave components. 

To see this consider the behavior of the basis functions $f_{l,p}(\bm{k}_0)$ defined in Eq.\ 15-16 in the Methods section of the main text, which determine the $\bm{k}_0$-dependence of the gap amplitude projected onto the band basis. These depend on $\bm{k}_0$ through functions $f_{1,2}$ defined below Eq.\ 3 of the main text, which leads to explicit expressions
\begin{equation}
    \begin{aligned}
f_{1, p}(\bm{k}_0) &= \sum_{m,j} \Lambda_{m,p}(\bm{k}_0) \sin(\bm{k}_0\cdot\bm{R}_j + \bm{\tilde{G}}_m\cdot\bm{R}_j), \\
f_{2, p}(\bm{k}_0) &= \sum_{n,j} \Lambda_{n,p}(\bm{k}_0)\sin(\bm{k}_0\cdot\bm{\tilde{R}}_j + \bm{G}_n\cdot\bm{\tilde{R}}_j).
    \end{aligned}
\end{equation}
Now the first term in the argument of the sine function produces the naively expected slow dependence on the twist angle $\theta$. For $m,n \neq 0$, however, the second term is nonzero and gives a large $\theta$-dependent contribution which acts as a $\bm{k}_0$-independent phase shift. One can show that at exactly $\theta=30^\circ$ the phase shifts in $f_{1,p}$ and $f_{2,p}$ are identical, leading to a `maximal' relative rotation due to $\bm{R}_j$ and $\bm{\tilde{R}}_j$ being at a $30^\circ$ angle. Away from this high-symmetry configuration the phase shifts rapidly grow and cause the evolution toward in-phase $f$-wave components.

To explicitly demonstrate how the nontrivial $\theta$-dependence discussed above affects the angular range of chiral $f\pm if'$ phase, we plot in Fig.\ \ref{FIGD1} the basis functions $f_{1,p}(\bm{k}_0), f_{2,p}(\bm{k}_0)$ of the pairing projected onto band $p=2$ at a series of angles gradually deviating from $30^{\circ}$. Clearly, when plotted along the Fermi surface in $\bm{k}_0$-space  the two $f$-wave components become almost in-phase for $\theta \simeq 29^{\circ}$. As the presence of two nearly orthogonal out-of-phase components at the Fermi surface is required for the formation of the chiral $f\pm if'$ phase, this nontrivial $\theta$-dependence limits the chiral phase to within a relatively narrow region of $\theta =30^{\circ} \pm 0.3^{\circ}$ as shown in Fig.\ 3b of the main text.

\section*{Supplementary Note 7: BDI invariant for nodal topological $f$-wave superconductivity} \label{AppendixG}

Here, we present details of the BDI topological invariant which can characterize the nontrivial bulk topology of the nodal $f$-wave superconductivity. As we explained in the main text, the Hamiltonian of the system in the $\varphi = 0$ phase respects a chiral symmetry: $\mathcal{C} H_{\rm{BdG}}(\bm{k}_0) \mathcal{C}^{-1} = -H_{\rm{BdG}}(\bm{k}_0)$ with the chiral operator $\mathcal{C} = \tau_2$. On the other hand, the BdG Hamiltonian has the built-in particle-hole symmetry: $\mathcal{P} H_{\rm{BdG}}(\bm{k}_0) \mathcal{P}^{-1} = -H_{\rm{BdG}}^{\ast}(-\bm{k}_0)$ with $\mathcal{P} = \tau_1 \mathcal{K}$, where $\mathcal{K}$ denotes complex conjugation. Combining $\mathcal{C}$ and $\mathcal{P}$, one can define a time-reversal-like anti-unitary symmetry $\mathcal{T}' \equiv \mathcal{CP} = -\tau_3 \mathcal{K}$ with $\mathcal{T}'^{2} = +1$, such that $\mathcal{T}'H_{\rm{BdG}}(\bm{k}_0)\mathcal{T}'^{-1} = H_{\rm{BdG}}^{\ast}(-\bm{k}_0)$. Given the fact that $f_{x(x^2-3x^2y)}$-pairing is even in ${k}_{0,y}$ which implies $H_{\rm{BdG}}({k}_{0,x}, {k}_{0,y}) = H_{\rm{BdG}}({k}_{0,x}, -{k}_{0,y})$, we deduce that for each fixed ${k}_{0,y}$, $H_{\rm{BdG}}({k}_{0,x}, {k}_{0,y})$ respect the following symmetries involving ${k}_{0,x}$ only:
\begin{align}
\mathcal{C} H_{\rm{BdG}}({k}_{0,x}, {k}_{0,y}) \mathcal{C}^{-1} &= -H_{\rm{BdG}}({k}_{0,x}, {k}_{0,y}) \nonumber \\   
\mathcal{P} H_{\rm{BdG}}({k}_{0,x}, {k}_{0,y}) \mathcal{P}^{-1} &= -H^{\ast}_{\rm{BdG}}(-{k}_{0,x}, {k}_{0,y}) \\\nonumber  
\mathcal{T}' H_{\rm{BdG}}({k}_{0,x}, {k}_{0,y}) \mathcal{T}'^{-1} &= -H^{\ast}_{\rm{BdG}}(-{k}_{0,x}, {k}_{0,y}).
\end{align}
Thus, by viewing ${k}_{0,y}$ as a parameter, $H_{\rm{BdG}}({k}_{0,x})$ describes a 1D superconductor belonging to the BDI topological class \cite{Schnyder1}, and its bulk topology is characterized by the winding number $N_{\rm BDI}$, which counts the number of Majorana end modes. $N_{\rm BDI}$ can be constructed as follows: in the eigenbasis of $\mathcal{C}$, $H_{\rm{BdG}}(\bm{k}_{0})$ is brought to an off-diagonal form
\begin{equation}
H_{\rm{BdG}}(\bm{k}_{0}) = 
\begin{pmatrix}
0 & z(\bm{k}_0)\\
z^{\ast}(\bm{k}_0) & 0
\end{pmatrix},
\end{equation}
where $z(\bm{k}_0) = \xi(\bm{k}_0) - i \Delta(\bm{k}_0)$ with $\Delta(\bm{k}_0) = \Delta_0 (f_1(\bm{k}_0) + f_2(\bm{k}_0))$. The winding number for each ${k}_{0,y}$ is then given by 
\begin{equation}\label{eq:windingnumber}
N_{\rm BDI}({k}_{0,y}) = \frac{1}{2\pi} \int_0^{2\pi} d {k}_{0,x} \frac{d \theta_{{k}_{0,y}}({k}_{0,x})}{d {k}_{0,x}},    
\end{equation}
where $\theta_{{k}_{0,y}}({k}_{0,x}) = \arg(z(\bm{k}_0))$. $N_{\rm BDI}({k}_{0,y})$ of the nodal $f_{x(x^2-3y^2)}$-wave superconductor is calculated numerically by using Eq.\ \ref{eq:windingnumber} and shown in in-set of Fig.\ 5b. 

\section*{Supplementary Note 8: Possibility of chiral $p \pm ip'$-wave symmetry for SC2 phase of RTG} \label{AppendixH}

In the main text, we assume a dominant pairing interaction in the $f$-wave channel for the spin-triplet SC2 phase of rhomobodedral trilayer graphene (RTG) and Bernel bilayer graphene (BBG), as well as their composite twisted double-layer system. While such assumption is supported by proposals based on acoustic phonons \cite{Chou1, Chou2} and renormalization group calculations \cite{Roy}, other recent theoretical works suggest an alternative chiral $p \pm ip'$-wave pairing symmetry for the unconventional spin-triplet superconductivity in RTG/BBG \cite{Chatterjee, Berg, Levitov}. Moreover, as we demonstrate in the main text, the maximal angular twist with $\theta \simeq 30^{\circ}$ leads to renormalization of the band structure and Fermi surface (FS) and whether the $f$-wave interaction would dominate is not \textit{a priori} obvious. Here, we present a detailed analysis of the possible pairing channels in the maximally twisted double-layer system. In particular, we extensively discuss the leading competition from the chiral $p$-wave channel suggested in Refs.\ \cite{Chatterjee, Berg, Levitov}.

\subsection*{A. Suppression of spin-singlet pairing channels}

It is important to note that the SC2 superconducting phase of RTG \cite{Young2} and the superconductivity (SC) in BBG \cite{Young3} considered in the main text are incompatible with spin-singlet pairing. In particular, as discussed in the Introduction of the main text, the SC2 phase of RTG is borne out of a \textbf{spin-polarized valley-unpolarized} half metal \cite{Young2}. Under such unusual spin-polarized normal-state fermiology, an electron of momentum $\bm{k}$ and certain spin, \textit{e.g.}, $\sigma = \uparrow$, does not have a time-reversal partner with momentum $-\bm{k}$ and $\sigma = \downarrow$ available at the other side of the spin-polarized FS to form a spin-singlet pair. In this case, the only possible spinor wave function of the Cooper pairs would have to be an equal-spin state, \textit{i.e.}, a spin-triplet state. In addition, the observation of Pauli limit violation in the SC2 phase of RTG provides a strong piece of evidence for its spin-triplet pairing nature: for spin-singlet superconductors to survive under magnetic fields higher than the Pauli limit, strong spin-orbit coupling in the electronic band structures is generally required, and this possibility is ruled out by the negligible spin-orbit coupling in graphene systems. It is worth noting that, while the microscopic mechanisms and exact pairing symmetries of RTG and BBG are still under debate, there is a general consensus that the SC2 phase must be a spin-triplet phase \cite{Chou1, Chou2, Roy, Chatterjee, Berg, Levitov}. In addition, the emergence of SC under a nonzero magnetic field in BBG implies that $T_c$ is enhanced by the applied magnetic field \cite{Young3}. This again strongly hints at spin-triplet pairing because an external field tends to destroy spin-singlet Cooper pairs and is well-known to suppress $T_c$ in a spin-singlet superconductor. 

Here, we present a general analysis on the suppression of spin-singlet pairing channels in the \textbf{twisted double-layer} based on the phenomenology of the SC2 phase in RTG. The fact that the SC2 phase in RTG is borne out of a spin-polarized half metal implies that a strong ferromagnetic (FM) order is established in the first place and sets the stage for the SC2 phase. This suggests the low-energy physics within each layer is dominated by the magnetic exchange interaction which can arise from general repulsive interactions (\textit{e.g.}, the Coulomb interaction) 
\begin{eqnarray}
\hat{H}_J = - \sum_{\bm{k}_1, \bm{k}_2, \bm{q}} J(\bm{k}_1, \bm{k}_2, \bm{q}) \sum_{\alpha \beta \gamma \delta} c^{\dagger}_{\bm{k}_1 + \bm{q}, \alpha}(\vec{\sigma})_{\alpha\beta} c_{\bm{k}_1, \beta} \cdot   c^{\dagger}_{\bm{k}_2 - \bm{q}, \gamma} (\vec{\sigma})_{\gamma\delta}c_{\bm{k}_2, \delta},
\end{eqnarray}
where $J(\bm{k}_1, \bm{k}_2, \bm{q}) = A^{-1}V(\bm{q})\braket{u_{\bm{k}_1 + \bm{q}}| u_{\bm{k}_1}} \braket{u_{\bm{k}_2 - \bm{q}}| u_{\bm{k}_2}}$ is the exchange integral with $V(\bm{q}) \sim 1/q$ being the Coulomb interaction strength in momentum-space, and the FM order implies $J(\bm{k}_1, \bm{k}_2, \bm{q}) >0$. $\ket{u_{\bm{k}}}$ is the periodic part of the Bloch state at $\bm{k}$, $\alpha, \beta,\gamma, \delta$ label the spin indices. 

Using the Pauli matrix identity: $\bm{\sigma}_{\alpha\beta} \cdot \bm{\sigma}_{\gamma\delta} = 2\delta_{\alpha\delta} \delta_{\beta\gamma} - \delta_{\alpha\beta}\delta_{\gamma\delta}$, and focusing on terms with scattering between zero-momentum pairs: $\bm{k}_1 + \bm{q} = -(\bm{k}_2-\bm{q}) \equiv \bm{k}$, $\bm{k}_1 = -\bm{k}_2 \equiv \bm{k}'$ with $\bm{q} \equiv \bm{k} - \bm{k}'$, the relevant interaction for zero-momentum pairing within each isolated layer can be written as
\begin{eqnarray}\label{eq:effectivePair}
\hat{H}_{I} = - \sum_{\bm{k}, \bm{k}'} g(\bm{k}, \bm{k}') \sum_{\alpha \beta} c^{\dagger}_{\bm{k}, \alpha} c^{\dagger}_{-\bm{k}, \beta} c_{-\bm{k}', \alpha}   c_{\bm{k}', \beta}
\end{eqnarray}
where $g(\bm{k}, \bm{k}') = 2A^{-1} V(\bm{k} - \bm{k}') \braket{u_{\bm{k}}| u_{\bm{k}'}} \braket{u_{-\bm{k}}| u_{-\bm{k}'}} \equiv 2J(\bm{k}_1 \equiv \bm{k}', \bm{k}_2 \equiv -\bm{k}', \bm{q} \equiv \bm{k}-\bm{k}') > 0$ and $g(\bm{k}', \bm{k}) = g^{\ast}(\bm{k}, \bm{k}')$ required by the Hermiticity of $\hat{H}_{I}$. We note that microscopically, the interaction in Eq.\ \ref{eq:effectivePair} arises from the exchange interaction between two electrons with opposite momenta: $\bm{k}_1 \simeq \bm{k}'$ and $\bm{k}_2 \simeq -\bm{k}'$, and for RTG, the relevant interaction near the FS would be given by two electrons near $\bm{k}_1 \simeq \bm{K}$ and $\bm{k}_2 \simeq -\bm{K}$. Thus, the origin of the attractive interaction in Eq.\ \ref{eq:effectivePair} can be thought of as an \textbf{inter-valley Hund's coupling}. Given $V(\bm{q}) = V(-\bm{q})$ for a general two-body interaction, we have $g(-\bm{k}, -\bm{k}') = 2A^{-1} V(\bm{k}'- \bm{k}) \braket{u_{-\bm{k}}| u_{-\bm{k}'}} \braket{u_{\bm{k}}| u_{\bm{k}'}} = g(\bm{k}, \bm{k}')$.

Upon stacking two layers of RTG at maximal twist, the effective interaction in the moir\'{e} bands of the twisted double-layer can be derived by projecting $\hat{H}_I$ in each layer to the band basis, following similar procedures described in Eq.\ 14 of the main text (subsection B of Methods section). By rewriting the fermionic operators in terms of creation operators for electrons in the moir\'{e} bands labeled by $p$: $c^{\dagger}(\bm{k}_m) = \sum_{p} u^{\ast}_{mp}(\bm{k}_0)a^{\dagger}_p(\bm{k}_0)$, $c^{\dagger}(\bm{\tilde{k}}_n) = \sum_{p} u^{\ast}_{np}(\bm{k}_0)a^{\dagger}_p(\bm{k}_0)
$, the effective interaction from $\hat{H}_I$ in band $p$ can be written as
\begin{eqnarray}
\hat{H}_{p, I} = - \sum_{\bm{k}_0, \bm{k}'_0} \tilde{g}(\bm{k}_0, \bm{k}'_0) \sum_{\alpha \beta} a^{\dagger}_{p, \bm{k}_0, \alpha} a^{ \dagger}_{p, -\bm{k}_0, \beta} a_{p, -\bm{k}'_0, \alpha}   a_{p, \bm{k}'_0, \beta},
\end{eqnarray}
where the renormalized interaction $\tilde{g}(\bm{k}_0, \bm{k}'_0) = \tilde{g}_1(\bm{k}_0, \bm{k}'_0) + \tilde{g}_2(\bm{k}_0, \bm{k}'_0)$ has contributions from both layer 1 and 2, which are given by:
\begin{eqnarray}
\tilde{g}_1(\bm{k}_0, \bm{k}'_0) &=& \sum_{m,m'} u^{\ast}_{m,p}(\bm{k}_0) u^{\ast}_{-m,p}(-\bm{k}_0) u_{-m',p}(-\bm{k}'_0) u_{m',p}(\bm{k}'_0) g(\bm{k}_0 + \bm{\tilde{G}}_m, \bm{k}'_0 + \bm{\tilde{G}}_m') \\\nonumber
&=& \sum_{m,m'} |u^{\ast}_{m,p}(\bm{k}_0)|^2 | u_{m',p}(\bm{k}'_0)|^2 g(\bm{k}_0 + \bm{\tilde{G}}_m, \bm{k}'_0 + \bm{\tilde{G}}_m') > 0, \\\nonumber
\tilde{g}_2(\bm{k}_0, \bm{k}'_0) &=& \sum_{n,n'} u^{\ast}_{n,p}(\bm{k}_0) u^{\ast}_{-n,p}(-\bm{k}_0) u_{-n',p}(-\bm{k}'_0) u_{n',p}(\bm{k}'_0) g(\bm{k}_0 + \bm{G}_n, \bm{k}'_0 + \bm{G}_n') \\\nonumber
&=& \sum_{n,n'} |u^{\ast}_{n,p}(\bm{k}_0)|^2 | u_{n',p}(\bm{k}'_0)|^2 g(\bm{k}_0 + \bm{G}_n, \bm{k}'_0 + \bm{G}_n') >0,
\end{eqnarray}
where we used the relations $u^{\ast}_{-m,p}(-\bm{k}_0) = u_{m,p}(\bm{k}_0)$, $u^{\ast}_{-n,p}(-\bm{k}_0) = u_{n,p}(\bm{k}_0)$ imposed by the spinless time-reversal symmetry $\mathcal{T}' \equiv \mathcal{K}$. It is straightforward to show that given $g(\bm{k}, \bm{k}') = g^{\ast}(\bm{k}', \bm{k}) $ and $g(-\bm{k}, -\bm{k}') = g(\bm{k}, \bm{k}') $, the renormalized interaction strengths $\tilde{g}_1(\bm{k}_0, \bm{k}'_0), \tilde{g}_2(\bm{k}_0, \bm{k}'_0)$ and $\tilde{g}(\bm{k}_0, \bm{k}'_0)$) also satisfy similar properties: $\tilde{g}(\bm{k}_0, \bm{k}'_0) = \tilde{g}^{\ast}(\bm{k}'_0, \bm{k}_0)$, $\tilde{g}(-\bm{k}_0, -\bm{k}'_0) = \tilde{g}(\bm{k}_0, \bm{k}'_0)$. 

To examine the possible pairing interactions derived from $\hat{H}_{p, I}$, we note that the most general pairing interactions can always be decomposed into all possible spin-singlet and spin-triplet pairing channels
\begin{eqnarray}\label{eq:pairing_interaction}
\hat{H}_{pair} = \sum_{l} g^{(l)}_s \hat{\Psi}^{(l),\dagger}_p \hat{\Psi}^{(l)}_p + g^{(l)}_t \hat{D}^{(l),\dagger}_p \hat{D}^{(l)}_p,
\end{eqnarray}
where $\hat{\Psi}^{(l),\dagger}_p = \sum_{\bm{k}_0, \alpha\beta}\psi^{(l)}_{\alpha\beta}(\bm{k}_0) a^{\dagger}_{p,\bm{k}_0,\alpha} a^{\dagger}_{p,-\bm{k}_0,\beta}$, $\hat{\Psi}^{(l)}_p = \sum_{\bm{k}'_0, \alpha\beta}\psi^{(l),\ast}_{\alpha\beta}(\bm{k}'_0) a_{p,-\bm{k}'_0,\beta} a_{p,\bm{k}'_0,\alpha}$ with even angular momentum channels $l=0,2,...$ as the creation/annihilation operators of spin-singlet pairs. Note that $\psi^{(l)}(\bm{k}_0) = \psi^{(l)}(-\bm{k}_0)$ (even spatial parity) and $ \psi^{(l)}_{\alpha\beta}(\bm{k}_0) = -\psi^{(l)}_{\beta\alpha}(\bm{k}_0)$ (anti-symmetry under exchange of spins) for spin-singlet pairing. On the other hand, $\hat{D}^{\dagger, (l)}_p = \sum_{\bm{k}_0, \alpha\beta}D^{(l)}_{\alpha\beta}(\bm{k}_0) a^{\dagger}_{p,\bm{k}_0,\alpha} a^{\dagger}_{p,-\bm{k}_0,\beta}$, $\hat{D}^{(l)}_p = \sum_{\bm{k}'_0, \alpha\beta}D^{ (l),\ast}_{\alpha\beta}(\bm{k}'_0) a_{p,-\bm{k}'_0,\beta} a_{p,\bm{k}'_0,\alpha}$ with odd angular momentum channels $l=1,3,...$ are the creation/annihilation operators of spin-triplet pairs, and $D^{(l)}(\bm{k}_0) = -D^{(l)}(-\bm{k}_0)$ (odd spatial parity) and $D^{(l)}_{\alpha\beta}(\bm{k}_0) = D^{(l)}_{\beta\alpha}(\bm{k}_0)$ (symmetry under exchange of spins) for spin-triplet pairing. The coupling constants $g^{(l)}_s$, $g^{(l)}_t$ denote the interaction strength in the spin-singlet and spin-triplet pairing channels with angular momentum $l$, and the sign of $g$ determines whether the interaction is attractive ($g<0$) or repulsive ($g>0$). Notably, because the interaction $\tilde{g}(-\bm{k}_0, -\bm{k}'_0) = \tilde{g}(\bm{k}_0, \bm{k}'_0)$ has even spatial parity, cross terms such as $(\hat{\Psi}\hat{D}^{\dagger} + h.c.)$ that mix different spatial parities are forbidden and no pair-scattering process occurs between spin-singlet and spin-triplet channels.

To make a direct comparison between $\hat{H}_{pair}$ (Eq.\ \ref{eq:pairing_interaction}) and $\hat{H}_{p, I}$ (Eq.\ \ref{eq:effectivePair}), we rewrite the pair annihilation operators as: 
\begin{eqnarray}
\centering
\hat{\Psi}^{(l)}_p &=& \sum_{\bm{k}'_0, \alpha\beta}\psi^{(l),\ast}_{\alpha\beta}(\bm{k}'_0) a_{p,-\bm{k}'_0,\beta} a_{p,\bm{k}'_0,\alpha} \\\nonumber
(\text{Fermion exchange}) &=& -\sum_{\bm{k}'_0, \alpha\beta}\psi^{(l),\ast}_{\alpha\beta}(\bm{k}'_0) a_{p,\bm{k}'_0,\alpha} a_{p,-\bm{k}'_0,\beta} \\\nonumber
(\text{Relabel $\bm{k}'_0$ by $-\bm{k}'_0$ }) &=& -\sum_{\bm{k}'_0, \alpha\beta}\psi^{(l),\ast}_{\alpha\beta}(-\bm{k}'_0) a_{p,-\bm{k}'_0,\alpha} a_{p,\bm{k}'_0,\beta} \\\nonumber
(\text{Even parity $\psi^{(l)}(\bm{k}_0) = \psi^{(l)}(-\bm{k}_0)$ }) &=& -\sum_{\bm{k}'_0, \alpha\beta}\psi^{(l),\ast}_{\alpha\beta}(\bm{k}'_0) a_{p,-\bm{k}'_0,\alpha} a_{p,\bm{k}'_0,\beta}, \\\nonumber
\hat{D}^{(l)}_p &=& \sum_{\bm{k}'_0, \alpha\beta}D^{(l),\ast}_{\alpha\beta}(\bm{k}'_0) a_{p,-\bm{k}'_0,\beta} a_{p,\bm{k}'_0,\alpha} \\\nonumber
(\text{Fermion exchange}) &=& -\sum_{\bm{k}'_0, \alpha\beta}D^{(l),\ast}_{\alpha\beta}(\bm{k}'_0) a_{p,\bm{k}'_0,\alpha} a_{p,-\bm{k}'_0,\beta} \\\nonumber
(\text{Relabel $\bm{k}'_0$ by $-\bm{k}'_0$ }) &=& -\sum_{\bm{k}'_0, \alpha\beta}D^{(l),\ast}_{\alpha\beta}(-\bm{k}'_0) a_{p,-\bm{k}'_0,\alpha} a_{p,\bm{k}'_0,\beta} \\\nonumber
(\text{Odd parity $D^{(l)}(\bm{k}_0) = -D^{(l)}(-\bm{k}_0)$ }) &=& \sum_{\bm{k}'_0, \alpha\beta}D^{(l),\ast}_{\alpha\beta}(\bm{k}'_0) a_{p,-\bm{k}'_0,\alpha} a_{p,\bm{k}'_0,\beta}. 
\end{eqnarray}
It is straightforward to see that due to the different parities in singlet and triplet wave functions, the pairing interaction $\hat{H}_{pair}$ (Eq.\ \ref{eq:pairing_interaction}) derived from $\hat{H}_{p, I}$ (Eq.\ \ref{eq:effectivePair}) must be associated with coupling constants $g^{(l)}_s > 0$ and $g^{(l)}_t < 0$. This reveals that the magnetic exchange interactions, which is expected to be strong for establishing the normal-state FM order in each layer, gives rise to effective \textbf{repulsive} interactions in all spin-singlet channels and \textbf{attractive} interactions in all spin-triplet channels in the twisted double-layer. Thus, we rule out the possibility of spin-singlet pairing and consider exclusively the spin-triplet pairing channels in the following discussions. 

We further note that in our spinless fermion model considered in the main text, we assume the spins in the two layers are aligned in the same direction. While spins in each free-standing monolayer may not be locked in a particular direction, given the ferromagnetic (half-metallic) nature of the normal state ($T>T_c$) \cite{Young2}, upon coupling the two layers we expect inter-layer Hund's coupling between the two layers, \textit{e.g.}, the exchange coupling induced by inter-layer Coulomb repulsion, would stabilize the configuration with a uniform spin orientation throughout the twisted double-layer. In addition, given both the ferromagnetic nature of the normal-state and perfect SU(2)-symmetry in graphene systems, in practice one can always apply a weak Zeeman field in the first place to align the spins of both half-metallic layers to a uniform direction at $T>T_c$, then retrieve the applied field before cooling down the sample to $T<T_c$. In this way, the spin orientation can be set to be uniform in the normal-state, and as the system enters the superconducting phase, the spin-triplet $\bm{d}$-vectors in both layers would be naturally aligned in the same direction.

\subsection*{B. Chiral $p \pm ip'$-wave versus chiral $f \pm if'$-wave channels in maximally twisted double-layer}

An exhaustive investigation of spin-triplet superconductivity in twisted double-layer would in principle require consideration of all possible pairing channels with $l = 1,3,5,7,...$. However, we note that many such channels (\textit{e.g.}, $l=5$ and $l=7$) are related to non-crystallographic rotation symmetries and thus unlikely to arise in the hexagonal lattice systems of our interest. Moreover, higher angular momentum channels would necessarily cause rapid sign changes or phase oscillations in the superconducting order parameter along the Fermi surface, which are energetically costly. These observations motivate us to focus on the $f$-wave ($l=3$) and chiral $p$-wave ($l=1$) channels, which are also the two leading candidates for the pairing symmetries of the spin-triplet SC2 phase \cite{Chou1, Chou2, Roy, Chatterjee, Berg, Levitov}.

\begin{figure}
\centering
\includegraphics[width=0.9\textwidth]{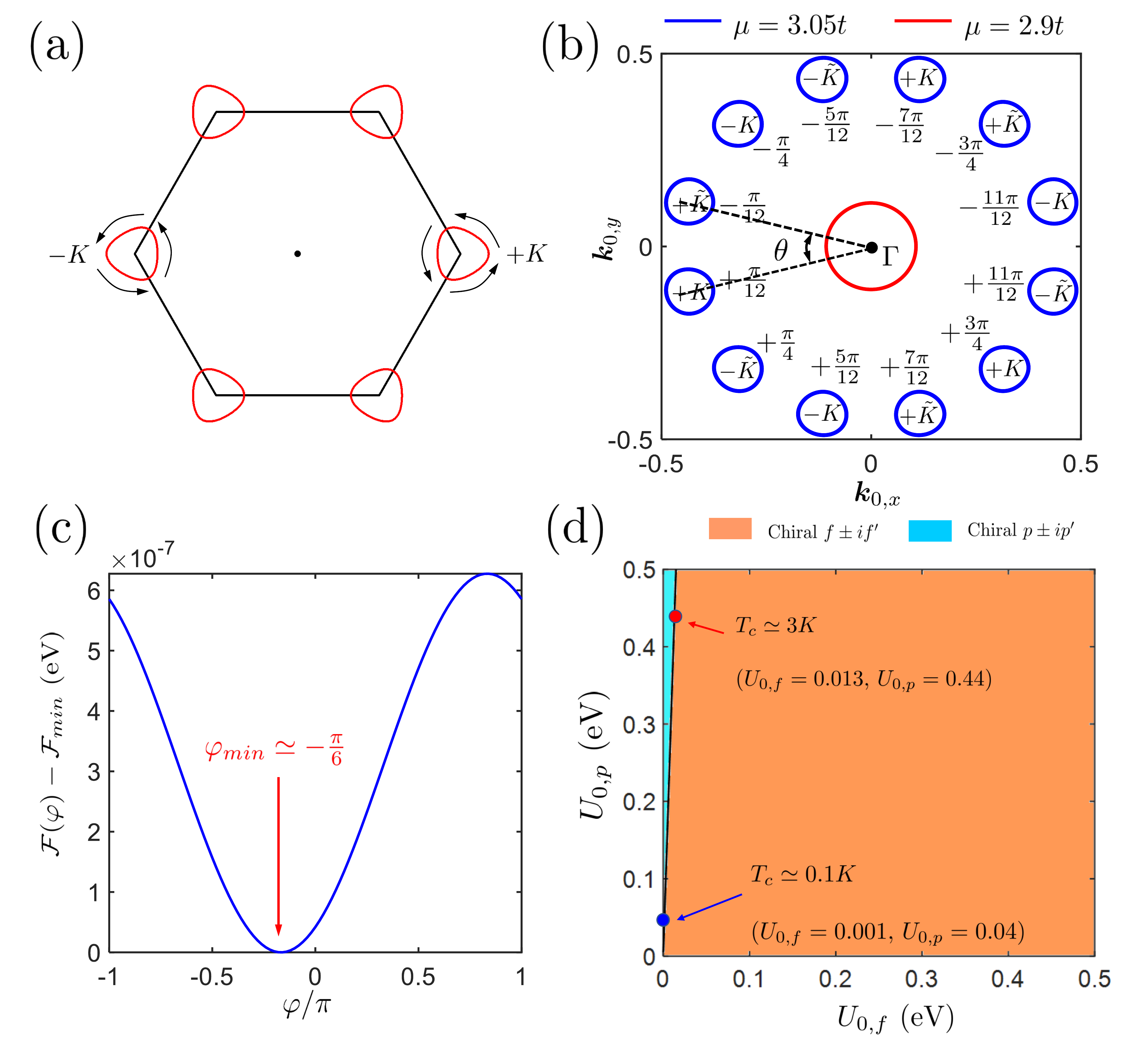}
\caption{(a) Schematic of phase winding of the chiral $p + ip'$-wave gap function around the two valleys $+K$ and $-K$ in an isolated monolayer. (b) Fermi surface of the twisted double-layer in the $\bm{k}_0$-space at $\mu = 3.05t$ (blue solid lines) and $\mu = 2.9t$ (red solid lines) as in Fig.\ 2 of the main text. The numbers $-\frac{11\pi}{12}, -\frac{3\pi}{4}, ..., +\frac{11\pi}{12}$ indicate schematically the phase of the complex $p+ip'$ gap function near a momentum-space point on the Fermi surface with both layers having the same $p$-wave chirality $c = +$. Note that due to the relative rotation of $\theta = 30^{\circ}$ between two layers, there is a phase offset of $\pi/6$ between layer 1 ($\pm K$) and layer 2 ($\pm \tilde{K}$) generated by the angular momentum $m_z = +1$ carried by $p+ ip'$ Cooper pairs.  (c) Phase dependence of the superconducting free energy $\mathcal{F}(\varphi)$ of the maximally twisted double-layer ($\theta = 30^{\circ}$) assuming chiral $p$-wave interactions (Eq.\ \ref{eq:intralayerinteractionP}). The normal-state model parameters are set to be the same as in Fig.\ 3c (red line at $\theta= 30^{\circ}$) of the main text. $\mathcal{F}(\varphi)$ is minimized at $\varphi \equiv \varphi_2 - \varphi_1 \simeq -\pi/6$, which compensates the phase offset in (b). (d) Phase diagram of the maximally twisted-double layer obtained by solving the linearized gap equations Eq.\ \ref{eq:LinearizedGapEqnF} and Eq.\ \ref{eq:LinearizedGapEqnC} under chiral $p$-wave and $f$-wave interactions, where we consider the most energetically favored $\varphi_{min} = -\pi/6$ for the chiral $p$-wave channel and $\varphi_{min} = \pm \pi/2$ for the $f$-wave channel (see Fig.\ 3c of the main text). $U_{0,f}>0$ and $U_{0,p}>0$ denote the attractive coupling strength in the $f$-wave and chiral $p$-wave channels, respectively (see sign convention of $U_{0,p}$ and $U_{0,f}$ in Eq.\ \ref{eq:intralayerinteractionP}). The chiral $p$-wave phase is favored only within a narrow region with $U_{0,p} \gg U_{0,f}$ in the phase diagram. }
\label{FIGS1}
\end{figure}

The momentum-space pairing interaction in the $p$-wave channel for each isolated layer has the general form
\begin{equation}
    \begin{aligned}\label{eq:intralayerinteractionP}
\mathcal{V}_p^{(1)} &= -U_{0,p} \sum_{\bm{k},\bm{k}'} \big(p_{+,1}(\bm{k}) p^{\ast}_{+,1}(\bm{k}') + p_{-,1}(\bm{k}) p^{\ast}_{-,1}(\bm{k}')\big) c^{\dagger}(\bm{k})c^{\dagger}(-\bm{k}) c(-\bm{k}')c(\bm{k}'),\\
\mathcal{V}_p^{(2)} &= -U_{0,p} \sum_{\bm{\tilde{k}},\bm{\tilde{k}}'} \big(p_{+,2}(\bm{k}) p^{\ast}_{+,2}(\bm{k}') + p_{-,2}(\bm{k}) p^{\ast}_{-,2}(\bm{k}')\big) c^{\dagger}(\bm{\tilde{k}})c^{\dagger}(-\bm{\tilde{k}}) c(-\bm{\tilde{k}}')c(\bm{\tilde{k}}'),
    \end{aligned}
\end{equation}
where the minus sign indicates the attractive nature of the interaction, $U_{0,p} > 0$ denotes the attractive interaction strength in the $p$-wave channel, and $p_{\pm, 1,2}$ are the basis functions characterizing the chiral $p$-wave pairing symmetry with $\pm$ denoting the two opposite chiralities. In the concrete setting of a triangular lattice as in RTG/BBG, these basis functions take the form of
\begin{eqnarray}
p_{1, \pm}(\bm{k}) &=& \sin(\bm{k}\cdot\bm{R}_1) + \omega_{\pm}\sin(\bm{k}\cdot\bm{R}_3) +  \omega^2_{\pm}\sin(\bm{k}\cdot\bm{R}_5), \\\nonumber
p_{2, \pm}(\bm{\tilde{k}}) &=&  \sin(\bm{\tilde{k}}\cdot\bm{\tilde{R}}_1) + \omega_{\pm} \sin(\bm{\tilde{k}}\cdot\bm{\tilde{R}}_3) + \omega^2_{\pm} \sin(\bm{\tilde{k}}\cdot\bm{\tilde{R}}_5),
\end{eqnarray} 
where $\omega_{\pm} \equiv e^{ \pm i 2\pi/3} = -\frac{1}{2} \pm \frac{\sqrt{3}}{2}i$ are the complex cubic roots of unity. Under $C_{3z}: \bm{R}_j \mapsto \bm{R}_{j+2}$ ($j=1,3,5$) (see Fig.\ 2a of the main text), these basis functions transform as: $p_{\pm}(\bm{k}) \mapsto e^{ \mp i 2\pi/3} p_{\pm}(\bm{k})$, reflecting the total angular momentum of $m_z = \pm 1$ for the chiral $p$-wave Cooper pairs, and $p_{\pm}(\bm{k})$ form a Kramers pair of the spinless time-reversal $\mathcal{T}'$ as $p^{\ast}_{\pm}(\bm{k}) = p_{\mp}(\bm{k})$. Notably, in the vicinity of $\pm K$, the chiral $p$-wave order parameter has the same phase winding around the two different $+K$ and $-K$ valleys (shown schematically in Fig.\ \ref{FIGS1}a). This can be seen by expanding $p_{1, \pm}(\bm{k} = \bm{p} + \xi \bm{K})$ ($\xi = \pm$: valley index) in terms of a small momentum $\bm{p}$ measured from $\pm \bm{K}$: $p_{1, \pm}(\bm{k} = \bm{p} + \xi \bm{K}) \simeq -\frac{3a}{4} (p_x \pm i p_y)$.

To examine the pairing instability in the chiral $p$-wave channel in the twisted double-layer, we follow the procedure outlined in Eqs.\ 14-16 in subsection B of the Methods section to obtain the effective pairing interaction in the  basis of the moir\'{e} bands of the twisted double-layer
\begin{eqnarray}\label{eq:totVP}
\mathcal{V}_{\rm {eff}, p} = -U_{0,p} (P^{\dagger}_{1,+}P_{1,+} + P^{\dagger}_{1,-}P_{1,-} + P^{\dagger}_{2,+}P_{2,+} + P^{\dagger}_{2,-}P_{2,-}), 
\end{eqnarray}
where $P^{\dagger}_{l, c} \equiv \sum_{\bm{k}_0, q} p_{l, q, c}(\bm{k}_0) a_q^{\dagger}(\bm{k}_0) a_q^{\dagger}(-\bm{k}_0)$ are the pair creation operators in band $q$ for layer $l = 1,2$ and chirality $c = \pm$, and their corresponding basis functions are given by:
\begin{equation}
\begin{aligned}\label{eq:basisfuncP}
p_{1, q, c=\pm}(\bm{k}_0) &= \sum_{m} \Lambda_{m,q}(\bm{k}_0) p_{1,c}(\bm{k}_0 + \bm{\tilde{G}}_m), \\
p_{2, q, c=\pm}(\bm{k}_0) &= \sum_{n} \Lambda_{n,q}(\bm{k}_0)p_{2, c}(\bm{k}_0 + \bm{G}_n),
\end{aligned}
\end{equation}
which characterize the projected chiral $p$-wave pairings in band $q$ for layer $l=1,2$ and chirality $c=\pm$, with $\Lambda_{m,q}, \Lambda_{n,q}$ being the form factors defined in Eq.\ 16 of the main text. Notably, the effective $p$-wave interaction in Eq.\ \ref{eq:intralayerinteractionP} leads to 4 independent $p$-wave order parameters associated with the 4 basis functions in Eq.\ \ref{eq:basisfuncP}: $\Delta_{l,q, \pm} \equiv -U_0 \braket{P_{l,q, \pm}}$ with $l=1,2$ and $c=\pm$. The general solution of the $p$-wave gap function is then a linear combination of the 4 basis functions: $\Delta_{p}(\bm{k}_0) = \Delta_0 \{s_{1,+}p_{1, q, +}(\bm{k}_0) + s_{1,-}p_{1, q, -}(\bm{k}_0) + s_{2,+}p_{2, q, +}(\bm{k}_0) + s_{2,-}p_{2, q, -}(\bm{k}_0)\}$ with $\Delta_{l, q, c} = \Delta_0 s_{l, c}$, and the total chiral $p$-wave order parameter can be represented by the vector $\vec{s} \equiv (s_{1,+}, s_{1,-}, s_{2,+}, s_{2,-})^{T}$. The $T_c$ and the coefficients $s_{l, c}$ of the $p$-wave solutions can be obtained by solving the linearized gap equation for the twisted double-layer (for analogy, see Eq.\ \ref{eq:LinearizedGapEqnF} for the $f$-wave channel),
\begin{eqnarray}\label{eq:LinearizedGapEqn} 
\vec{s} = U_{0,p} \hat{\chi}_{SC}(T) \vec{s},
\end{eqnarray}
where $\hat{\chi}_{SC}(T_c)$ is the $4\times 4$ \textbf{pairing susceptibility} matrix with elements given by the Matsubara sum
\begin{eqnarray}\label{eq:PSelements}
\chi_{SC, ll', cc'}(T) = -\frac{1}{\beta} \sum_{n, \bm{k}_0}  \frac{p_{l, q, c}(\bm{k}_0) p^{\ast}_{l', q, c'}(\bm{k}_0)}{[i\omega_n -\xi(\bm{k}_0)] [i\omega_n +\xi(-\bm{k}_0)] }.
\end{eqnarray}

Before solving the linearized gap equation, we note that the $C_{3z}$-symmetry implies the following relations: given $\bm{\tilde{G}}_{m'} = \hat{C}_{3z} \bm{\tilde{G}}_m$, we have (i)  $\Lambda_{m',p} (\hat{C}_{3z}\bm{k}_0) = \Lambda_{m,p} (\bm{k}_0)$, and (ii) $p_{1,\pm}(\hat{C}_{3z}\bm{k}_0 + \bm{\tilde{G}}_{m'}) = \omega_{\pm} p_{1,\pm}(\bm{k}_0 + \bm{\tilde{G}}_m)$. These relations allow us to deduce that $\hat{C}_{3z}: p_{1, q, \pm}(\bm{k}_0) \mapsto e^{\mp i 2\pi/3} p_{1, q, \pm}(\bm{k}_0)$, and similarly $\hat{C}_{3z}: p_{2, q, \pm}(\bm{k}_0) \mapsto e^{\mp i 2\pi/3} p_{2, q, \pm}(\bm{k}_0)$. This reveals that the projected chiral $p$-wave basis functions $p_{l, q, \pm}(\bm{k}_0)$ in Eq.\ \ref{eq:basisfuncP} transform exactly as the usual chiral $p$-wave functions under $C_{3z}$, and $P^{\dagger}_{l, \pm}$ create $m_z = \pm 1$ Cooper pairs in the twisted double-layer. Notably, since states with opposite chiralities correspond to different angular momenta $m_z = \pm 1$ and distinct eigenvalues of $C_{3z}$, we deduce that under $C_{3z}$: $\chi_{SC, ll', +-} \mapsto e^{-i4\pi/3}\chi_{SC, ll', +-}$, $\chi_{SC, ll', -+} \mapsto e^{i4\pi/3}\chi_{SC, ll', -+}$. For $\chi_{SC}$ to respect $C_{3z}$-symmetry, we must have $\chi_{SC, ll', +-} = e^{-i4\pi/3}\chi_{SC, ll', +-}$ and $\chi_{SC, ll', -+} = e^{i4\pi/3}\chi_{SC, ll', -+} \Rightarrow \chi_{SC, ll', +-} = \chi_{SC, ll', -+} =0$. Physically, this implies $p$-wate states with opposite chiralities do not mix in the pairing susceptibility and the energetically favored solution (the one with highest $T_c$) is given by the channel with \textbf{same chiralities} in both layers. In this case, the complex order parameter is non-vanishing on the Fermi surface (see Fig.\ \ref{FIGS1}b) and the twisted double-layer is expected to possess a full chiral superconducting gap. In fact, this result can be inferred by observing that states with opposite chiralities on two different layers, \textit{e.g.}, $\Delta_p(\bm{k}_0) = \Delta_0 \{(s_{1,+}p_{1,q,+}(\bm{k}_0) + s_{2,-}p_{2,q,-}(\bm{k}_0)\}$, are not eigenstates of $C_{3z}$. Instead, they transform as $p_x$- or $p_y$-wave functions with point nodes in the $p$-wave superconducting gap, and thus are less energetically favored than the fully gapped chiral state with same chiralities on both layers. 

The symmetry analysis above allows us to decouple the two chirality sectors in the linearized gap equation (Eq.\ \ref{eq:LinearizedGapEqn}). Besides, since states with  two opposite chiralities form Kramers pairs of the spinless $\mathcal{T}'$-symmetry, they must provide two degenerate solutions with the same $T_c$ and it suffices to consider the $c = +$ case without loss of generality. The linearized gap equation for $c = +$ is given by
\begin{eqnarray}\label{eq:LinearizedGapEqnC}
\begin{pmatrix}
s_{1,+}\\
s_{2,+}
\end{pmatrix} 
= U_{0,p} 
\begin{pmatrix}
\chi_{SC, 11, ++}(T) & \chi_{SC, 12, ++}(T)\\
\chi_{SC, 21, ++}(T) & \chi_{SC, 22, ++}(T)
\end{pmatrix}
\begin{pmatrix}
s_{1,+}\\
s_{2,+}
\end{pmatrix},
\end{eqnarray}
with matrix elements $\chi_{SC, ll', ++}$ defined in Eq.\ \ref{eq:PSelements}. Due to the $C_{2y}$ symmetry that swaps the two layers, we have $\chi_{SC, 11, ++}(T) = \chi_{SC, 22, ++}(T) = \chi_{0,+}(T)$. Also, it is clear from Eq.\ \ref{eq:PSelements} that $\chi_{SC, 12, ++}(T) = \chi^{\ast}_{SC, 21, ++}(T)$. The matrix equation in Eq.\ref{eq:LinearizedGapEqnC} is essentially an eigenvalue equation, which can be easily solved by diagonalizing the $2\times 2$ pairing susceptibility matrix. The two eigenvalues of the matrix are given by: $\chi_{p, I, +}(T) = \chi_0(T) + |\chi_{SC,12,++}(T)|, \chi_{p, II, +}(T) = \chi_0(T) - |\chi_{SC,12,++}(T)|$ and the larger eigenvalue $\chi_{p, I, +}(T)$ determines the maximal $T_c$ of the chiral $p$-wave channel. 

The eigenvector of Eq.\ \ref{eq:LinearizedGapEqnC} has the general form of $(s_{1,+}, s_{2,+})^T = s (1, e^{i \varphi})^{T}$, where $s \equiv|s_{1,+}| = |s_{2,+}|$ and $\varphi \equiv \varphi_2 - \varphi_1$ is exactly the phase difference between the two chiral $p$-wave order parameters in two different layers. Interestingly, we find that when the disconnected small Fermi surfaces (FSs) around $\pm K (\tilde{K})$ coalesce into a single connected FS (\textit{e.g.}, red solid line at $\mu = 2.9t$ in Fig.\ \ref{FIGS1}b), $\varphi$ corresponding to $\chi_{p, I, +}(T)$ is given by $\varphi \simeq - \pi/6 = -30^{\circ}$, which matches almost exactly with the twist angle $\theta = 30^{\circ}$ between the two layers. We further confirm this by calculating the $\varphi$-dependence of the superconducting free energy in the twisted double-layer at $\mu = 2.9t$ under chiral $p$-wave interactions (Eq.\ \ref{eq:intralayerinteractionP}), following similar calculations performed for the $f$-wave interaction in the main text (Fig.\ 3). Evidently, the free energy in the chiral $p$-wave channel is minimized at $\varphi_{min} \simeq - \pi/6 = \theta$ (Fig.\ \ref{FIGS1}c). Therefore, the energetically favored chiral $p + ip'$-wave solution in the twisted double-layer corresponds to the following gap function:
\begin{eqnarray}\label{eq:PWaveGapFunc}
\Delta_{p,+}(\bm{k}_0) = \Delta_0 \big(p_{1,+}(\bm{k}_0) + e^{i \varphi_{min}} p_{2,+}(\bm{k}_0)\big),
\end{eqnarray}
with $\varphi_{min} = -\pi/6$. 

Here, we point out that this intriguing agreement between $\varphi_{min}$ and twist angle $\theta$ in the chiral $p$-wave channel can be understood physically as a result of the $m_z = +1$ angular momentum carried by chiral $p$-wave Cooper pairs: upon making an angular twist of $\theta = 30^{\circ}$, \textit{i.e.}, a relative rotation of $\pi/6$ about the principal $z$-axis of the two layers, the nonzero $m_z = +1$ carried by the chiral $p$-wave Cooper pairs in both layers acquire an offset in the phases of their superconducting order parameters (Fig.\ \ref{FIGS1}b). In particular, from the vantage point of layer 2 (1), the phase in layer 1 (2) is advanced (retarded) by $\pi/6$, and the two complex order parameters are no longer aligned in their complex phase. By adjusting the phase difference to be $\varphi = \varphi_2 - \varphi_1 = -\pi/6$, the twisted double-layer compensates the phase offset caused by the angular twist and maximizes the magnitude of its complex superconducting gap (minimizes the free energy). 

By solving the linearized gap equations Eq.\ \ref{eq:LinearizedGapEqnF} under $f$-wave interactions and Eq.\ \ref{eq:LinearizedGapEqnC} under chiral $p$-wave interactions, we obtain the full superconducting phase diagram of the twisted double-layer at $\mu = 2.9t$ as a function of the coupling constants $U_{0,f}$ and $U_{0,p}$ in both channels (Fig.\ \ref{FIGS1}d). It is evident that the chiral $f \pm if'$ phase takes up the vast majority of the phase diagram, while the chiral $p$-wave phase is favored only within a narrow region with $U_{0,p} \gg U_{0,f}$. In particular, we find that at the phase boundary where the two phases correspond to the same $T_c$, the coupling constant in the chiral $p$-wave channel needs to be 30-40 times larger than its counterpart in the $f$-wave channel (see values of $U_{0,f}, U_{0,p}$ corresponding to the blue and red dots in the phase diagram in Fig.\ \ref{FIGS1}d with $T_c = 0.1 K$ and $T_c = 3 K$). These results suggest that the chiral $f$-wave is the phase most likely favored in the twisted double-layer and lend strong support to our assumption of a dominant $f$-wave interaction made in the main text.

\subsection*{C. Inter-layer Josephson effect near maximal twist as a novel experimental probe for pairing symmetry}

While the phase diagram in Fig.\ \ref{FIGS1}d strongly supports the scenario of a dominant chiral $f$-wave pairing in the twisted double-layer, one cannot completely rule out the possibility of a dominant chiral $p$-wave interaction $U_{0,p} \gg U_{0,f}$ under which the chiral $p$-wave phase is favored. Indeed, the possibility of dominant chiral $p$-wave interactions mediated by fluctuations in inter-valley coherence order is suggested by a recent study to account for the spin-triplet SC2 phase of RTG \cite{Chatterjee}. We note that the exact pairing symmetry of the SC2 phase of RTG, whether it be $f$-wave or chiral $p$-wave, is bound to details of the microscopic mechanisms being considered and thus nearly impossible to be fully settled by theory alone. While debates are going on and further experimental evidence is clearly desired, we point out here that inter-layer Josephson effects at the maximal twist of $\theta = 30^{\circ}$ provides a novel and unambiguous way to probe the pairing symmetry of the SC2 phase in RTG.

\begin{figure}
\centering
\includegraphics[width=0.9\textwidth]{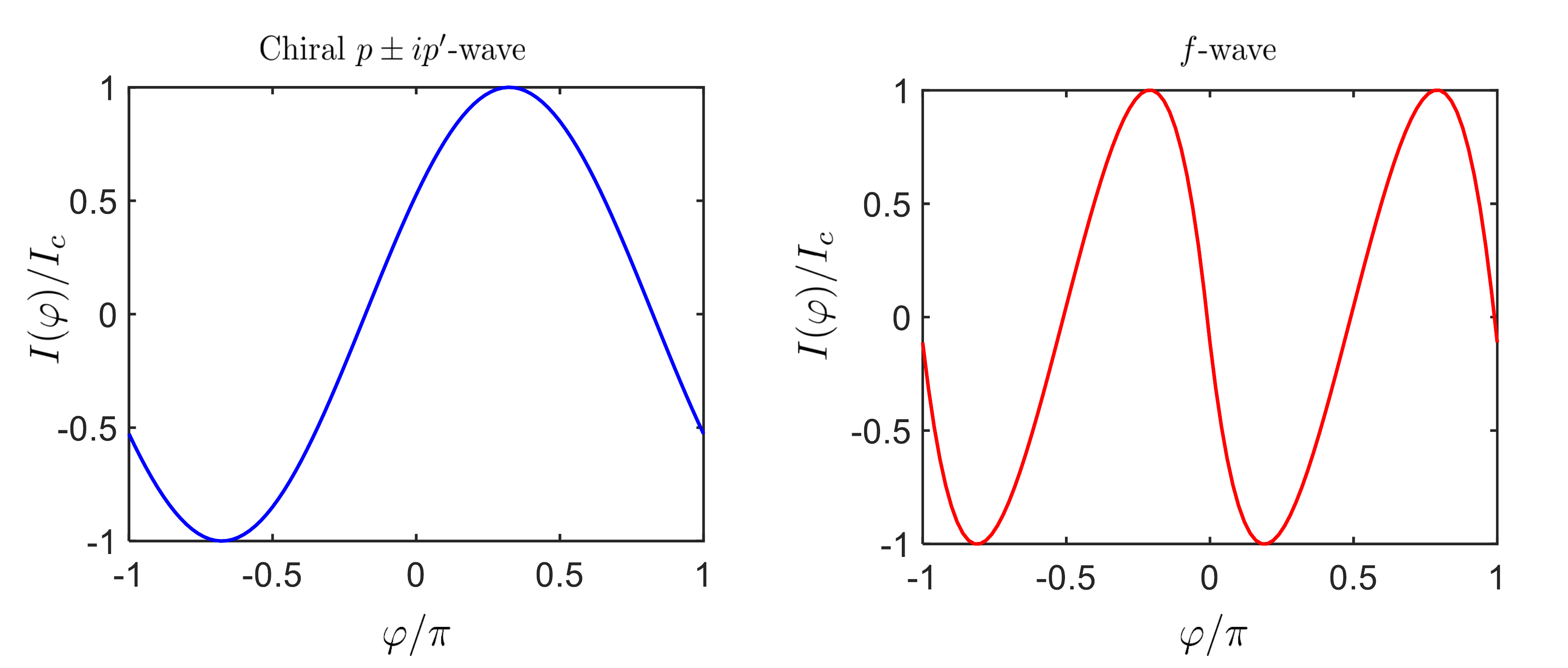}
\caption{Inter-layer Josephson currents $I_J(\varphi) = (2e/\hbar) \partial \mathcal{F}/\partial \varphi$ through a maximally twisted double-layer at $\theta = 30^{\circ}$ under chiral $p$-wave (a) and $f$-wave (b) interactions in each individual layer. The Josephson current-phase relations show a usual $2\pi$ periodicity under the $p$-wave interaction, while exhibit an anomalous $\pi$-periodicity under the $f$-wave interaction. }
\label{FIGS2}
\end{figure}

It is important to note that the $\varphi$-dependence of the free energy $\mathcal{F}(\varphi)$ under chiral $p$-wave interaction in Fig.\ \ref{FIGS1}c has only one minimum at $\varphi = -\pi/6$ and an overall periodicity of $T_{\varphi} = 2\pi$. In contrast, as we discuss in details in the main text, at the maximal twist of $\theta = 30^{\circ}$, $\mathcal{F}(\varphi)$ under $f$-wave interaction has a two-minima structure at $\varphi = \pm \pi/2$ with periodicity of $T_{\varphi} = \pi$. As such, the inter-layer Josephson current $I_J(\varphi) = (2e/\hbar) \partial \mathcal{F}/\partial \varphi$ under the two different chiral $p$-wave and $f$-wave pairing interactions would also exhibit different periodicities as shown in Fig.\ \ref{FIGS2}. These contrasting features in inter-layer Josephson currents can thus be used to distinguish the two different pairing symmetries, namely, chiral $p$-wave and $f$-wave, that are under current debates.

We further point out that the anomalous $\pi$-periodic inter-layer Josephson effect at the maximal twist $\theta = 30^{\circ}$ relies on the unique symmetry properties of the two orthogonal $f$-wave components formed at $\theta = 30^{\circ}$, which cannot occur for any other pairing symmetries such as $s$-wave, $p$-wave, or $d$-wave symmetries. Thus, the $\pi$-periodic Josephson effect serves as a distinctive experimental signature for $f$-wave pairing symmetry in each constituent layer, which applies in principle to all superconductors with $f$-wave symmetries. In particular, while $f$-wave pairing symmetry was proposed for superconductivity in doped ZrNCl and supported by the unusual doping dependence of the superconducting gap structure found in early specific heat measurements \cite{Iwasa0}, there is no direct experimental evidence on the exact pairing symmetry of the superconducting state. Our results on the $\pi$-periodic current through maximally twisted Josephson junction provides a novel and reliable way to test the proposed $f$-wave pairing symmetry of superconducting ZrNCl. 

\subsection*{D. Non-Abelian topological superconductivity in twisted double-layer under dominant chiral $p$-wave interactions}

\begin{figure}
\centering
\includegraphics[width=0.9\textwidth]{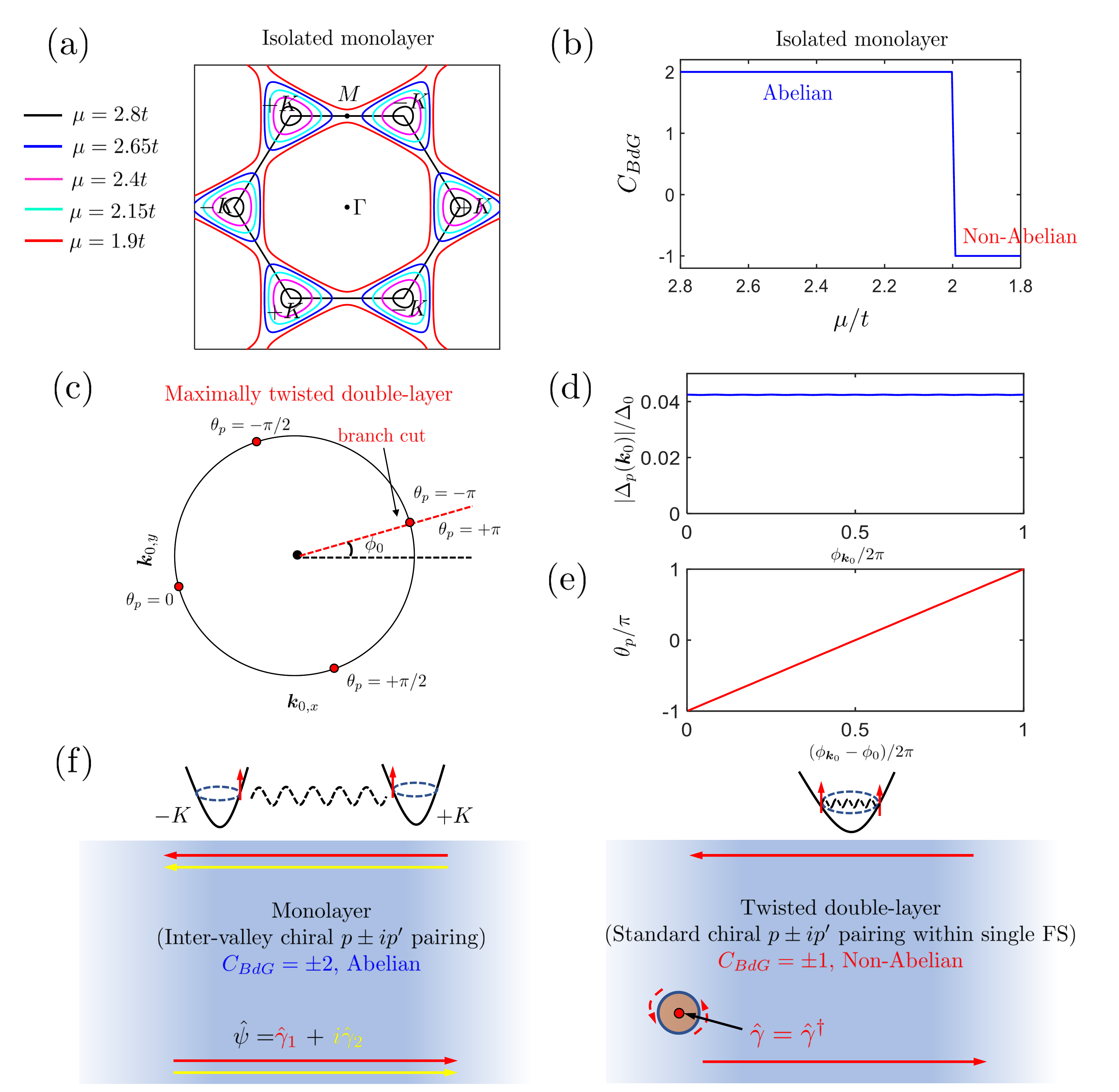}
\caption{(a) Fermi surface (FS) evolution in an isolated monolayer upon increasing doping from $\mu = 2.8t$ to $\mu = 1.9t$, during which a Lifshiftz transition happens at $\mu = 2t$ where the two disconnected pockets around $\pm K$ merge at the time-reversal-invariant $M$-point in the Brillouin zone. (b) Chern number $C_{BdG}$ in an isolated monolayer under chiral $p+ ip'$-pairing on a triangular lattice (Eq.\ \ref{eq:basisfuncP}) as a function of chemical potential $\mu$. $C_{BdG}$ stays at $+2$ as FSs around $\pm K$ remains disconnected. A topological phase transition happens right at the Lifshiftz transition point $\mu = 2t$ where the two disconnected pockets coalesce into a single connected FS around $\Gamma$ as shown in (a). The system then becomes a standard $p + i p'$ superconductor which supports non-Abelian MZMs. (c) Schematic of phase winding of the chiral $p$-wave gap function $\Delta_{p, +}(\bm{k}_0)$ defined in Eq.\ \ref{eq:PWaveGapFunc} along the single connected Fermi surface at $\mu = 2.9t$ (red solid line in Fig.\ \ref{FIGS1}b) in the maximally twisted double-layer. $\phi_0$ denotes the azimuthal angle in $\bm{k}_0$-space along which the branch cut of the complex logarithmic function is defined. (d)-(e) Amplitude $|\Delta_{p, +}(\bm{k}_0)|$ (d) and phase $\theta_p$ (e) of $\Delta_{p, +}(\bm{k}_0) = |\Delta_{p, +}(\bm{k}_0)| e^{i \theta_p(\bm{k}_0)}$ along the single connected FS in (c) with $\varphi_{min} = -\pi/6$ (Eq.\ \ref{eq:PWaveGapFunc}). Clearly, $|\Delta_{p, +}(\bm{k}_0)|$ remains almost a constant along the Fermi surface while $\theta_p$ winds by $2\pi$, revealing the standard chiral $p+ip'$ symmetry of $\Delta_{p, +}(\bm{k}_0)$ formed within a single connected Fermi surface in the twisted double-layer. (f) Schematics of the differences in topological nature between a monolayer RTG under inter-valley chiral $p + ip'$ pairing (left panel) and the standard chiral $p + ip'$ pairing formed within a single Fermi surface in the twisted double-layer (right panel). Due to the two disconnected $\pm K$-pockets, the chiral $p + ip'$ phase in an isolated monolayer is characterized by Chern number $C_{BdG} = +2$ as shown in (b), which corresponds to a superposition of two different Majorana modes $\hat{\gamma}_{1}$ and $\hat{\gamma}_{2}$ (\textit{c.f.} analysis for small-angle twisted graphene/TMDs in Supplementary Note 1). These two Majorana modes always combine into a single fermionic mode $\hat{\psi}$ that obeys the usual fermionic statistics.}
\label{FIGS3}
\end{figure}

As discussed in detail in the main text, under dominant $f$-wave interactions, each layer of RTG in the SC2 phase is topologically trivial, while the maximal twist of $\theta = 30^{\circ}$ between two layers turns the system into a chiral $f \pm if'$ topological superconductor hosting non-Abelian Majorana zero modes. Here, we study the topological property of the twisted double-layer RTG under dominant chiral $p$-wave interactions suggested by Refs.\cite{Chatterjee, Berg}. Importantly, we point out that while each layer of RTG would already be topologically nontrivial under chiral $p$-wave interactions, the two-valley structure in the Fermi surface of a monolayer (shown schematically in Fig.\ \ref{FIGS1}a) would necessarily imply that the system only supports \textbf{Abelian} excitations. On the other hand, we show that upon stacking the two layers under dominant chiral $p$-wave interactions at maximal twist, the system becomes the standard spinless chiral $p \pm ip'$ superconductor when a \textbf{single} connected Fermi surface forms (see red solid line in Fig.\ \ref{FIGS1}b and Fig.\ 2 in the main text). Thus, the twisted double-layer system is yet again a \textbf{non-Abelian} topological superconductor. We further explain why the scheme of maximal twist considered in this work provides arguably the most effective way to turn an inter-valley-paired chiral $p$-wave superconductor into a non-Abelian topological superconductor. Without loss of generality, we consider the case of positive chirality $c = +$ in the following discussions.

To examine the topological property of the unusual inter-valley-paired chiral $p + ip'$-wave pairing (Fig.\ \ref{FIGS1}a), we note that the effective Bogoliubov-de-Gennes (BdG) Hamiltonian for a small momentum $\bm{p} \equiv \bm{k} - \xi \bm{K} (\xi = \pm)$ displaced from $\pm K$-points can be written in the Nambu basis $\Psi_{\bm{p}} = (c_{+K}(\bm{p}), c_{-K}(\bm{p}), c^{\dagger}_{+K}(-\bm{p}), c^{\dagger}_{-K}(-\bm{p}))^T$ as $\mathcal{H}_{BdG, p} = \sum_{\bm{p}} \Psi^{\dagger}_{\bm{p}} H_0(\bm{p}) \Psi_{\bm{p}}$, where
\begin{eqnarray}\label{eq:effectiveHChiralP}
H_0(\bm{p}) = 
\begin{pmatrix}
\xi_{K}(\bm{p}) & 0 & 0 & \Delta_{p} (p_x + i p_y)\\
0 & \xi_{-K}(\bm{p})&  \Delta_{p} (p_x + i p_y) & 0\\
0 &  \Delta_{p} (p_x - i p_y) & -\xi_{K}(-\bm{p}) & 0\\
\Delta_{p} (p_x - i p_y) & 0 & 0 & -\xi_{-K}(-\bm{p})
\end{pmatrix}.
\end{eqnarray}
Note that a momentum $\bm{k} = \bm{p} + \bm{K}$ near valley $+K$ is paired with momentum $-\bm{k} = -\bm{p} - \bm{K}$ near $-K$. In the small $\bm{p}$ limit, the chiral $p$-wave function (Eq.\ \ref{eq:basisfuncP}) takes the same form of $p_x + ip_y$ around $\pm K$ as we discussed in Subsection B above. In this case, the phase of the chiral $p$-wave order parameter winds by $2\pi$ around both $+K$ and $-K$ valleys, and one expects the total Chern number (defined in the BdG formalism) of the system to be $C_{BdG} = +2$ instead of $C_{BdG} = +1$ as in a standard chiral $p$-wave superconductor with a single Fermi surface around the $\Gamma$-point. This result is confirmed numerically in Fig.\ \ref{FIGS3}a-b by the triangular lattice model for an isolated monolayer. In fact, as long as the two valleys remain disconnected, $C_{BdG}$ stays at $+2$ (Fig.\ \ref{FIGS3}b) until the two pockets assess the $M$-point of the Brillouin zone and get reconnected into a single FS at $\mu = 2t$ (Fig.\ \ref{FIGS3}a). At the Lifshiftz transition point $\mu = 2t$, the system undergoes a topological transition from $C_{BdG} = +2$ to $C_{BdG} = -1$ (Fig.\ \ref{FIGS3}b).

The reason that the inter-valley-paired chiral $p$-wave phase with $C_{BdG} = +2$ can only support Abelian excitations can be understood easily by following a similar line of reasoning we presented for the small-angle twisted graphene/TMDs in Supplementary Note 1 (note that the form of the effective BdG Hamiltonian in Eq.\ \ref{eq:effectiveHChiralP} is the same as Eq.\ \ref{eq:BdGSmallAngle}). Due to the inter-valley pairing nature, the chiral edge modes of the $C_{BdG} = +2$ phase are always formed by a superposition of two different Majorana modes (depicted schematically in the left panel of Fig.\ \ref{FIGS3}f) which combine to form a usual Abelian fermionic mode. This reveals that even if the spin-triplet SC2 phase of RTG turned out to be a topological chiral $p$-wave pairing phase, it still cannot host non-Abelian excitations.

Moreover, the results in Fig.\ \ref{FIGS3}a-b suggest that to create a non-Abelian Majorana mode in an isolated monolayer, the only possible way is to increase the doping level such that the system becomes a standard chiral $p$-wave superconductor. However, this requires the Fermi momentum $k_F$ measured from $\pm K$ to match the distance between $M$-point and $\pm K$-points: $k_F a = 2\pi/3 \simeq 2$. As we discussed in the Experimental relevance and signatures section of the main text, the relatively low doping level accessible in experiments on RTG corresponds to a Fermi momentum $k_F a \sim 0.1$. As $k_F \propto \sqrt{n}$ ($n$: carrier density) in 2D, this implies the doping level needs to be at least two orders of magnitude higher than $n \sim 10^{12} cm^{-2}$ achieved in current experiments. Thus, it is experimentally challenging to realize non-Abelian excitations within an isolated monolayer even if the system has the chiral $p$-wave pairing symmetry.

On the other hand, as we demonstrated in Fig.\ 2 of the main text, thanks to the band folding effects introduced by the large-angle moir\'{e} physics at maximal twist, the Fermi surface in the twisted double-layer has already become connected at the experimentally relevant doping level with $k_F a \sim 0.1$ in each individual layer. This establishes the crucial prerequisite for non-Abelian topological superconductivity. As we demonstrated in Subsection B, under dominant $p$-wave interactions, the twisted double-layer favors the configuration with same chiralities on two layers with an energetically favored $\varphi_{min} = -\pi/6$ (Fig.\ \ref{FIGS1}c). To unveil the topological nature of this chiral $p$-wave phase in the twisted double-layer, we consider the single connected Fermi surface at $\mu = 2.9 t$ (shown schematically in Fig.\ \ref{FIGS3}c), and plot the magnitude (Fig.\ \ref{FIGS3}d) and phase angle $\theta_p$ (Fig.\ \ref{FIGS3}e) of the pairing $\Delta_{p,+}(\bm{k}_0) $ in Eq.\ \ref{eq:PWaveGapFunc} along the Fermi surface with $\varphi_{min} = -\pi/6$ (Fig.\ \ref{FIGS1}c). The magnitude of $\Delta_{p,+}(\bm{k}_0)$ remains almost a constant along the Fermi surface, while its phase $\theta_p$ winds by $2\pi$ around the $\Gamma$-point. This confirms that under dominant $p$-wave interactions the twisted double-layer becomes a standard spinless chiral $p \pm ip'$-wave superconductor with a single connected Fermi surface ($C_{BdG} = \pm 1)$, which is known to host a non-Abelian Majorana mode at its vortex core (shown schematically in \ref{FIGS3}f). 

The results above lead us to the conclude that the maximal twist between two layers of RTG in the spin-triplet SC2 phase, regardless of whether its pairing symmetry being $f$-wave \cite{Chou1, Chou2, Roy} or chiral $p$-wave \cite{Chatterjee, Berg, Levitov}, turns the composite system from an Abelian superconductor into a non-Abelian topological superconductor. This further fortifies the connection between the large-angle moir\'{e} physics and non-Abelian TSC as we discussed in the Conclusions and outlook section of the main text.

\section*{Supplementary Note 9: Effect of inter-layer spatial displacement} \label{AppendixI}

As shown in Refs. \cite{Wu}, inter-layer spatial displacement $\bm{d}_0$ can induce nonzero phases in the inter-layer tunneling terms. As we mentioned in subsection A of the Methods section in the main text, these phases due to nonzero $\bm{d}_0$ do not enter the total phase of the Cooper pairs and thus having no effect on the inter-layer phase difference $\varphi_p$ between the superconducting order parameters $\psi_{1,p}$ and $\psi_{2,p}$. 

To demonstrate this result explicitly, we derive the general form of the inter-layer tunneling term for the dual-momentum space sites in the DMSTB model with $\bm{d}_0 \neq 0$. Without loss of generality, we assume that the rotation center of the twisted double-layer coincides with an atomic site in layer 1, and the nearest atomic site in layer 2 is displaced by $\bm{d}_0$. The atomic sites in layer 1 and layer 2 are given by $\bm{R}$ and $\tilde{\bm{R}} + \bm{d}_0$, respectively, and we denote the Wannier orbitals as $\ket{\bm{R}, 1}$ and $\ket{\bm{\tilde{R}} +\bm{d}_0, 2}$. For a given pair of Bloch states with $\bm{k}_m = \bm{k}_0 + \tilde{\bm{G}}_m$ in layer 1 and $\bm{\tilde{k}}_n = \bm{k}_0 + \bm{G}_n$ in layer 2 (see definition of $\bm{k}_m, \bm{\tilde{k}}_n$ in Methods section and Supplementary Note 3), the tunneling amplitude, which is now a function of $\bm{d}_0$, can be written as:
\begin{eqnarray}
t_{\perp}(\bm{k}_m,\bm{\tilde{k}}_n)(\bm{d}_0) &=& \braket{\bm{k}_m, 1| \mathcal{H}_{T} | \bm{\tilde{k}}_n, 2} \\\nonumber
&=& \frac{1}{N} \sum_{\bm{R},\bm{\tilde{R}}} e^{-i \bm{k}_m \cdot \bm{R} } e^{i \bm{\tilde{k}}_n \cdot (\bm{\tilde{R}} + \bm{d}_0) } \braket{\bm{R}, 1| \mathcal{H}_{T} | \bm{\tilde{R}} + \bm{d}_0 , 2} \\\nonumber
(\text{two-center approximation) }&=&  -\frac{1}{N} \sum_{\bm{R},\bm{\tilde{R}}} e^{-i \bm{k}_m \cdot \bm{R} } e^{i \bm{\tilde{k}}_n \cdot (\bm{\tilde{R}} + \bm{d}_0) } t_{\perp} (\bm{R} - \bm{\tilde{R}} -\bm{d}_0) \\\nonumber
&=&  -\frac{1}{N} \sum_{\bm{R},\bm{\tilde{R}},\bm{q}} e^{-i \bm{k}_m \cdot \bm{R} } e^{i \bm{\tilde{k}}_n \cdot (\bm{\tilde{R}} + \bm{d}_0) } \frac{t_{\perp} (\bm{q})}{A} e^{i \bm{q}\cdot (\bm{R} - \bm{\tilde{R}} -\bm{d}_0) } \\\nonumber
&=&  -\frac{1}{N} \sum_{\bm{q}} \frac{t_{\perp} (\bm{q})}{A}e^{i (\bm{\tilde{k}}_n -\bm{q}) \cdot \bm{d}_0} \big(\sum_{\bm{R}} e^{i (\bm{q}-\bm{k}_m) \cdot \bm{R} }\big)  \big(\sum_{\bm{\tilde{R}}} e^{i (\bm{\tilde{k}}_n -\bm{q}) \cdot \bm{\tilde{R}}  } \big) \\\nonumber
&=&  - N\sum_{\bm{q}} \frac{t_{\perp} (\bm{q})}{A} e^{i (\bm{\tilde{k}}_n -\bm{q}) \cdot \bm{d}_0} \big(\sum_{\bm{G}} \delta_{\bm{q}, \bm{k}_m + \bm{G}}\big)  \big(\sum_{\bm{\tilde{G}}} \delta_{\bm{q}, \bm{\tilde{k}}_n +\bm{\tilde{G}}} \big) \\\nonumber 
&=&  - \frac{t_{\perp} (\bm{k}_0 + \bm{\tilde{G}}_m +\bm{G}_n)}{\Omega} e^{-i \bm{\tilde{G}}_m  \cdot \bm{d}_0} 
\end{eqnarray}
Here, $t_{\perp}(\bm{q}) = \int d^2\bm{r} t_{\perp}(\bm{r}) e^{-i\bm{q}\cdot\bm{r}}$ is the Fourier transform of $t_{\perp}(\bm{r})$, $A$ is the total area of the system, $\Omega = A/N$ is the area of the primitive unit cell. In the last two steps of the derivation above, we made use of the following results: (i) $\sum_{\bm{R}} e^{i (\bm{q}-\bm{k}) \cdot \bm{R} } =N \sum_{\bm{G}} \delta_{\bm{q}, \bm{k}+\bm{G}}$, and (ii) the sums over $\bm{G}, \bm{\tilde{G}}$ are non-vanishing only for $\bm{q} = \bm{k}_m + \bm{G} = \bm{\tilde{k}}_n + \bm{\tilde{G}}$, which holds only for $\bm{G} = \bm{G}_n$ and $\bm{\tilde{G}} = \bm{\tilde{G}}_m$ (see Supplementary Note 3) and implies $\bm{q} = \bm{k}_0 + \bm{\tilde{G}}_m +\bm{G}_n$. It is evident that the nonzero $\bm{d}_0$ introduces an extra phase of $e^{-i \theta_m}$ with $\theta_m = \bm{\tilde{G}}_m \cdot \bm{d}_0$ to the inter-layer terms as compared to Eq.\ 13 in the main text, which amounts to a transformation $c^{\dagger}_1(\bm{k}_m) \mapsto e^{-i \theta_m}c^{\dagger}_1(\bm{k}_m)$ for the fermionic operator that creates the electron on the dual-space lattice sites $\bm{k}_m$ in layer 1, $m = 0,1,2,...,12$. Notably, due to the spinless time-reversal symmetry $\mathcal{T}' \equiv \mathcal{K}$, one must have $c^{\dagger}_{1}(-\bm{k}_m) = \mathcal{T}'c^{\dagger}_{1}(\bm{k}_m)\mathcal{T}'^{-1} \rightarrow e^{i \theta_{m}}c^{\dagger}_{1}(-\bm{k}_m)$.

Due to the extra $\theta_m$, the fermionic operators in the band basis in Eq.\ 14 of the main text (section B of the Methods section) need to be rewritten as:
\begin{equation}
a^{\dagger}_p(\bm{k}_0) = \sum_{m} u_{pm}(\bm{k}_0) e^{-i \theta_{m}}c^{\dagger}(\bm{k}_m) + \sum_{n} u_{pn}(\bm{k}_0) c^{\dagger}(\bm{\tilde{k}}_n).
\end{equation}
It is worth to note that the phase factor $e^{-i \theta_m}$ can be incorporated in the new form factors in Eq.\ 16: $\tilde{\Lambda}_{m,p}(\bm{k}_0) \equiv e^{i \theta_m} u^{\ast}_{m,p}(\bm{k}_0) e^{-i \theta_m} u^{\ast}_{-m,p}(-\bm{k}_0) = u^{\ast}_{m,p}(\bm{k}_0) u^{\ast}_{-m,p}(-\bm{k}_0) = \Lambda_{m,p}(\bm{k}_0)$. Therefore, the mean-field gap equations in Eq.\ 14-16 remain unchanged even by including the phase factors $\theta_n$. This reveals that the extra factor of $e^{-i \theta_m}$ due to nonzero $\bm{d}_0$ does not affect the total phase of the Cooper pairs.


\begin{thebibliography}{99}
\bibitem{Note} The form factors $\Lambda_{m,p}(\bm{k}_0)$ and $\Lambda_{n,p}(\bm{k}_0)$ involved in $f_{1,p}(\bm{k}_0)$ and $f_{2,p}(\bm{k}_0)$ are gauge-dependent quantities in general. Throughout our calculations, we choose a smooth gauge such that all pairing functions in the band basis are smoothly defined.
\end{thebibliography}
\end{document}